\documentclass[aps,twocolumn,groupedaddress,floatfix,nofootinbib]{revtex4}

\begin{document}

\title{Cascades and Collapses, Great Walls and Forbidden Cities:\\
        Infinite Towers of Metastable Vacua in Supersymmetric Field Theories}
\author{Keith R. Dienes and Brooks Thomas}
\affiliation{Department of Physics, University of Arizona, Tucson, AZ  85721 USA}

\begin{abstract}
    In this paper, we present a series of supersymmetric models 
    exhibiting an entirely new vacuum structure:  
    towers of metastable vacua with higher and higher energies.  
    As the number of vacua grows towards infinity,
    the energy of the highest vacuum remains fixed while  
    the energy of the true ground state tends towards zero. 
    We study the instanton-induced tunneling dynamics associated with
    such vacuum towers, and find that many distinct decay patterns 
    along the tower are possible:  these include not only 
    regions of vacua experiencing direct collapses and/or tumbling cascades,
    but also other regions of vacua whose stability is protected by ``great walls'' 
    as well as regions of vacua populating ``forbidden cities'' 
    into which tunnelling cannot occur. 
    We also discuss possible applications 
    of this setup for the cosmological-constant problem,
    for studies of the string landscape, for supersymmetry breaking, and
    for $Z'$ phenomenology.
    Finally, we point out that a limiting case of our setup
    yields theories with yet another new vacuum structure:  
    infinite numbers of degenerate vacua.
    As a result, the true ground states of such theories are Bloch waves,
    with energy eigenvalues approximating a continuum and giving rise to 
    a vacuum ``band'' structure.  
\end{abstract}

\pacs{12.60.Jv,11.27.+d,14.70.Pw,11.25.Mj}

\maketitle

%========================================================================
%          KEYSROKE-SAVING MACROS, nothing complicated 
%========================================================================
\newcommand{\newc}{\newcommand}
\newc{\gsim}{\lower.7ex\hbox{$\;\stackrel{\textstyle>}{\sim}\;$}}
\newc{\lsim}{\lower.7ex\hbox{$\;\stackrel{\textstyle<}{\sim}\;$}}

\def\vac#1{{\bf \{{#1}\}}}

\def\beq{\begin{equation}}
\def\eeq{\end{equation}}
\def\beqn{\begin{eqnarray}}
\def\eeqn{\end{eqnarray}}
\def\calM{{\cal M}}
\def\calV{{\cal V}}
\def\calF{{\cal F}}
\def\half{{\textstyle{1\over 2}}}
\def\quarter{{\textstyle{1\over 4}}}
\def\ie{{\it i.e.}\/}
\def\eg{{\it e.g.}\/}
\def\etc{{\it etc}.\/}

%     The following macros are to create the "blackboard bold"
%     characters for "R" (set of real numbers),
%     "C" (set of complex numbers), and "Q" (set of rational numbers).

\def\inbar{\,\vrule height1.5ex width.4pt depth0pt}
\def\IR{\relax{\rm I\kern-.18em R}}
 \font\cmss=cmss10 \font\cmsss=cmss10 at 7pt
\def\IQ{\relax{\rm I\kern-.18em Q}}
\def\IZ{\relax\ifmmode\mathchoice
 {\hbox{\cmss Z\kern-.4em Z}}{\hbox{\cmss Z\kern-.4em Z}}
 {\lower.9pt\hbox{\cmsss Z\kern-.4em Z}}
 {\lower1.2pt\hbox{\cmsss Z\kern-.4em Z}}\else{\cmss Z\kern-.4em Z}\fi}
%========================================================================

\input epsf

%========================================================================
%========================================================================
%               MAIN TEXT BEGINS HERE
%========================================================================

%========================================================================

\section{Introduction\label{sec:Introduction}}

%  \indent
%  There's a whole lotta vacuum goin' on! 

The vacuum structure of any physical theory plays a significant 
and often crucial role in determining the physical properties of that theory. 
Indeed, critical issues such as the presence or absence of spontaneous
symmetry breaking often depend entirely on the vacuum structure
of the theory in question.

Likewise, it may happen that a given model contains
not only a true ground state, but also a metastable vacuum
state above it.  Such models are also of considerable interest,
for even when the true ground
state preserves the apparent symmetries of a model,
the physical properties associated with the metastable
vacua can often differ markedly from those of the ground
state. In such situations, the resulting phenomenology
of the model might be determined by the properties of
a metastable vacuum rather than by those of the true
ground state.

In recent years,
models containing metastable vacua have captured
considered attention.
This is true for a variety of reasons.
For example, metastable vacua can serve as 
a tool for breaking supersymmetry~\cite{preISS,ISS} in the context 
of certain supersymmetric non-Abelian gauge theories
which are otherwise known to contain supersymmetric ground states.
In addition, theories with large numbers of vacua 
have also been exploited in various ways as a means of 
addressing the cosmological-constant problem~\cite{BoussoPolchinski,banks,Gordy,tye} 
and obtaining de~Sitter vacua in string compactifications~\cite{Gary}.
Furthermore, the possibility of phase transitions in 
theories with multiple (meta)stable vacua leads 
to a number of implications for cosmology.  

Such ideas provide ample motivation to investigate whether there
might exist 
relatively simple field theories which give rise to additional,
heretofore-unexplored vacuum structures.
If so, such structures could potentially provide new ways
of addressing a variety of unsolved questions about the universe we inhabit.   

In this and a subsequent companion paper~\cite{toappear}, we will demonstrate that two new
non-trivial vacuum structures are possible in relatively simple supersymmetric
field theories.
Moreover, as we shall see, the models which give rise to 
these non-trivial vacuum structures are not esoteric;
they are, in fact, simple generalizations of $U(1)$ quiver gauge theories. 
\begin{itemize}
\item First, we shall demonstrate through an explicit construction
     that certain supersymmetric field theories can give rise 
     to large (and even infinite) towers of metastable vacua with 
    higher and higher energies.  
     The emergence and analysis of this vacuum structure will be the primary focus 
    of the present work. 
    As we shall see, as the number of vacua grows towards infinity in such models,
    the energy of the highest vacuum remains fixed while  
    the energy of the true ground state tends towards zero. 
    We shall study the instanton-induced tunneling dynamics associated with
    such vacuum towers, and find that many distinct decay patterns 
    along the tower are possible:  these include not only 
    regions of vacua experiencing direct collapses and/or tumbling cascades,
    but also other regions of vacua whose stability is protected by ``great walls'' 
    as well as regions of vacua populating ``forbidden cities'' 
    into which tunnelling cannot occur. 
    Furthermore, as we shall see,
    these vacua are phenomenologically distinct from one another in terms 
    of their mass spectra and effective interactions.
\item Second, we shall also show that there exists a limiting case of the
     above construction in which all of these infinite metastable vacua become degenerate,
     and in which there emerges a shift symmetry relating one vacuum to the next.
     As a result, the true ground states of such theories are nothing but Bloch waves
     across these degenerate ground states,
     with energy eigenvalues approximating a continuum and giving rise to 
     a vacuum ``band'' structure.  
     In this paper, we will merely sketch how such a vacuum structure emerges;
     the complete analysis of such a structure will be the subject of a subsequent
    companion paper~\cite{toappear}.
\end{itemize}

This paper is organized as follows.  In Sect.~II, we present 
the framework on which our model is based.  As we shall see,
our model is essentially nothing more than an Abelian quiver gauge theory, 
expanded to allow kinetic mixing between the various $U(1)$ factors.  
We shall then proceed to discuss the corresponding vacuum structure 
which emerges from this framework,
including all stable vacua and all saddle-point barriers between them.  
We shall also discuss radiative corrections to this vacuum structure, and demonstrate 
that these corrections can be kept under control.  
In Sect.~III, we shall then discuss the decay dynamics along these metastable vacuum towers,
and examine the different sorts of instanton-induced tunneling
decay patterns which are possible.
In Sect.~IV, we then analyze the particle spectra in each vacuum of the tower,
and demonstrate how these spectra evolve as our system tumbles down the vacuum tower.
In Sect.~V we shift gears briefly, and consider the limiting case of 
our scenario in which our infinite
towers of metastable vacua become an infinite series of degenerate ground states.
Thus, in this limit, the true ground states of such theories are Bloch waves. 
Finally, in Sect.~VI, we enumerate the potential physical applications of 
our vacuum towers, including possible new ideas
for the cosmological-constant problem,
    for studies of the string landscape, and for $Z'$ phenomenology.

We emphasize that our primary goal in both papers is the demonstration that such
non-trivial vacuum structures can emerge in relatively simple supersymmetric field
theories.  Although there exist numerous implications and applications of these ideas 
(some of which will be discussed in Sect.~VI),  
our primary goal in these papers will be the study of the emergence and properties of 
these vacuum structures themselves.

%=====================================================================================================

\section{The General Framework\label{sec:framework}}

We begin by presenting our series of supersymmetric 
models which give rise to infinite towers of metastable vacua.  
Specifically, for each $N>1$, we shall present
a model which contains not only
a stable vacuum ground state but also a tower of
$N-2$ metastable vacua above it.
Our model consists of 
$N$ different $U(1)$ gauge group factors, denoted $U(1)_a$ ($a=1,...,N$),
as well as $N+1$ different chiral superfields,
denoted $\Phi_i$ ($i=1,...,N+1$). 
These superfields carry the $U(1)$ charge assignments
shown in Table~\ref{tab:U1Chgs}, and follow the well-known orbifolded ``moose'' pattern
wherein each field $\Phi_i$ with $2\leq i\leq N$ simultaneously carries both
a positive unit charge under $U(1)_{i-1}$
and a negative unit charge under $U(1)_{i}$. 
By contrast, the fields $\Phi_1$ and $\Phi_{N+1}$ sit at the orbifold endpoints
of the moose, and are charged only under the corresponding 
endpoint gauge groups respectively.
We shall assume that each $U(1)_a$ gauge field has a corresponding gauge coupling $g_a$,
and for simplicity we shall further assume that $g_a\equiv g$ for all $a$.
Note that 
the only non-vanishing gauge anomalies inherent in this charge configuration are mixed $U(1)$
anomalies proportional to $\sum_i^{N+1} Q_{ai}^2 Q_{bi}$, which can 
be canceled by the variant of the
Green-Schwarz mechanism~\cite{GreenSchwarz} discussed in Ref.~\cite{DudasDecon}.

%==================== TABLE BEGINS HERE ====================================
\begin{table}[t!]
\begin{center}
\begin{tabular}{||c|ccccccc||}
       \hline
       \hline
      $~$ & $U(1)_1$ & $U(1)_2$ & $U(1)_3$ & $U(1)_4$ & \ldots & $U(1)_{N-1}$ & $U(1)_{N}$ \\ 
       \hline
     $\Phi_{1}$ & $-1$ & $0$ & $0$ & $0$ & $\ldots$ & $0$ & $0$ \\
     $\Phi_{2}$ & $+1$ & $-1$ & $0$ & $0$ & $\ldots$ & $0$ & $0$ \\
     $\Phi_{3}$ & $0$ & $+1$ & $-1$ & $0$ & $\ldots$ & $0$ & $0$ \\
     $\Phi_{4}$ & $0$ & $0$ & $+1$ & $-1$ & $\ldots$ & $0$ & $0$ \\
     $\vdots$ & $\vdots$ & $\vdots$& $\vdots$ & $\vdots$ & $\ddots$ & $\vdots$ & $\vdots$ \\
     $\Phi_{N-1}$ & $0$ & $0$ & $0$ & $0$ & $\ldots$ & $-1$ & $0$ \\ 
     $\Phi_{N}$ & $0$ & $0$ & $0$ & $0$ & $\ldots$ & $+1$ & $-1$ \\
     $\Phi_{N+1}$ & $0$ & $0$ & $0$ & $0$ & $\ldots$ & $0$ & $+1$ \\
       \hline
       \hline
\end{tabular}
\end{center}
\caption{The field content and charge assignments for the chiral superfields in the 
models under consideration.}
\label{tab:U1Chgs}
\end{table}
%==================== TABLE ENDS HERE ====================================

To this core model we then add three critical ingredients, each of which
is vital for the emergence of our metastable vacuum towers. 
First, given the field content of each model, we see that the most general superpotential 
that can be formed in each case consists of a single Wilson-line operator
\begin{equation}
            W~=~\lambda\, \prod_{i=1}^{N+1}\Phi_i~.
\label{eq:W}
\end{equation}
We shall therefore assume that this operator is turned on for each value of $N$.
Note that the coupling $\lambda$ has mass dimension $2-N$ and is 
therefore non-renormalizable for all $N>2$.

Our second and third ingredients both exploit the Abelian nature of our gauge groups.
The second ingredient is to introduce non-zero Fayet-Iliopoulos terms $\xi_1$ and $\xi_N$
for the ``endpoint'' gauge groups $U(1)_1$ and $U(1)_N$ respectively.
While all of our $U(1)_a$ gauge groups could in principle have corresponding 
non-zero Fayet-Iliopoulos terms $\xi_a$,
we shall see that turning on only $\xi_1$ and $\xi_N$ 
will be sufficient for our purposes.  Indeed, we shall prune
our model further by taking $\xi_1=\xi_N\equiv \xi$.

Finally, our third ingredient is a simple one:  kinetic mixing~\cite{Holdom}.  
It is well-known that the field strength tensor $F^{\mu\nu}$ for an 
Abelian gauge group is gauge invariant by itself.  Thus, in
a theory involving multiple $U(1)$ gauge groups, nothing forbids terms 
proportional to $F_a^{\mu\nu}F_{b,\mu\nu}$ from appearing as kinetic terms in the Lagrangian, 
where $a\not= b$.  Similarly, in a supersymmetric model, mixing between the 
field-strength superfields $W^a_{\alpha}$ 
is permitted,
whereupon the gauge-kinetic part of the Lagrangian may take the generic form~\cite{SUSYKineticMixing}
\begin{equation}
    {\cal L}~\ni~\frac{1}{32}\int d^2\theta ~~ W_{a\alpha} \, X_{ab} \, W_b^{\alpha}~
\label{eq:KMixLagrange}
\end{equation}
with a general (symmetric) kinetic-mixing matrix $X_{ab}$:
\begin{equation}
       X_{ab}\equiv \pmatrix{
             1 & -\chi_{12} & -\chi_{13} & \ldots & -\chi_{1N} \cr
                 -\chi_{12} & 1 & -\chi_{23} & \ldots & -\chi_{2N} \cr
                 -\chi_{13} & -\chi_{23} & 1 & \ldots & -\chi_{3N} \cr
                 \vdots & \vdots & \vdots & \ddots & \vdots \cr
                 -\chi_{1N} & -\chi_{2N} & -\chi_{3N} & \ldots & 1 \cr}~.
\label{eq:XabArbitraryMix} 
\end{equation}              
As long as $X_{ab}$ is non-singular, with positive real eigenvalues, 
there exists a matrix $M_{ab}$ which transforms
the $U(1)$ gauge groups into a basis in which their gauge-kinetic terms are 
diagonal and canonically normalized.  
Specifically, one can write 
$W_a^T X_{ab} W_b= (\hat W_a)^T (\hat W_a)$
where $\hat W_a \equiv  M_{ab} W_b$.
In general, such a matrix $M$ takes the form $M= {\cal SO}$
where ${\cal S}$ is a diagonal rescaling matrix
whose entries are the square roots of the eigenvalues of $X$
and where ${\cal O}$ is an orthogonal rotation matrix diagonalizing $X$.
After this diagonalization process,
the new $U(1)_a$ charge assignments $\hat Q_{ai}$  
for our fields $\Phi_i$ and the  new Fayet-Iliopoulos 
parameters $\hat{\xi}_a$ for our gauge groups $U(1)_a$
are given in terms of the quantities 
$Q_{ai}$ and $\xi_a$ in the original basis: 
\begin{eqnarray}
      \hat{Q}_{ai}&=& [(M^{-1})^T]_{ab} ~ Q_{bi}\nonumber\\
      \hat{\xi}_{a} &=& [(M^{-1})^T]_{ab} ~ \xi_{b}~.
\label{hatted}
\end{eqnarray}
In this vein, it is important to note 
that the matrix $M$ corresponding to each kinetic-mixing matrix $X$ is 
not unique.  
Any matrix of the form $M'=V M$, where $V$ is an orthogonal matrix,
also yields the correct normalization for the gauge-kinetic terms. 
Different choices for $V$ correspond to 
different orthogonal choices for the final basis of $U(1)$'s.
Ultimately, the physics is insensitive
to which basis is chosen.
By contrast, the rescaling matrix ${\cal S}$ is unique, and it is
this matrix which carries the physical effects of 
kinetic mixing.

In general, any of the $\chi_{ab}$ parameters in Eq.~(\ref{eq:XabArbitraryMix})
may be non-zero.
However, it will be sufficient for our purposes to restrict our attention
to the case in which only ``nearest-neighbor'' $U(1)$'s experience mixing.
Thus we shall assume that $\chi_{ab}\not=0$ if and only if $b=a+1$. 
For simplicity, we shall further assume that all non-zero $\chi_{ab}$ are
equal, so that $\chi_{ab}= \chi \delta_{a+1,b}$. 
While more general kinetic-mixing parameters may be chosen,
we shall see that these simplifications enable us to expose the existence
of our metastable vacuum towers most directly.

Needless to say, it would have been possible to construct our models entirely without
kinetic mixing by postulating highly non-trivial choices
for $\hat Q_{ai}$ and $\hat \xi_a$ right from the beginning.
However, we have found that it is easier to begin with the simpler
assignments $Q_{ai}$ shown in Table~\ref{tab:U1Chgs}
and the Fayet-Iliopoulos terms $\xi_1=\xi_N\equiv \xi$ described above, 
and to bundle all of the remaining complexities in terms of a single 
kinetic-mixing parameter $\chi$.

\begin{figure}[t!]
\centerline{
   \epsfxsize 3.0 truein \epsfbox {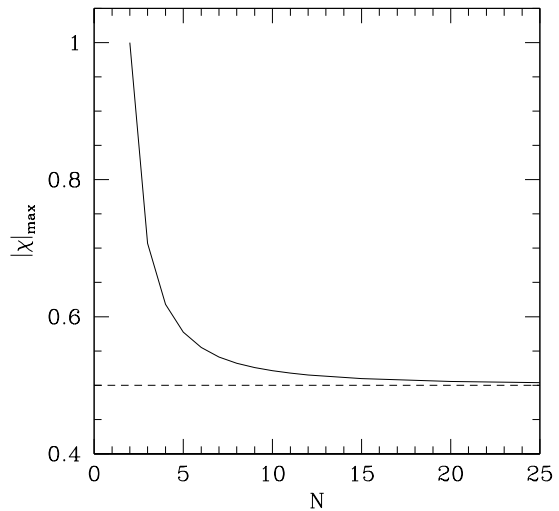} }
\caption{The maximum allowed value of $|\chi|$, plotted as a function of $N$.
      As $N\to \infty$, we see that $|\chi|_{\rm max}$ asymptotically
      approaches $1/2$ from above.  } 
\label{fig:ChiMax}
\end{figure}

Not all values of the parameter $\chi$ lead to 
self-consistent theories, however;  
we must also ensure that the kinetic-mixing matrix $X_{ab}$
in Eq.~(\ref{eq:XabArbitraryMix}) is invertible with positive (real) eigenvalues.
For $N=2$, we find that this requires $|\chi|<1$, while for
$N=3$, $4$, and $5$ this requires $|\chi|< 1/\sqrt{2}$,
$|\chi|< 2/(1+\sqrt{5})$,
and 
$|\chi|< 1/\sqrt{3}$ respectively.
The behavior of the maximum allowed value of $|\chi|$ as a function of $N$ is shown in 
Fig.~\ref{fig:ChiMax}.

%================== FIGURE ============================================
%======================================================================== 

For arbitrary $N$, we see from Fig.~\ref{fig:ChiMax}
that the maximum allowed value of $|\chi|$ always exceeds $1/2$, and asymptotically
approaches $1/2$ as $N\to\infty$.
Moreover, we find that negative values of $\chi$ do not lead to the 
metastable vacuum towers which are our main interest in this paper.
As a result, we shall simplify matters by restricting our attention to the range
\beq
              0 ~ < ~\chi~ < ~1/2~.
\label{range1}
\eeq
(Indeed, only in Sect.~V shall we consider the $\chi=1/2$ limit.)
Likewise, 
our orbifold moose structure for any length $N$ possesses a reflection symmetry under which
the combined transformations $\xi_a\rightarrow -\xi_{N+1-a}$, $g_a\rightarrow g_{N+1-a}$, 
$\xi_2\rightarrow-\xi_2$,
and $\Phi_j\rightarrow\Phi_{N-j+2}$ leaves the physics invariant.  
This means that the scalar potential in a theory of given $N$ with $\xi<0$
will be identical to that with $\xi>0$, save that the role played by $\Phi_1$ in the former
is played by $\Phi_{N+1}$ in the latter, and so forth.
As a result, we will restrict our attention to situations with 
\beq
                 \xi~>~0~.
\label{range3}
\eeq
Finally, our model also has a reflection symmetry under $\lambda\to -\lambda$, as a result
of which we can further restrict to $\lambda>0$.  However, for each $N$, we 
shall find that there is actually a minimum positive value $\lambda_N^\ast$
which is needed in order for our entire tower of $N-1$ vacua to be (meta)stable.
The derivation and interpretation of this critical value $\lambda_N^\ast$ will be
discussed further below. 
We shall therefore actually restrict to the range  
\beq
                  \lambda~>~\lambda_N^\ast~
\label{range2}
\eeq
in much of what follows.

\begin{figure}[thb!]
\centerline{
   \epsfxsize 3.3 truein \epsfbox {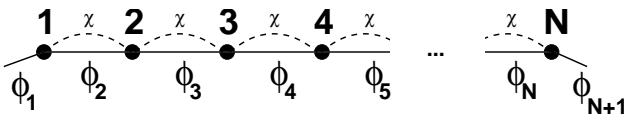}
 }
\caption{ The ``moose'' diagram for our series of models.  
  The $N$ sites [each representing a $U(1)$ gauge group]  
  are connected by links corresponding to the
  chiral superfields $\Phi_i$, and experience nearest-neighbor kinetic mixing governed
  by a universal parameter $\chi$.}
\label{fig:moose}
\end{figure}

Thus, to summarize, our models are defined in terms of $N$ different $U(1)_a$ gauge
groups and $N+1$ different chiral superfields $\Phi_i$, 
with charges indicated in Table~\ref{tab:U1Chgs}. 
This structure can be indicated pictorially through the moose diagram in Fig.~\ref{fig:moose},
which shows not only the $U(1)_a$ gauge groups but also the $\Phi_i$ fields which provide
nearest-neighbor ``links'' between them
as well as the $\chi$ parameter which governs
their universal nearest-neighbor kinetic mixing.  
For each value of $N\geq 2$, our models are therefore 
governed by four continuous parameters, namely  
$g$, $\chi$, $\xi$, and $\lambda$, subject to the bounds
in Eqs.~(\ref{range1}), (\ref{range3}) and (\ref{range2}).
Note that a similar model, but with $\chi=0$, was considered
in Ref.~\cite{DDG}. 
%   as a toy model of the string landscape.

%================== FIGURE ============================================
%========================================================================

Our main interest in this paper is in the vacuum structure of these models.
This in turn is governed by their corresponding scalar potentials.  
In general, the scalar potential for such a supersymmetric
 gauge theory coupled to matter includes
both $D$-term and $F$-term contributions and can be written in the form
\begin{equation}
      V~=~ \half \sum_{a=1}^N  g^2 \hat D_a^2 + \sum_{i=1}^{N+1}  |F_i|^2~,
\label{Vdef}
\end{equation}
where each gauge-group factor has a common coupling $g$ and where
\begin{equation}
  \hat D_a = \hat \xi_a+\sum_{i=1}^{N+1} \hat Q_{ai}  |\phi_i|^2~,~~~~~~
       F_i =-\frac{\partial W^{\ast}}{\partial\phi^{\ast}_i}~.
\label{DFdefs}
\end{equation}
For any choice of parameters $\lbrace g,N,\chi,\xi,\lambda\rbrace$,
the extrema of the scalar potential
can then be obtained by solving the $N+1$ coupled simultaneous equations
\begin{equation}
          \frac{\partial V}{\partial\phi_i}=0~~~~~(i=1,\ldots,N+1)~.
\label{eq:ExistEqs}
\end{equation}
However, a solution is a local minimum only if the eigenvalues of the 
$2(N+1)\times 2(N+1)$ mass matrix
\begin{equation}
  {\mathcal M}^2~\equiv~ \pmatrix{
  \frac{\partial^2 V}{\partial\phi_i^\ast \partial\phi_j} &
  \frac{\partial^2 V}{\partial\phi_i^{\ast}\partial\phi_j^\ast} \cr
  \frac{\partial^2 V}{\partial\phi_i\partial\phi_j} &
  \frac{\partial^2 V}{\partial\phi_i\partial\phi_j^\ast} \cr}
\label{matrix}
\end{equation}
are all non-negative and the number of zero eigenvalues is precisely equal to the
number of Goldstone bosons eaten by the massive gauge fields.
(Indeed, additional
zeroes would indicate the presence of classical flat directions.)
In what follows, however, we will use the term ``vacuum'' loosely to refer to
any extremum of the potential and employ adjectives such as ``stable'' and ``unstable''
to distinguish the eigenvalues of the mass matrix.
Of course, a ``metastable'' vacuum exists only when two or more vacua exist and
are stable according to the above definitions;  all but the vacuum with lowest energy
are considered metastable.

Note that the scalar potential $V$ 
will in general be a function of only the absolute squares of fields.
As a result, we can take all non-zero vacuum expectation values 
$v_i\equiv \langle \phi_i \rangle$
to be real and positive without loss of generality.

Also note that while the $F$-terms are insensitive to the kinetic
mixing, the $D$-terms in Eq.~(\ref{DFdefs}) 
are calculated in terms of the 
charges $\hat Q_{ai}$ and Fayet-Iliopoulos coefficients 
$\hat \xi_a$ in the {\it orthonormalized}\/ basis 
given in Eq.~(\ref{hatted}).
It is apparent from the form of Eq.~(\ref{Vdef}) that
while $\hat D_a$ depends on the choice of ${\cal O}$ within the matrix $M$,
the scalar potential as a whole is insensitive to this choice.
By contrast, the rescaling matrix ${\cal S}$ within $M$ is physical, modifying
the $D$-terms in a non-trivial way.  It is in this manner that the effects
of kinetic mixing are felt in the vacuum structure of the theory. 

Finally, we emphasize that we shall deem an extremum of the scalar
potential $V$ to be (meta)stable only if there are neither 
flat directions nor negative eigenvalues in the mass matrix.
These two restrictions are rather severe, since most models tend to
give extrema which have either tachyonic modes or flat directions.  

In the rest of this paper, we shall simplify our analysis by scaling out 
the gauge coupling $g$ in the manner discussed in Ref.~\cite{Nest}.
Specifically, we shall define the rescaled quantities
$\xi'\equiv g\xi$, $\lambda'\equiv \lambda/g^{N/2}$, and $\Phi'_i\equiv \sqrt{g} \Phi_i$,
holding all other quantities fixed.
We shall then eliminate explicit dependence on $\xi'$ by further 
rescaling all dimensionful quantities by appropriate powers
of $\xi'$ in order to render them dimensionless.
Specifically, we shall define 
$\lambda''\equiv \lambda'/(\xi')^{1-N/2}$,  
$\Phi''_i\equiv \Phi'_i/\sqrt{\xi'}$,
and $V''\equiv V/(\xi')^2$.
In practical terms, the net effect of these two rescalings is 
that we simply rewrite all of our original expressions in terms of the
new variables $\lambda'' \equiv \xi^{N/2-1} \lambda/g$,
$V''\equiv V/(g\xi)^2$, and
$\Phi''_i\equiv \Phi_i/\sqrt{\xi}$,
and then drop the double primes.
Thus, for each $N\geq 2$, our models can be analyzed purely in terms of 
a single kinetic-mixing parameter
$\chi$ and the rescaled (dimensionless) Wilson-line coefficient $\lambda$ defined above;
the resulting vacuum energies $V$ and field VEV's $v_i\equiv \langle \Phi_i \rangle$
will then be dimensionless as well.
Finally, we shall adopt a notation (first introduced in Ref.~\cite{DDG}) 
wherein we describe a 
particular field configuration of VEV's $v_i$ 
as belonging to a class 
denoted $\vac{p,q,...}$ if the only non-zero VEV's for the vacuum solutions in this
class are $v_p$, $v_q$, and so forth.
For example, $\vac{1,2,4}$ will refer to a vacuum configuration in which
$v_1$, $v_2$, and $v_4$ are non-zero, with all other VEV's vanishing.

Our claim, then, is that for each $N$, our model gives 
rise to a tower consisting of $N-1$ vacuum solutions.  Specifically, 
for each $N$, we claim that the corresponding model will give rise
to a true stable ground state along with a series of $N-2$ metastable ground states 
with higher and higher vacuum energies.

\subsection{Example:  $N=3$\label{sec:Nis3}}

We shall begin the analysis of our models by focusing on a simple example: 
the $N=3$ special case, which consists of 
three $U(1)$ gauge groups and four chiral superfields.  
The scalar potential in this case
is given by $V= V_D+V_F$, where
\beqn
  V_D  &=& 
   {1\over 4} \left( |\phi_1|^2 - |\phi_2|^2 - |\phi_3|^2 + |\phi_4|^2 \right)^2
         \nonumber\\
&& ~+~
  {1\over {8(1-\sqrt{2}\chi)}} \biggl[
    2 - |\phi_1|^2 - (\sqrt{2}-1)|\phi_2|^2 \nonumber\\
         && ~~~~~~~~~~~~~~~~~+ (\sqrt{2}-1)|\phi_2|^2 +|\phi_4|^2 \biggr]^2 
         \nonumber\\
&& ~+~
  {1\over {8(1+\sqrt{2}\chi)}} \biggl[
    2 - |\phi_1|^2 + (\sqrt{2}+1)|\phi_2|^2 \nonumber\\
         && ~~~~~~~~~~~~~~~~~- (\sqrt{2}+1)|\phi_2|^2 +|\phi_4|^2 \biggr]^2 ~,
\eeqn
and 
\beq
  V_F~=~|\lambda|^2\, \sum_{i=1}^4\, \frac{|\phi_1|^2|\phi_2|^2|\phi_3|^2|\phi_4|^2}{|\phi_i|^2}~.
\eeq
In these expressions, of course,
$\phi_i$ denotes the (complex) scalar component of the chiral supermultiplet $\Phi_i$.

\begin{figure}[t!]
\centerline{
   \epsfxsize 3.0 truein \epsfbox {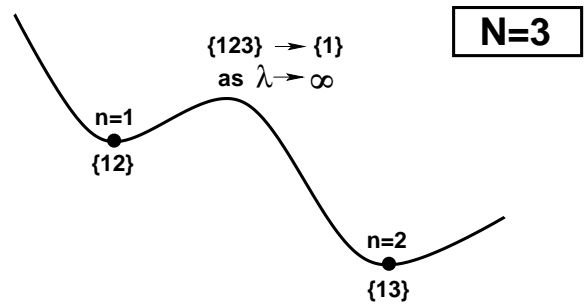}
 }
\caption{
      A sketch of the vacuum structure of the $N=3$ model.
      For $\lambda>\lambda_3^\ast$, the corresponding scalar potential 
      gives rise to two distinct minima: 
      a $\vac{13}$ vacuum which serves as the true ground state of the theory,
      and a $\vac{12}$ vacuum which serves as an additional, metastable vacuum.
      These two minima are separated by a saddle-point $\vac{123}$ extremum 
      which reduces to the $\vac{1}$ extremum in the formal $\lambda\to\infty$ limit.   
      Note that this sketch is actually a two-dimensional representation of  
      potential energy contours in a three-dimensional field space parametrized
      by $\lbrace v_1^2, v_2^2, v_3^2\rbrace$.} 
\label{fig:SketchNis3}
\end{figure}

Given the scalar potential, 
it is then a straightforward matter to calculate the vacuum structure of this potential.
Our results are as follows.
Defining
\beq
     \lambda_3^\ast ~\equiv~  {1\over \sqrt{\chi(1+\chi)}}~,
\label{lam3star}
\eeq
we find that there are two (meta)stable vacua in this model 
for all  $\lambda >\lambda_3^\ast$. 
The first vacuum state (which we shall call the $n=1$ vacuum)
has energy $V_1=1/2$ and corresponds to the solution with
\beq
       n=1:~~~~~
          v_1^2 = 1+\chi~,~~~~~ 
            v_2^2 = \chi~,~~~~~
            v_3^2=v_4^2=0~.
\label{3n=1}
\eeq
By contrast, the second vacuum state (which we shall call the $n=2$ vacuum)
has vacuum energy $V_2=\quarter(1-\chi)^{-1}$ and corresponds to the solution
with
\beq
       n=2:~~~~~
          v_1^2 =  {2-\chi\over 2-2\chi}~,~~~~~ 
            v_3^2 = \half~,~~~~~
            v_2^2=v_4^2=0~.
\label{3n=2}
\eeq
We thus see that the $n=1$ vacuum is of $\vac{12}$-type,
while the $n=2$ vacuum is of $\vac{13}$-type.
As emphasized above, both of these solutions are {\it stable}\/ (without
any flat or tachyonic directions) for all $\lambda>\lambda_3^\ast$;  
however, since $V_1>V_2$ for all $\chi<1/2$,
we see that the $n=2$ vacuum is the true ground state in this theory,
while the $n=1$ vacuum is only metastable.  
This vacuum configuration
is sketched in Fig.~\ref{fig:SketchNis3}.

%================== FIGURE ============================================
%======================================================================== 

\begin{figure*}[thb!]
\centerline{
   \epsfxsize 2.8 truein \epsfbox{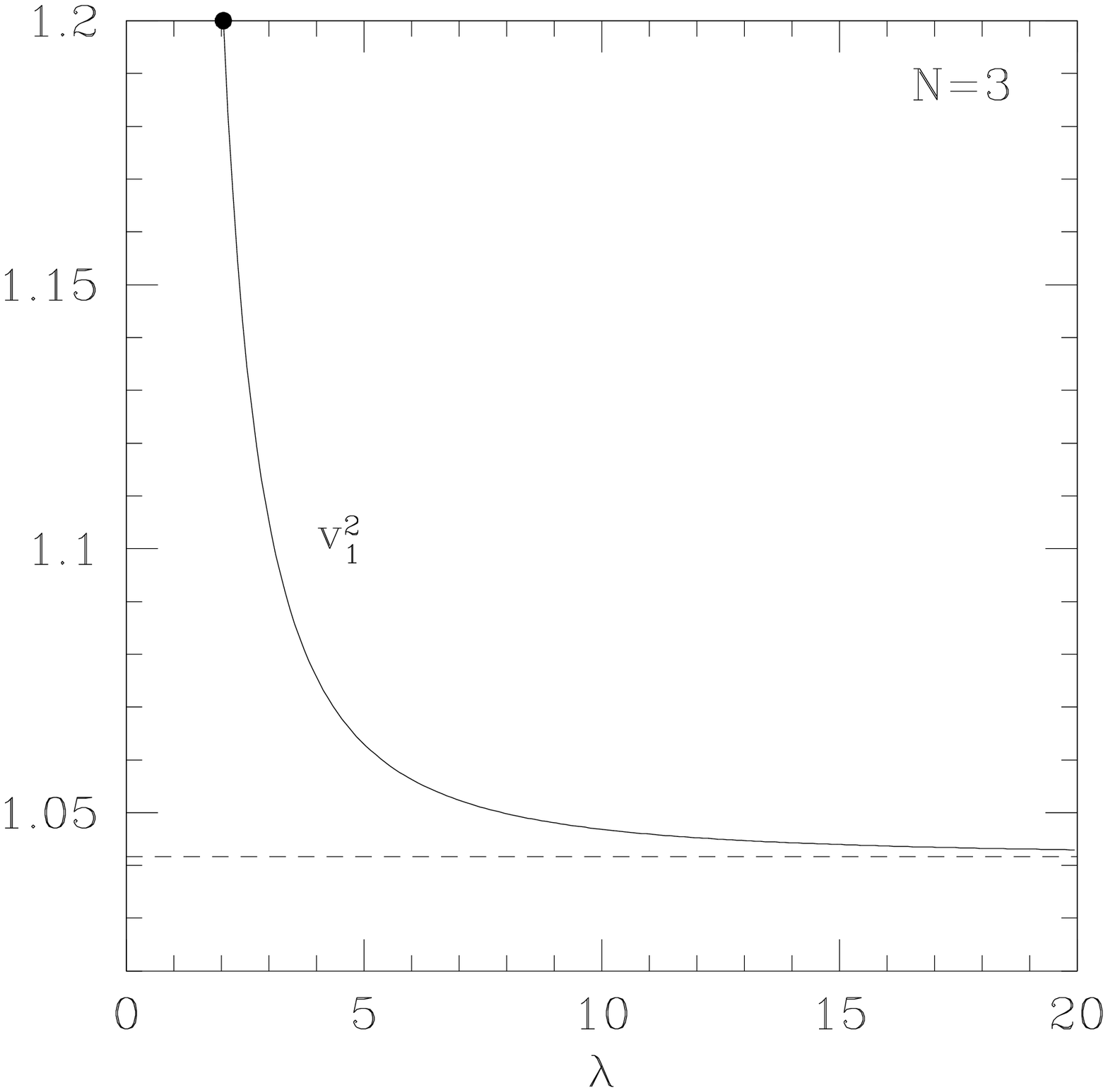}
   \epsfxsize 2.8 truein \epsfbox{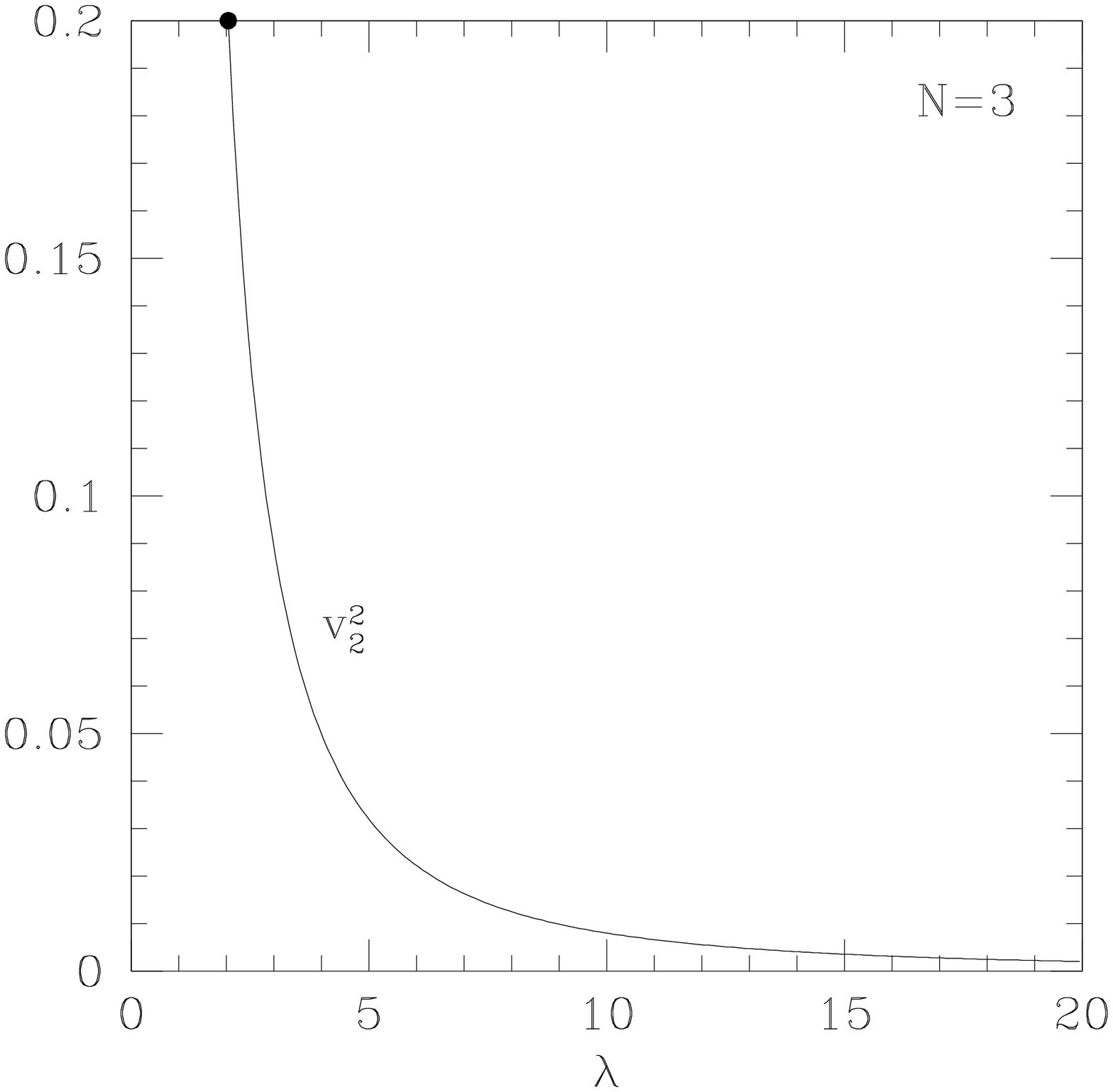}}
\centerline{
   \epsfxsize 2.8 truein \epsfbox{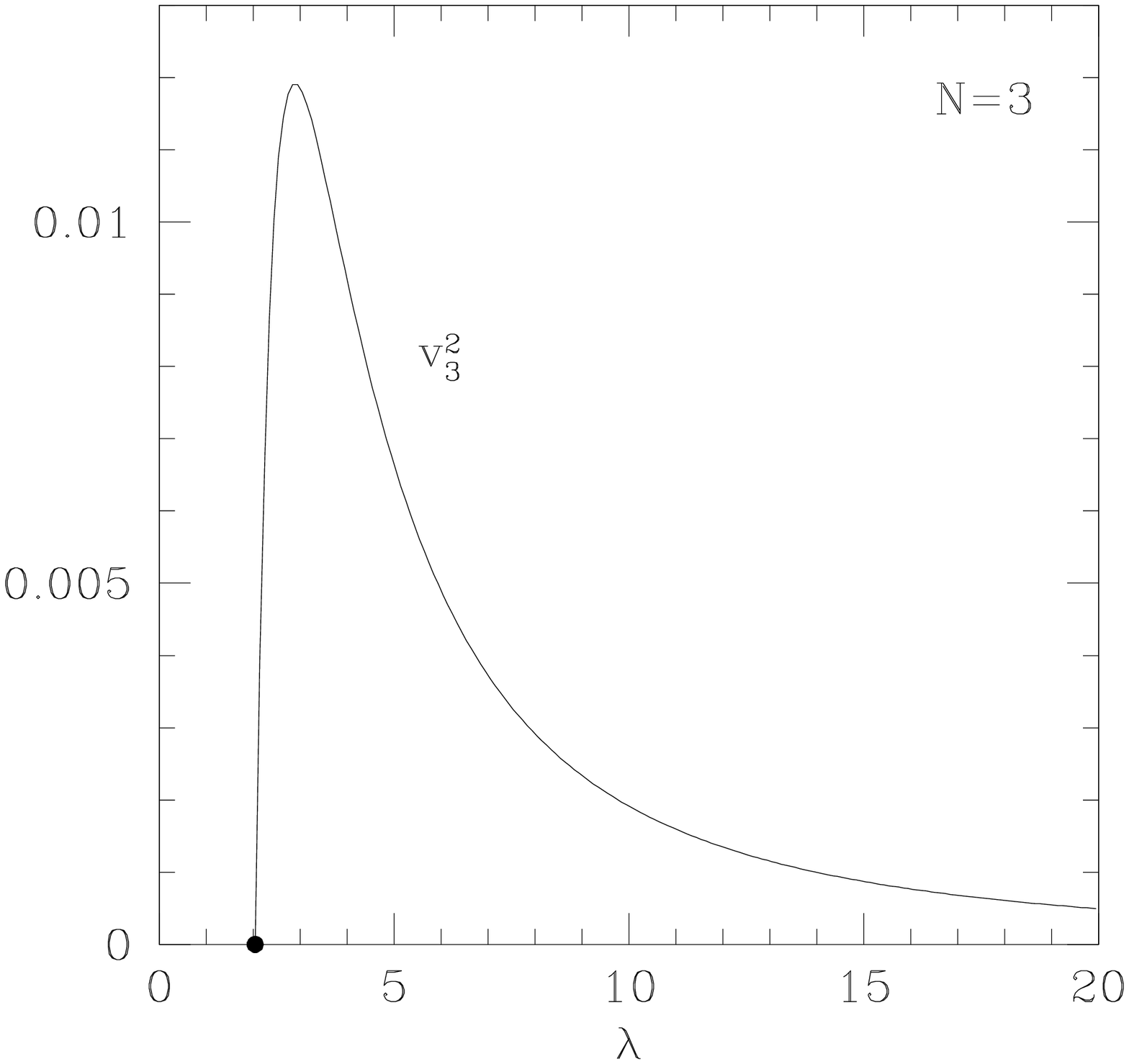}
   \epsfxsize 2.8 truein \epsfbox{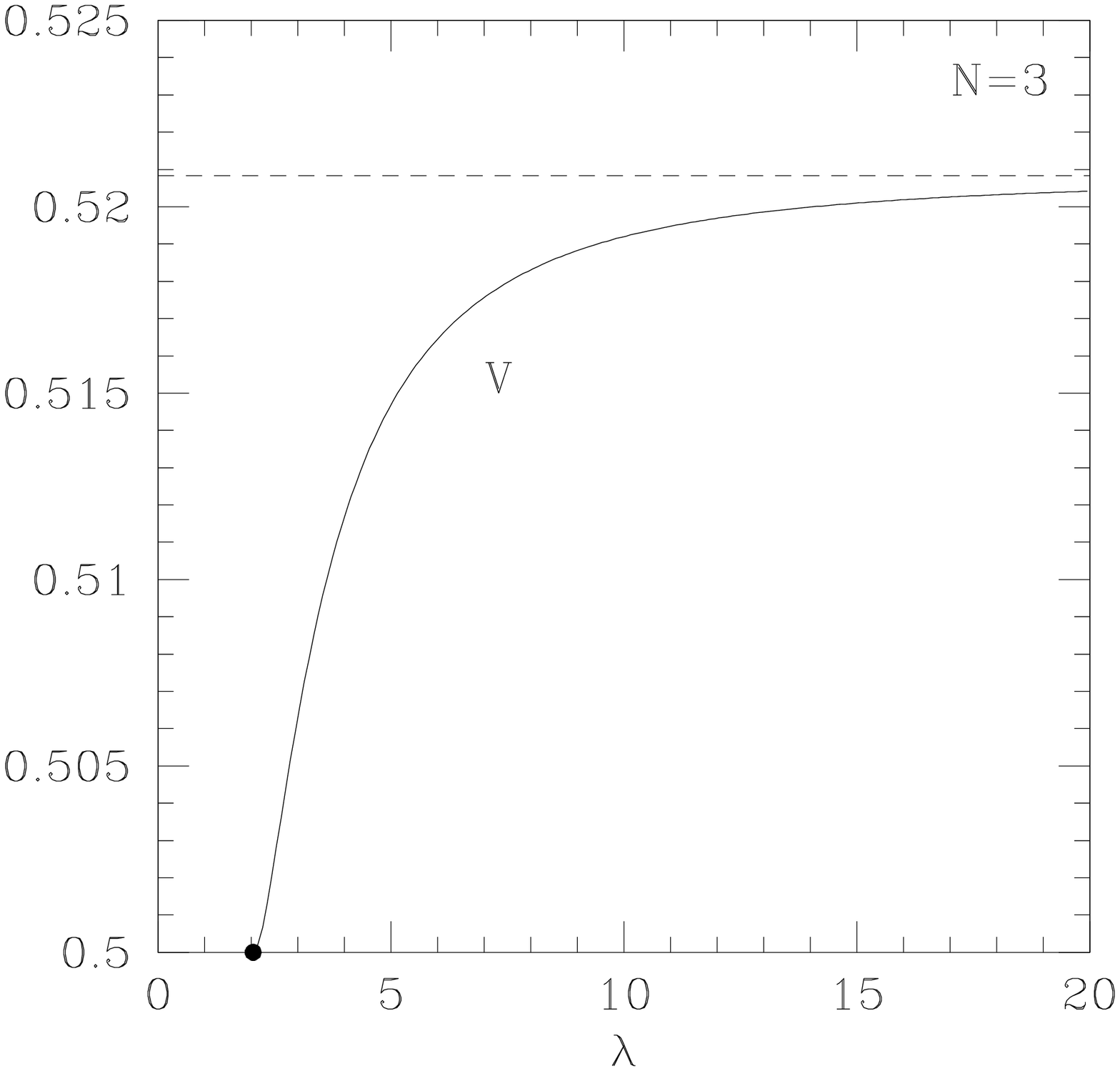}}
\caption{
  The $\vac{123}$ saddle-point solution for $N=3$ and $\chi=1/5$, 
 plotted as a function of $\lambda\geq \lambda_3^\ast=5/\sqrt{6}\approx 2.04$.
   Here we show the non-zero field VEV's $|v_1|^2$ (upper left plot),
       $|v_2|^2$ (upper right plot),
       and $|v_3|^2$ (lower left plot),
    as well as the corresponding scalar potential $V$ (lower right plot).
  While each of these quantities varies with $\lambda$,  they quickly 
   reach formal asymptotes as $\lambda\to\infty$.}
\label{fig:Saddle}
\end{figure*}

Note that for $\lambda> \lambda_3^\ast$, 
the $n=1$ and $n=2$ vacua are separated by a 
potential barrier whose 
lowest point is a $\vac{123}$ saddle-point extremum of the
scalar potential.
Unlike the field-space solutions for the $n=1$ and $n=2$ vacua,
which are $\lambda$-independent,
the field-space solution for this saddle point
depends quite strongly on $\lambda$. 
This is shown in Fig.~\ref{fig:Saddle},
where the explicit solutions for $\lbrace v_1^2, v_2^2, v_3^2\rbrace$
are plotted for $\lambda>\lambda_3^\ast$.

It will be useful to understand how this vacuum structure deforms
as a function of $\lambda$. 
Na\"ively, one might suspect that taking $\lambda\to\infty$
would cause the height of this saddle-point barrier to diverge.
However, we see from Fig.~\ref{fig:Saddle}
that this is not the case:  
the scalar potential $V$ and all of the field VEV's $v_i^2$ quickly reach
finite asymptotes.
Indeed, in the formal $\lambda\to\infty$ limit, we see
that $v_2^2,v_3^2\to 0$, whereupon our $\vac{123}$ saddle-point solution
reduces to the $\vac{1}$ solution
given by
\beq
        v_1^2= {1\over 1-\chi^2}~,~~~~~
        v_2^2= v_3^2 = v_4^2 = 0~~~~
\label{asymN=3}
\eeq
with $V_{12}= \half (1-\chi^2)^{-1}$.

%================== FIGURE ============================================
%========================================================================

The above results are valid for all $\lambda > \lambda_3^\ast$.
However, it will also be important for us to understand 
what happens as we reduce the value of $\lambda$ below $\lambda_3^\ast$.
Since $\lambda$ sets the scale for the barrier height between the two
vacua in Fig.~\ref{fig:SketchNis3}, reducing $\lambda$ 
has the effect of reducing the barrier height between the 
two vacua.  This in turn will destabilize our metastable vacuum.   
Specifically, we see from Fig.~\ref{fig:Saddle} that
$\lambda_3^\ast\equiv 1/\sqrt{\chi(1+\chi)}$ is
nothing but the critical value of $\lambda$ at which   
the barrier height becomes equal to $V_1$.
At this critical value, the $\vac{123}$ saddle-point 
solution shown in Fig.~\ref{fig:Saddle} 
actually merges with the metastable $n=1$ vacuum 
solution (which is of $\vac{12}$-type) and thereby destabilizes it.
Thus, as we take $\lambda$ below $\lambda_3^\ast$, we lose the $n=1$
vacuum.  
In order to emphasize that this is the critical
$\lambda$-value at which the $n=1$ vacuum is destabilized,
we shall also refer to $\lambda_3^\ast$ as $\lambda_{3,1}^\ast$. 

Taking $\lambda$ still lower, we ultimately reach a second 
critical value $\lambda_{3,2}^\ast\equiv \sqrt{2/ (2-\chi)}$ 
at which   
even the $n=2$ vacuum becomes destabilized.
In this case, a new $\vac{123}$ solution becomes stable and serves
as the ground state of the theory for all $\lambda<\lambda_{3,2}^\ast$.

%================== FIGURE ============================================
\begin{figure}[h!]
\centerline{
   \epsfxsize 3.0 truein \epsfbox {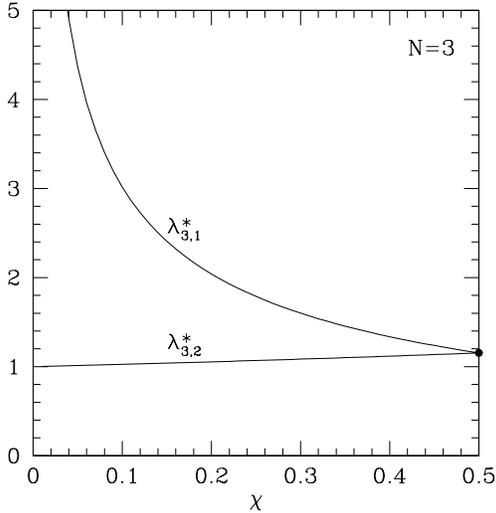}
 }
\vskip -0.2 truein
\caption{
      Critical $\lambda$-values for the $N=3$ model, plotted as functions of $\chi$. 
      Here $\lambda_3^\ast= \lambda_{3,1}^\ast= 1/\sqrt{\chi(\chi+1)}$ and
      $\lambda_{3,2}^\ast= \sqrt{2/(2-\chi)}$.
      For $\lambda>\lambda_{3,1}^\ast$, both the $n=1$ and $n=2$ vacua
      are stable, while for $\lambda_{3,2}^\ast<\lambda\leq \lambda_{3,1}^\ast$,
      the $n=1$ vacuum is destabilized and 
      only the $n=2$ vacuum is stable.
      For $\lambda \leq \lambda_{3,2}^\ast$,
      even the $n=2$ vacuum is destabilized;  in this range 
      a new $\vac{123}$ solution becomes stable and serves as the
      ground state of the theory.} 
\label{fig:lam3plot}
\end{figure}  
%======================================================================== 

These two critical values of $\lambda$ for the $N=3$ case
are plotted in Fig.~\ref{fig:lam3plot} as a function of the kinetic-mixing
parameter $\chi$.
Note, in particular, that $\lambda_{3,1}^\ast$ diverges as $\chi\to 0$.
This shows 
that the stability of the $n=1$ metastable vacuum
solution relies not only on having a sufficiently large value of $\lambda$,
but also on the existence of non-zero kinetic mixing.
Note also that 
\beq
         0 ~< ~ \lambda_{3,2}^\ast ~<~ \lambda_{3,1}^\ast ~~~~~~~ 
          {\rm for ~all~} ~0<\chi<1/2~. 
\eeq
This indicates that as reduce $\lambda$ below $\lambda_3^\ast=\lambda_{3,1}^\ast$,
our $N=3$ metastable ``tower'' destabilizes from the top down, with the
$n=1$ metastable vacuum destabilizing before the $n=2$ vacuum. 
It is for this reason that we can associate $\lambda_3^\ast$ (the
critical $\lambda$-value for stability for the entire tower) 
with $\lambda_{3,1}^\ast$ (the critical $\lambda$-value for stability of
the highest vacuum).

We see, then, that our $N=3$ model
gives rise to two vacuum solutions (one stable ground state and one metastable
 state above it)
for all $\lambda >\lambda_3^\ast=\lambda_{3,1}^\ast$.
The solutions for these vacua are $\lambda$-independent,
and depend only on $\chi$.
However, the barrier height between these two vacua (and hence the stability
of the metastable vacuum) depends 
intimately on both $\chi$ and $\lambda$, and formally reaches an asymptote as $\lambda\to\infty$.

%==================================================================

\subsection{Example: $N=4$\label{sec:Nis4}}

\begin{figure}[htb!]
\centerline{
   \epsfxsize 3.1 truein \epsfbox {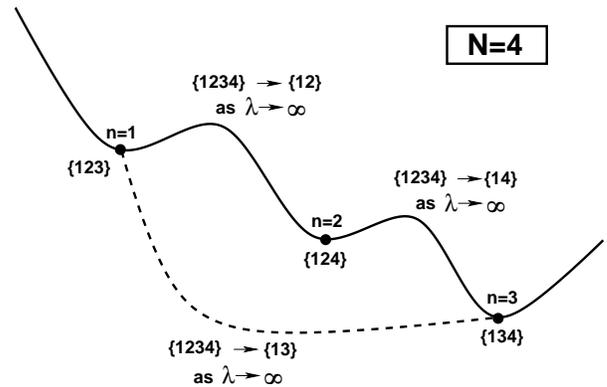}
 }
\caption{
      A sketch of the vacuum structure of the $N=4$ model.
      For $\lambda>\lambda_4^\ast$, the corresponding scalar potential 
      gives rise to three distinct minima: 
      a $\vac{134}$ vacuum which serves as the true ground state of the theory,
      and two additional metastable vacua of types $\vac{123}$ and $\vac{124}$
      above it.
      These three vacua are separated by three different saddle-point extrema
      which are each different solutions of $\vac{1234}$-type;
      in the formal $\lambda\to\infty$ limit,
      these reduce to the different two-VEV solutions shown.  
      Note that this sketch is actually a two-dimensional representation of  
      potential energy contours in a four-dimensional field space parametrized
      by $\lbrace v_1^2, v_2^2, v_3^2, v_4^2\rbrace$.  } 
\label{fig:SketchNis4}
\end{figure}

Having discussed the vacuum structure of the $N=3$ model,
we now turn to the $N=4$ model.  In this case,
there are four $U(1)$ gauge groups and five chiral superfields. 
Defining 
\beq
     \lambda_4^\ast~\equiv~ {1\over \chi \sqrt{1+\chi}}~,
\label{lam4star}
\eeq
we find that the vacuum structure now consists of {\it three} vacuum solutions,
each without tachyonic or flat directions,
for all $\lambda>\lambda_4^\ast$. 
The $n=1$ vacuum has energy $V_1=1/2$, just as in the $N=3$ case,
and corresponds to the solution with
\beq
 n=1:~~~~~\cases{
 v_1^2= 1+\chi~,~~~
 v_2^2=v_3^2= \chi~, & \cr
   &\cr
 v_4^2=v_5^2=0~, &\cr}
\label{4n=1}
\eeq
while the $n=2$ vacuum has energy $V=\quarter (1-\chi)^{-1}$,
again just as in the $N=3$ case, and now corresponds to the
solution with
\beq
 n=2:~~~~~\cases{
 v_1^2= {2-\chi\over 2-2\chi}~,~~~
 v_2^2= {\chi \over 2-2\chi}~,~~~& \cr
   &\cr
 v_4^2= \half~,~~~~
 v_3^2=v_5^2=0~.& \cr}
\label{4n=2}
\eeq
The solutions in Eqs.~(\ref{4n=1}) and (\ref{4n=2})
are clearly the $N=4$ generalizations of the 
corresponding $N=3$ solutions in Eqs.~(\ref{3n=1}) and
(\ref{3n=2}) respectively.
However, the crucial new feature for the $N=4$ case is the appearance of an
additional vacuum state of even lower energy.
This vacuum state, which we shall refer to as the $n=3$
vacuum, has energy $V_3=\half(3-4\chi)^{-1}$ and corresponds to
the solution with
\beq
 n=3:~~~~~\cases{
 v_1^2= {3(1-\chi)\over 3-4\chi}~,~~~
 v_3^2= {1-\chi\over 3-4\chi}~,~~~ 
         & \cr
   &\cr
 v_4^2= {2-3\chi\over 3-4\chi}~,~~~
 v_2^2=v_5^2=0~. &\cr}
\label{4n=3}
\eeq
This vacuum structure is sketched in Fig.~\ref{fig:SketchNis4}.
Note that just as in the $N=3$ case, 
these vacuum solutions are all $\lambda$-independent.

%================== FIGURE ============================================
%======================================================================== 

For $\lambda> \lambda_4^\ast$, 
the $n=1$, $n=2$, and $n=3$ vacua are separated by 
potential barriers whose 
lowest points are all $\vac{1234}$ saddle-point extrema of the
scalar potential.
Unlike the field-space solutions for the (meta)stable vacua,
the field-space solutions for these saddle points
depend quite strongly on $\lambda$. 
However, in the formal $\lambda\to\infty$ limit,
these solutions all quickly reach finite asymptotes.
For example,
the asymptotic saddle-point solution between the $n=1$ and $n=2$ vacua
is of $\vac{12}$-type and is given by
\beq
   (1,2):~~~~\cases{
    v_1^2 = {1\over 1-\chi^2}~,~~~ 
    v_2^2 = {\chi^2\over 1-\chi^2}~, & \cr
      &\cr
       v_3^2=v_4^2=v_5^2 = 0~ &\cr}
\eeq
with energy $V_{12}=\half(1-\chi^2)^{-1}$.
This is the  $N=4$ analogue
of the $N=3$ asymptotic saddle-point solution in Eq.~(\ref{asymN=3}),
and continues to have the same energy $V_{12}$. 
However, in the $N=4$ case there are also additional 
saddle-point solutions which involve the new $n=3$ vacuum.
Specifically, the  asymptotic saddle-point solution 
which lies directly between the $n=1$ and $n=3$ vacua
is of $\vac{13}$-type and is given by
\beq
   (1,3):~~~~\cases{
    v_1^2 = {2(1-\chi) \over 2-2\chi-\chi^2 }~,~~~ 
    v_3^2 = {\chi(1-\chi) \over 2-2\chi-\chi^2 }~, &\cr
      &\cr
    v_2^2=v_4^2=v_5^2 = 0~ &\cr}
\eeq
with energy $V_{13}= (1-\chi)/(2-2\chi-\chi^2)$,
while the asymptotic saddle-point solution between the $n=2$ and $n=3$ vacua
is of $\vac{14}$-type and is given by
\beq
   (2,3):~~~~\cases{
    v_1^2 = {2(1-\chi) \over 2-2\chi-\chi^2 }~,~~~ 
    v_4^2 = {1-\chi-\chi^2 \over 2-2\chi-\chi^2 }~, &\cr
      &\cr
    v_2^2=v_3^2=v_5^2 = 0~ &\cr}
\eeq
with energy $V_{23}= \half (2-2\chi-\chi^2)^{-1}$.

\begin{figure}[h!]
\centerline{
   \epsfxsize 3.0 truein \epsfbox {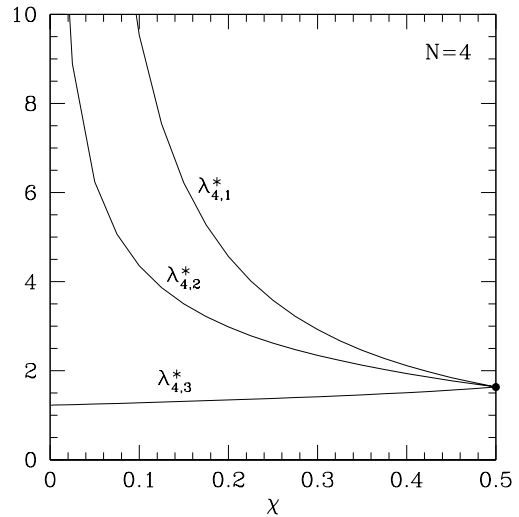}
 }
\caption{
      Critical $\lambda$-values for the $N=4$ model, plotted as functions of $\chi$. 
      For $\lambda>\lambda_4^\ast=\lambda_{4,1}^\ast$, all three of our vacua 
      are stable,
      while for $\lambda_{4,2}^{\ast}< \lambda\leq \lambda_{4,1}^\ast$,
      the $n=1$ vacuum is destabilized 
      and for 
      $\lambda_{4,3}^{\ast}<\lambda\leq \lambda_{4,2}^{\ast}$,
      both the $n=1$ and $n=2$ vacua are destabilized. 
      Finally, for $\lambda \leq \lambda_{4,3}^{\ast}$,
      even the $n=1$ vacuum is destabilized;  in this range 
      a new $\vac{1234}$ solution becomes stable and serves as the
      ground state of the theory.} 
\label{fig:lam4plot}
\end{figure}

This vacuum structure emerges for all $\lambda>\lambda_4^\ast$.
However, just as in the $N=3$ case, we find that reducing $\lambda$
below $\lambda_4^\ast$ tends to destabilize our vacuum tower.
Specifically, one finds that there are now {\it three}\/ critical
$\lambda$-values, denoted $\lambda_{4,1}^\ast$, $\lambda_{4,2}^\ast$,
and $\lambda_{4,3}^\ast$,
at which the $n=1$, $n=2$, and $n=3$ vacua are respectively destabilized.
These three critical values are given by
\beqn
      \lambda_{4,1}^\ast& \equiv &  {1\over \chi \sqrt{1+\chi}}~\nonumber\\
      \lambda_{4,2}^\ast& \equiv &  2\, \sqrt{\frac{(1-\chi)}{\chi(2-\chi)}}\nonumber\\
      \lambda_{4,3}^\ast& \equiv &  {3-4\chi\over 1-\chi} \, \sqrt{{1\over 3(2-3\chi)}}~,
\eeqn
and all three are plotted in Fig.~\ref{fig:lam4plot}
as functions of $\chi$.   We see from this figure that 
\beq
   0 ~<~ \lambda_{4,3}^\ast
   ~<~ \lambda_{4,2}^\ast
   ~<~ \lambda_{4,1}^\ast
\label{lamstar4orders}
\eeq
for all $0< \chi<1/2$.
This implies that just as in the $N=3$ case, 
our vacuum tower destabilizes from the top down
as we reduce $\lambda$ below $\lambda_{4,1}^\ast$.
As a result of Eq.~(\ref{lamstar4orders}), we see that 
$\lambda_4^\ast$ (\ie, the
critical $\lambda$-value for stabilizing the entire $N=4$ vacuum tower)
is nothing but $\lambda_{4,1}^\ast$ (the maximum of the individual
values $\lambda_{4,n}^\ast$ for stabilizing any
of the individual vacua in the tower).
It is this observation which
underlies the identification given in Eq.~(\ref{lam4star}).
We also observe from Fig.~\ref{fig:lam4plot} that
non-zero kinetic mixing is also required in order for the
stability of our metastable vacua.

%================== FIGURE ============================================
%======================================================================== 

It is clear that the $N=4$ case is a direct generalization 
of the $N=3$ case.  Once again, the vacuum solutions are $\lambda$-independent,
while the barrier heights (and hence the stability of these vacuum solutions)
are $\lambda$-dependent.
Moreover, the energies associated with the $n=1$ and $n=2$ vacua,
as well as the barrier between them, 
are unchanged in passing from the $N=3$ case to the $N=4$ case. 
Indeed, the primary new feature in passing from
the $N=3$ case to the $N=4$ case is the emergence of a new vacuum
solution, our so-called $n=3$ vacuum, which ``slides in'' {\it below}\/ the previous bottom
of the tower and becomes the new ground state of the theory for 
all $\lambda>\lambda_{4,3}^\ast$.
As a result, our previous $n=2$ ground state in the $N=3$ theory
now becomes the first-excited metastable state in the $N=4$ theory,
and our tower of metastable vacua has grown by one additional metastable vacuum.
We stress again that all of these vacua are either strictly stable 
or strictly metastable.  In particular, they contain neither tachyonic masses
nor flat directions.

\subsection{Results for general $N$}

We now turn to the case with general $N$.
As might be expected, the pattern we have seen in passing from 
the $N=3$ case to the $N=4$ case continues without major alteration.
For general $N$ and for $\lambda$ exceeding a critical value $\lambda_N^\ast$,
we find that our model has a vacuum structure
consisting of a tower of $N-1$ stable vacua:  
a single stable ground state, and $N-2$ metastable vacua above it.
As in the $N=3$ and $N=4$ cases discussed above,
we shall number these vacua from the top down with an index $n$,
so that the $n=1$ vacuum sits at the top of the tower
and the $n=N-1$ vacuum (the true ground state of the theory)
sits at the bottom.  

We then find that the $n$-vacuum has energy
\beq
         V_n ~=~   \half \left( 1\over \chi R_n\right)~,~~~~~~~{1\leq n \leq N-1}
\label{Vn}
\eeq
where
\beq
            R_n~\equiv~ \left({1\over \chi} - 2\right)n + 2~,
\label{Rndef}
\eeq
and corresponds to the solution with
\beqn
v_j^2 = \cases{
         1+ 1/R_n  &  for $j=1$\cr
    1/R_n        & for $2\leq j \leq N-n$\cr
    0            & for $j=N-n+1$\cr
   (R_{j-N+n-1}-1)/R_n & for $N-n+2\leq j \leq N$ \cr
    0             &  for $j=N+1$~.\cr}\nonumber\\
        ~~~~~~~~
\label{vacsolns}
\eeqn
As evident from these VEV's, this is clearly a 
solution of $\vac{1,...,N-n,N-n+2,...,N}$-type.
It is easy to verify that these results reduce
to those already quoted for the $N=3$ and $N=4$
special cases.
Note that the vacuum energies $V_n$ along
the tower are independent of $N$, as expected;
indeed, all that depends on $N$ are the number of such
vacua and their precise field VEV's.
Also note that $V_{N-1}\to 0$ as $N\to \infty$.

As in the previous special cases with $N=3$ and $N=4$,
any two vacua
$n$ and $n'$ are separated 
by a saddle-point solution.  In the formal
$\lambda\to\infty$ limit, we find that this
barrier height asymptotes to the value
\beq
     V_{nn'}~=~ {1\over 2\chi}\,     
        \left({ R_{n'-n} \over 
        R_n R_{n'-n} -1}\right)  
\eeq
where $R_n$ is defined in Eq.~(\ref{Rndef})
and where we have taken $n'>n$.
Note that $V_{nn'}$ is also independent of $N$.
The corresponding asymptotic saddle-point
solutions are given by
\beq
      v_j^2 = \cases{
     {R_n R_{n'-n}  \over R_n R_{n'-n} -1} & for $j=1$\cr
      &\cr
     {1            \over R_n R_{n'-n} -1} & for $2\leq j \leq N-n'$\cr
      &\cr
     0                                    & for $j=N-n'+1$\cr
      &\cr
     {R_{j-N+n'-1} -1  \over R_n R_{n'-n} -1} & for $N-n'+2 \leq j \leq N-n$ \cr
      &\cr
     0                                    & for $j=N-n+1$\cr
      &\cr
     {R_{n'-n} (R_{j-N+n-1} -1)-1  \over R_n R_{n'-n} -1} 
                                          & for $N-n+2 \leq j \leq N$ \cr
      &\cr
     0                                    & for $j=N+1$~.\cr}
\label{saddlesolns}
\eeq

\begin{figure}[t!]
\centerline{
   \epsfxsize 3.0 truein \epsfbox {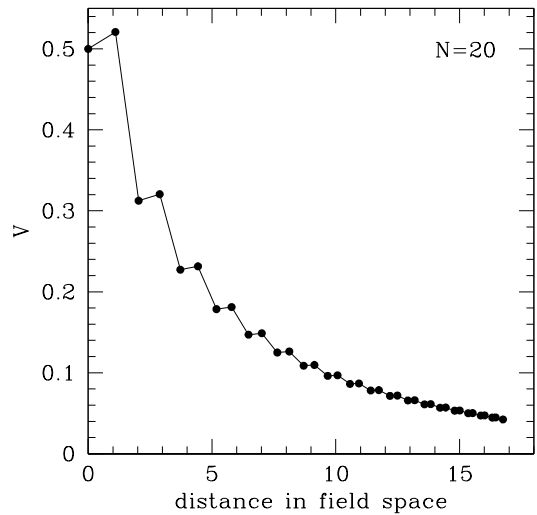}
 }
\caption{The vacuum structure of the $N=20$ model, plotted for $\chi=1/5$.
      Each of the local minimum points corresponds to a (meta)stable vacuum state, 
      while each local maximum point corresponds to the saddle-point configuration
      which directly connects the nearest-neighbor vacuum configurations on either side.
      For the purposes of this plot, we have then simply connected these points 
      sequentially with straight lines.
      The vertical axis indicates the energies of 
      these vacua or saddle points (the latter in their asymptotic limits),
      while the horizontal axis indicates the cumulative distances in field space
      along a trajectory which begins at the $n=1$ vacuum and then proceeds
      along straight-line path segments to the $(1,2)$ saddle point, then to
      the $n=2$ vacuum, then to the $(2,3)$ saddle point, and so forth.}
\label{fig:VPlot}
\end{figure}

These results are plotted in Fig.~\ref{fig:VPlot} for the $N=20$ model.
Clearly, the vacuum structure of this model consists of  
a ground-state vacuum along with a tower of 18 metastable vacua above it.
In Fig.~\ref{fig:VPlot}, we have shown the vacuum energies of these
vacua, along with the asymptotic energies of the saddle-point barriers which connect 
``nearest-neighbor'' vacua, as functions of a cumulative distance in field space
along a path that winds through each vacuum configuration and over each saddle
point as it comes down the tower, vacuum by vacuum. 
In essence, then, this figure forms a linear ``picture'' of the tower.

%================== FIGURE ============================================
%======================================================================== 

\begin{figure}[htb!]
\centerline{
   \epsfxsize 3.0 truein \epsfbox {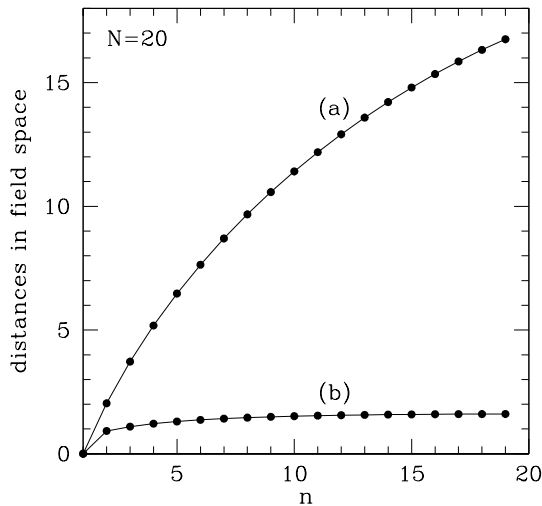}
 }
\caption{The geometry of our metastable vacuum towers in field space, here illustrated 
   for the $N=20$ model with $\chi=1/5$.  
   For each $n$-vacuum ($n=1,...,19$) in the $N=20$ vacuum tower,
   we have plotted two different notions of corresponding field-space distance.
   In curve~(a), we have plotted the  cumulative distance (as used in 
    Fig.~\protect\ref{fig:VPlot})
   that accrues as we wind our way from the top of the tower down to the $n$-vacuum,
    passing through nearest-neighbor saddle points along the way.
   By contrast, in curve~(b), we have plotted the direct straight-line distance
   between each $n$-vacuum and the reference $n=1$ vacuum at the top of the tower. 
   The dramatic difference between these two curves is a reflection of the fact
   the vacua in our vacuum tower actually lie along a ``spiral'' or ``helix''
   in the full $20$-dimensional field space.  }
\label{fig:VPlot2}
\end{figure}

Despite the relative simplicity of Fig.~\ref{fig:VPlot}, 
it is worth emphasizing that the geometry of the metastable vacuum tower in
the full $(N+1)$-dimensional field space is rather non-trivial.
For example, 
although 
the vacuum and saddle-point energies are plotted 
in Fig.~\ref{fig:VPlot}
versus a cumulative, integrated distance as we wind our way
down the vacuum tower, we could have just as easily defined
the field-space distance associated with any vacuum or saddle point in terms
of its straight-line distance directly back to a reference vacuum 
(such as the $n=1$ vacuum at the top of the tower).
The difference between these two different notions of distance is
shown in Fig.~\ref{fig:VPlot2} for the $N=20$ case.  
Indeed, as evident from the actual solutions given
Eq.~(\ref{vacsolns}), 
the vacua in our vacuum tower actually lie along a ``spiral'' or ``helix''
in the $N$-dimensional $v_{N+1}=0$ subspace of our full $(N+1)$-dimensional
field space.

%================== FIGURE ============================================
%======================================================================== 

\begin{figure}[htb!]
\centerline{
   \epsfxsize 3.0 truein \epsfbox {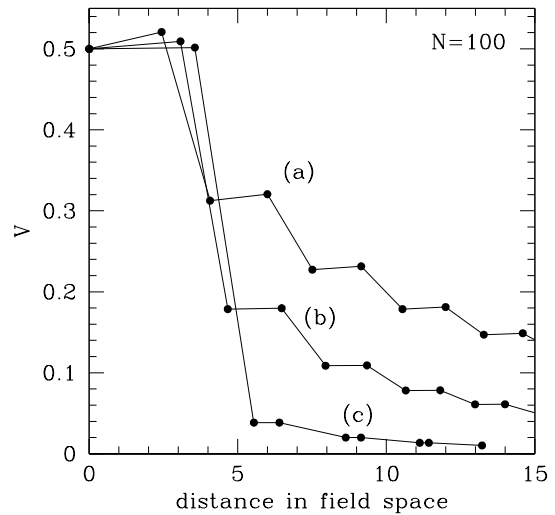}
 }
\caption{The upper portions of the $N=100$ vacuum tower
  with $\chi=1/5$,
   plotted for saddle-point ``hops'' of fixed magnitudes
   (a) $\Delta n= 1$,
   (b) $\Delta n=3$, and
   (c) $\Delta n=20$ 
   respectively.
   Note that the energy of a given vacuum
   is independent of how it is reached;  for example,
   we see that the $n=4$ vacuum has the same energy whether it
   is realized as the fourth minimum on the $\Delta n=1$
   curve or the second minimum on the $\Delta n=3$ curve. 
   However, the corresponding field-space distances 
   relative to the top of the tower 
     are generally smaller when larger
   hops are utilized when descending.}
\label{fig:FirstHops}
\end{figure}

It is also important to emphasize
that Fig.~\ref{fig:VPlot} shows 
only those
saddle-point barriers which exist between nearest-neighbor
vacua along the vacuum tower.
In actuality, however, 
there are saddle-point barriers which exist directly between 
any two vacua $(n,n')$. 
As a result, it is possible to imagine descending through the vacuum tower 
taking ``hops'' with different values of $\Delta n$ at each step. 
Fig.~\ref{fig:FirstHops} provides an illustration of 
the effects of different possiblities.

Finally, we turn to the one remaining issue:
the critical value $\lambda_{N,n}^\ast$ at which the $n^{\rm th}$ vacuum
in the tower is destabilized.
In general, for any $N\geq 2$ and $1\leq n\leq N-1$,
it turns out that $\lambda_{N,n}^\ast$ is given by
\beq
    \lambda_{N,n}^{\ast 2} ~=~   
           y^n ~ {\Gamma(y)\over \Gamma(n+y)}
                ~{R_n^{N-2} \over \chi (1+R_n)}
\label{lambdastarNn}
\eeq
where $y\equiv \chi/(1-2\chi)=n/(R_n-2)$ and where
$\Gamma(z)$ is the Euler $\Gamma$-function [for which
$\Gamma(z)= (z-1)!$ when $z\in\IZ^+$]. 
It is straightforward
to verify that these expressions reduce to the corresponding
expressions for the $N=3$ and $N=4$ cases plotted in Figs.~\ref{fig:lam3plot}
and \ref{fig:lam4plot} respectively. 
For example, using the result in Eq.~(\ref{lambdastarNn}), we find
that
\beq
           \lambda_{N,1}^\ast~\equiv~ \sqrt{1\over 1+\chi}
              ~\chi^{1-N/2}~.
\eeq

%================== FIGURE ============================================
%======================================================================== 

\begin{figure}[htb!]
\centerline{ \epsfxsize 2.9 truein \epsfbox {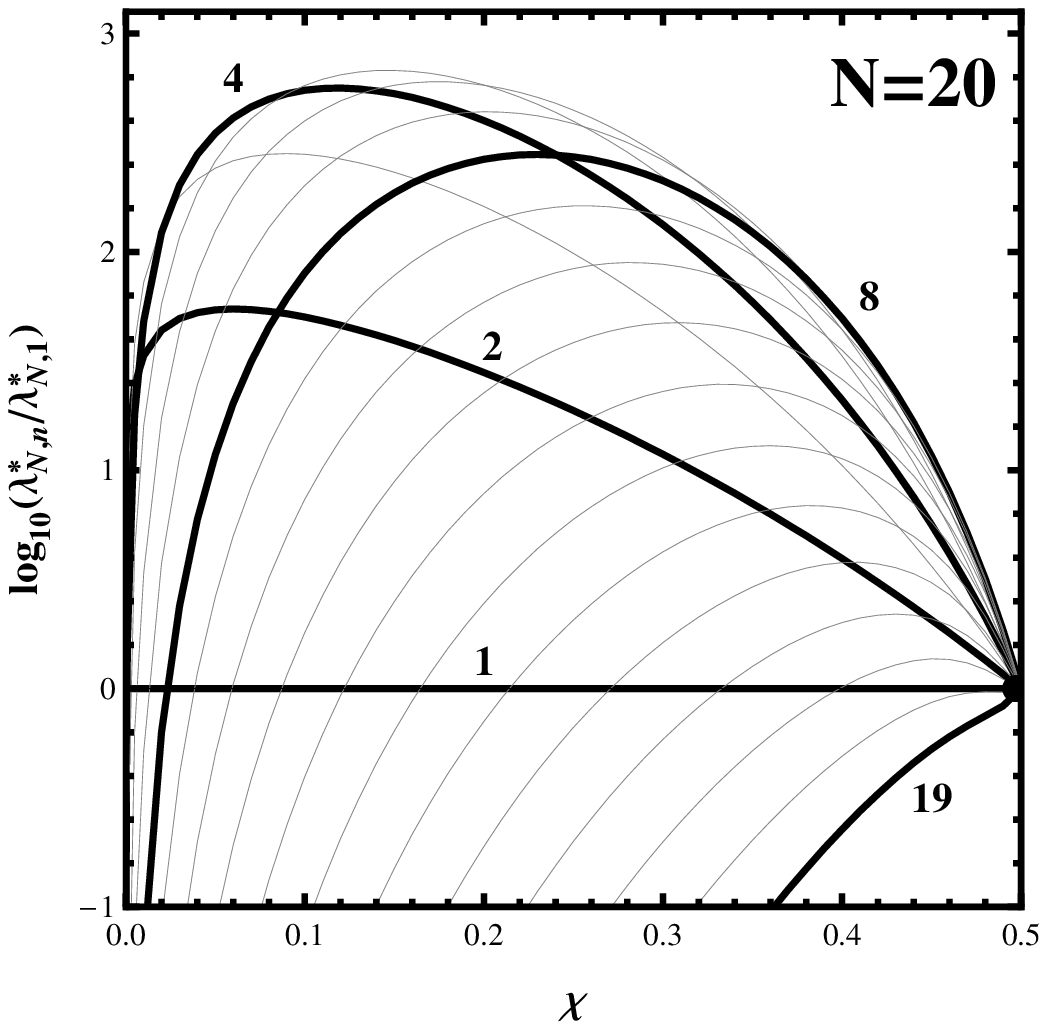} }
\caption{Values of $\lambda_{N,n}^\ast/\lambda_{N,1}^\ast$, plotted as functions of $\chi$
    for $N=20$ and $1\leq n\leq 19$.  The curves corresponding to $n=1,2,4,8,19$
    are highlighted and labeled;  the rest are not highlighted but proceed in sequential
    order relative to those which are.  For $\chi\approx 0.05$, we
    see that the $n=2$ vacuum has the highest
    value of $\lambda_{N,n}^\ast$, while 
     it is the $n=4$
     vacuum which has this property
     for $\chi\approx 0.45$. 
      The upper ``envelope'' of all of these curves
    corresponds to the critical stability value $\lambda_N^\ast$ 
    for the vacuum tower as a whole.
    Note that all of these curves converge to $\lambda_{N,n}^\ast=
         \sqrt{2^{N-1}/3}$ at $\chi=1/2$.}
\label{crossings}
\vskip 0.4truein
\centerline{ \epsfxsize 2.9 truein \epsfbox {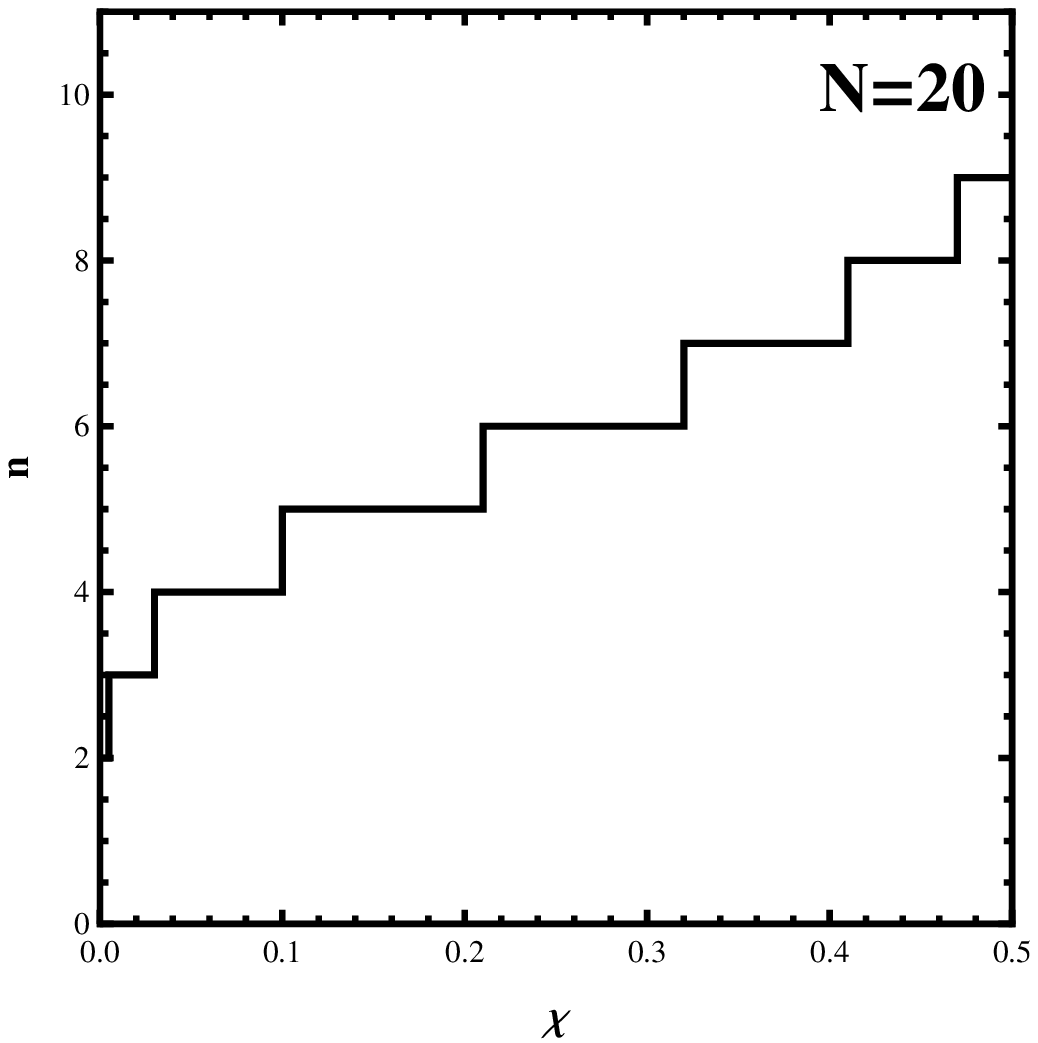} }
\caption{The $n$-index of the vacuum with the largest corresponding critical 
       value $\lambda_{N,n}^\ast$,
     plotted as a function of $\chi$.  This index specifies which vacuum in the tower
     is the first to destabilize as $\lambda$ is lowered from infinity.
     In general, for large $N$, this index shifts from 
     $n=1$ to $n=[N/2]$ (where [x] is the greatest integer $\leq x$)     
     as $\chi$ increases within the range $0 < \chi < 1/2$.}
\label{criticalns}
\end{figure}

In Fig.~\ref{crossings},
we plot the values of $\lambda_{N,n}^\ast$ as functions of $\chi$ 
for $N=20$ and $1\leq n\leq 19$.
Unlike the simpler $N=3$ and $N=4$ cases, however,
we see that it is no longer true that $\lambda_{N,n}^\ast > \lambda_{N,n'}^\ast$
for any $n< n'$.
Instead, as $N$ increases, we see that a complicated ``crossing'' pattern
develops as a function of $\chi$.  As a result of this crossing pattern,
the value of $n$ which results in the maximum value of $\lambda_{N,n}^\ast$
itself varies with $\chi$, as shown in Fig.~\ref{criticalns}.
Nevertheless, for any value of $\chi$, we see that our entire vacuum tower 
will be stable if $\lambda>\lambda_N^\ast$, where
\beq
   \lambda_N^\ast ~\equiv~ {\displaystyle \max_{1\leq n\leq N-1}}   \, \lambda_{N,n}^\ast~.
\eeq
Thus, $\lambda_N^\ast$ corresponds to the upper ``envelope'' of the
curves shown in Fig.~\ref{crossings}.

%================== FIGURE ============================================
%================== FIGURE ============================================

The crossing pattern shown in Fig.~\ref{crossings} implies
that our tower of metastable vacua will experience a non-trivial
destabilization pattern as $\lambda$ is reduced from infinity.
For any value of $\chi$ in the range $0<\chi< 1/2$, the 
first vacuum to be destabilized is
indicated in Fig.~\ref{criticalns}.    
This serves as the initial destabilization point on the tower. 
Reducing $\lambda$ still further
then induces a destabilization of the vacua immediately above
and below this point,
and further reductions in $\lambda$ result in a destabilization ``wave''
which simultaneously runs both up and down the vacuum tower from 
this initial point until
ultimately all vacua are destabilized.

%================== FIGURE ============================================
\begin{figure}[htb!]
\centerline{
   \epsfxsize 3.1 truein \epsfbox {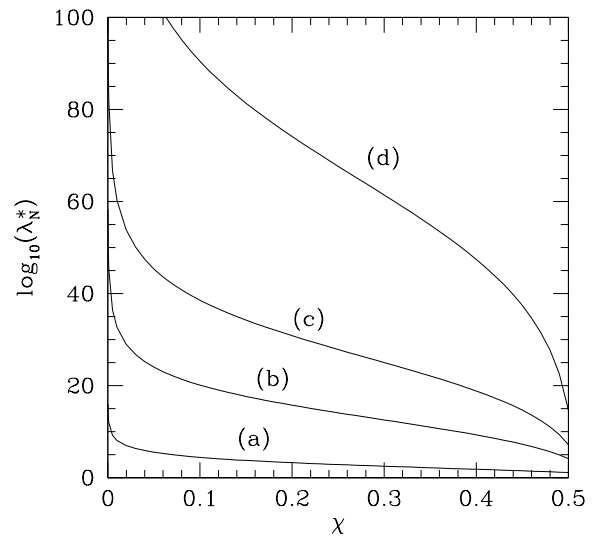}
 }
\caption{
    The critical values $\lambda_N^\ast$, plotted as functions of $\chi$
    for (a) $N=10$, (b) $N=30$, (c) $N=50$, and (d) $N=100$.  
    We observe that $\lambda_N^\ast$ generally grows with $N$,
    diverging as $\chi\to 0$ and asymptoting to $\sqrt{2^{N-1}/3}$ 
    as $\chi\to 1/2$.  We stress that the $\lambda$-values plotted here
    are the {\it rescaled, dimensionless}\/ versions of our original Wilson-line
    coefficients.  As such, there is no conflict with the perturbativity of
    our model.}
\label{lamstarN}
\end{figure}  
%================== FIGURE ============================================

Finally, the dependence of $\lambda_N^\ast$ on $N$ is shown in Fig.~\ref{lamstarN}.
We see that $\lambda_N^\ast$ generally grows rather quickly with $N$.
Moreover, as already anticipated from the $N=3$ and $N=4$ special cases,
$\lambda_N^\ast$ always diverges as $\chi\to 0$ 
and asymptotes to $\lambda_N^\ast\to \sqrt{2^{N-1}/3}$ 
    as $\chi\to 1/2$.  
The fact that 
$\lambda_N^\ast$ diverges as $\chi\to 0$ for every $N$
indicates that kinetic mixing plays a critical role in keeping
our metastable vacuum tower stable.

We also stress that the $\lambda$-values plotted in Fig.~\ref{lamstarN}
are the {\it rescaled, dimensionless}\/ versions of our original Wilson-line
coefficients.  
Indeed, they are only rescaled variables, not to be confused with our primordial
Lagrangian couplings;
indeed, as discussed earlier, this rescaling absorbs powers of the
underlying gauge coupling $g$ and the Fayet-Iliopoulos coefficient $\xi$.
This rescaling is thus partly responsible for the rise in $\lambda_N^\ast$
as a function of $N$.  However, having such large values of $\lambda$ naturally
begs the question as to whether the stability of our vacuum towers is in conflict
with the assumed perturbativity of our model.
However, as we shall see below, there is no conflict between the two, and indeed
such large values of the rescaled $\lambda$ do not 
in and of themselves undermine the validity of our 
tree-level calculations.

%======================================================================================================
\subsection{Perturbativity and Mass Scales}

The results obtained thus far have rested on the assumption that
the physics of our model is perturbative at all relevant scales and is therefore
accurately approximated by tree-level calculations.  
However, the true vacua of our theory 
are those field configurations which minimize the {\it full}\/ effective 
potential $V_{\mathrm{eff}}(\phi_i)$, and this includes radiative corrections.  
Thus, the tower of metastable vacua which we have presented
above is guaranteed to be an accurate description of the actual vacuum structure 
of our model if and only if such corrections are small 
and $V_{\mathrm{eff}}(\phi_i)\approx V_{\mathrm{tree}}$, with $V_{\mathrm{tree}}$ given 
in Eq.~(\ref{Vdef}).  Indeed, this must hold within the vicinity of the 
solution to the classical potential for each applicable vacuum index $n$.

We will now demonstrate that there is no problem
satisfying all applicable perturbativity constraints, provided the gauge coupling
$g$ is taken to be sufficiently small.  Moreover, since small $g$ is also beneficial for 
stabilizing the vacua in the tower,
we shall find that there is no conflict between the constraints 
stemming from perturbativity and those stemming from vacuum 
stability --- even as $N\to\infty$.

Note that in this subsection only, we shall revert 
back to our original {\it un}\/rescaled dimensionful parameters 
in order to expose the explicit dependence of our physical quantities on
the gauge coupling $g$.
This will restore factors of $g$ and $\xi$ in many of our previous expressions.
For example, in terms of the original unrescaled 
dimensionful quantities $\lambda$ and $\xi$,
our expression for $\lambda_{N,n}^\ast$ in Eq.~(\ref{lambdastarNn})
takes the form
\beq
    \lambda_{N,n}^{\ast 2} ~=~   
           y^n ~ {\Gamma(y)\over \Gamma(n+y)}
                ~{R_n^{N-2} \over \chi (1+R_n)}
                  ~ g^2 \, \xi^{2-N}~.
\label{truelambdastar}
\eeq
Similar modifications to other expressions occur as well.

It turns out that a great deal of information about the effective 
potential can be gleaned from non-renormalization theorems~\cite{Grisaru}.  
For example, in supersymmetric field theories, the superpotential is not 
renormalized by perturbative effects, except via wavefunction renormalization.  
Moreover, this holds true even when the superpotential includes 
non-renormalizable operators~\cite{Weinberg}.  As a result of such
theorems, we 
expect corrections to $V_F$ to arise only at scales near or below the 
supersymmetry-breaking scale $g\sqrt{\xi}$.  
However, these are typically the energy scales in which we are interested.
Moreover, corrections to the 
K\"{a}hler potential do not, in general, vanish in supersymmetric theories.  
Thus, it will be necessary to discuss both sorts of corrections.

%================== FIGURE ============================================
\begin{figure}[thb!]
\centerline{
   \epsfxsize 3.0 truein \epsfbox {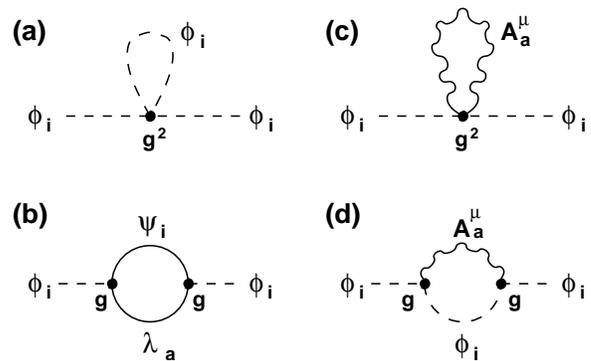} }
\caption{Diagrams contributing to the wavefunction renormalization 
of $\phi_i$ at one loop.  Each of these diagrams is proportional to $g^2$.
  Diagram (a) arises
due to the quartic couplings in the $D$-term potential, while diagrams
(b), (c), and (d) arise due to gauge-interaction terms for the scalars
and the supersymmetrization thereof.}
\label{props}
\end{figure}  
%======================================================================== 

We begin by addressing radiative corrections to the K\"{a}hler potential.  
These arise due to diagrams which contribute to wavefunction renormalization 
of the various fields in the theory, and depend on $g$.
For our purposes, it will be sufficient to focus on one-loop radiative 
corrections to the $\phi_i$ propagator;  the contributing diagrams are then shown 
in Fig.~\ref{props}.  Let us begin by considering the contribution from diagrams 
with scalars running in the loop.  There are $N+1$ such diagrams, and they
result in a net contribution 
\begin{equation}
     \sum_j^{N+1}\frac{g^2}{16\pi^2}\, T_{ij} \, m_j^2\, \ln\left(\frac{m_j^2}{\mu^2}\right)~
\label{eq:ScalDiagSum}
\end{equation}
where $m_{j}$ is the mass of $\phi_j$, where $\mu$ is 
an arbitrary renormalization scale, and where
\begin{equation}
       T_{ij}~\equiv~\sum_{a=1}^{N}\, \hat{Q}_{ai}\hat{Q}_{aj}~.
\end{equation}
If $T_{ij}$ were an arbitrary matrix,
the expression in Eq.~(\ref{eq:ScalDiagSum}) 
would scale roughly as $N+1$ for large $N$, and the theory would rapidly become 
non-perturbative.  

This is not the case, however, due to certain properties of $T_{ij}$ 
which are essentially consequences of the moose structure of the model.  
In particular, all entries along the diagonal of this matrix are positive and 
$\mathcal{O}(1)$.  Likewise, all 
elements with $i\neq j$ satisfy $-1<T_{ij}<0$, and the sum of elements in any row or 
column of $T$ vanishes.  Together, these properties imply that the contribution to the 
renormalization of the K\"{a}hler potential from scalar loops is 
essentially independent of $N$.  Moreover, each 
of the additional diagrams in Fig.~\ref{props} yields a contribution 
to the two-point function for $\phi_i$
which is roughly of the 
order of $T_{ii}$ (no sum on $i$). 
This is also essentially independent of $N$.  Consequently, the renormalization of the 
K\"{a}hler potential is under control, and radiative corrections of this sort can 
be safely neglected as long as $g \ll 4\pi$ --- even for very large $N$.

%================== FIGURE ============================================
\begin{figure}[thb!]
\centerline{
   \epsfxsize 3.0 truein \epsfbox {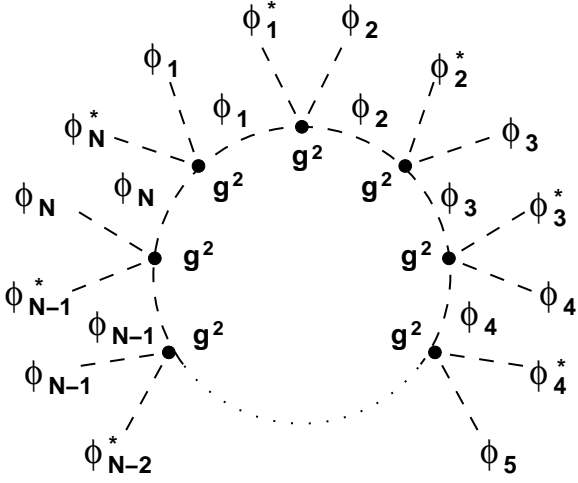} }
\caption{A representative of the class of diagrams which contribute to the
renormalization of the effective $2N$-scalar vertex at one loop in the
broken phase of the theory, due to quartic $D$-term interactions.
The diagram contains $N$ vertices (each proportional to $g^2$)
and $N$ scalar propagators (with masses
proportional to $g^2\xi$);  the overall contribution is thus
proportional to $g^2$.  }
\label{circle}
\end{figure}  
%======================================================================== 

The second class of diagrams we must consider are corrections to the $2N$-field 
couplings in $V_F$ which come from the terms in $V_D$.  In other words, these are corrections
to the superpotential coupling $\lambda$ which arise from non-zero $g$.
The leading such contribution arises at 
one-loop order from diagrams of the sort depicted in Fig.~\ref{circle}, along with 
additional diagrams in which gauge bosons and gauginos run in the loop.  
Note that similar diagrams were examined in Ref.~\cite{DeconOperatorCorrex}.
In the limit of unbroken supersymmetry, of course, these contributions would sum to zero.  

Each of the 
contributing diagrams of the sort pictured in Fig.~\ref{circle} contains 
$N$ vertices, and each of these vertices contributes a factor of $g^2T_{ij}$ 
as well as $N$ scalar 
propagators.  In any $n$-vacuum, the scalar masses are expected to be 
$\sim \mathcal{O}(g^2\xi R_n^{-1})$, so each diagram is proportional to $g^2$.  There 
are $N!\sim N^N$ such diagrams, but each is proportional to 
$\mathrm{Tr}[T_{i_1i_2}T_{i_2i_3}\ldots T_{i_{N-1}T_{N}}]$ and hence suppressed 
by a factor of $T_{ij}^N\sim (1/N)^{N}$, where $i\neq j$.  Thus, as was the case 
with the wavefunction-renormalization calculation above, the $N$-dependence 
essentially cancels.  Thus, as long as $g\ll 4\pi$, this contribution too can be neglected ---
regardless of the value of $N$.  By the same token, contributions to other 
effective operators which involve couplings of various numbers of scalar fields 
can also be safely neglected.

%================== FIGURE ============================================
\begin{figure}[thb!]
\centerline{
   \epsfxsize 3.0 truein \epsfbox {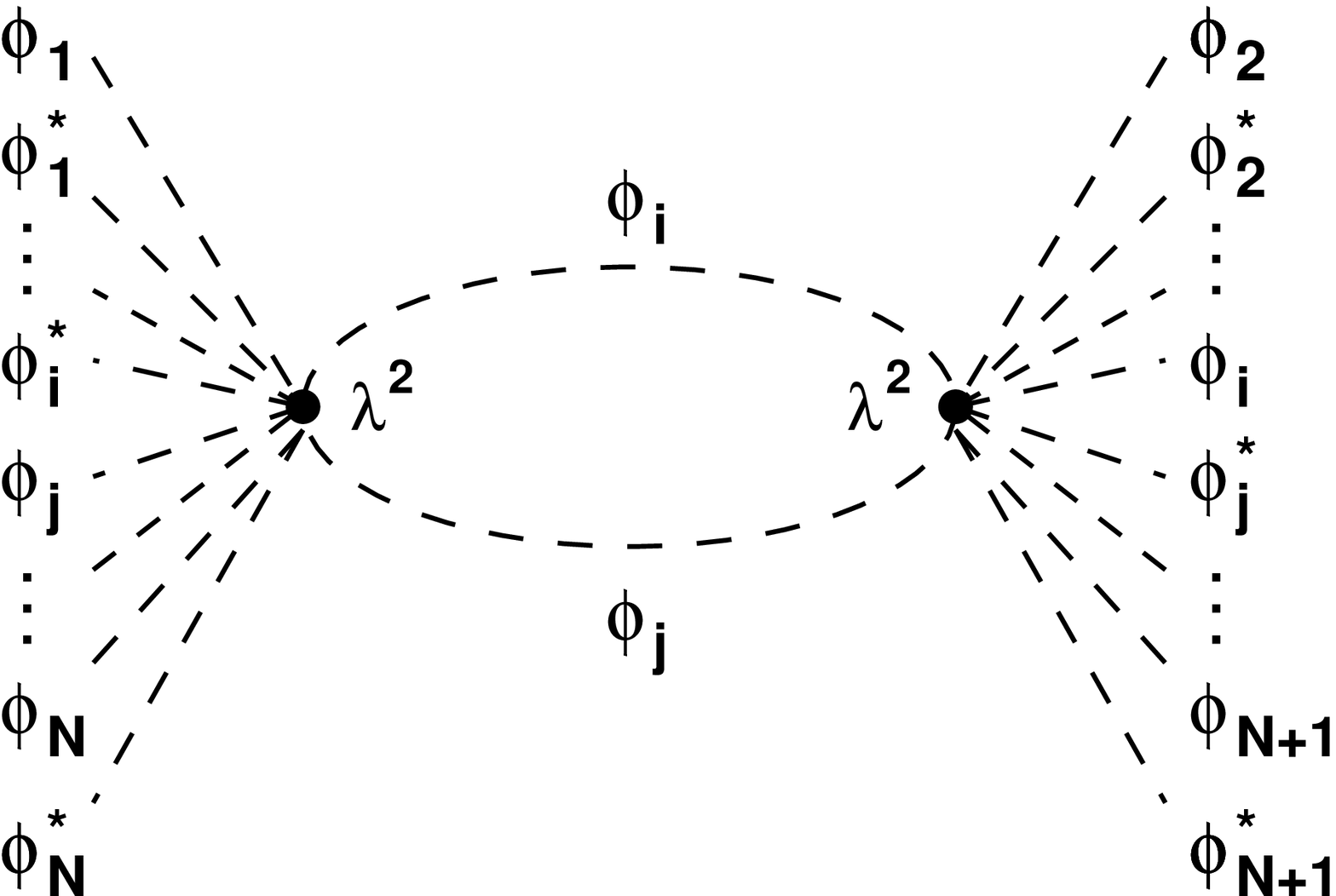} }
\caption{A representative of the class of diagrams which contribute to the
renormalization of the effective $2N$-scalar vertex at one loop in the
broken phase of the theory, due to $F$-term interactions.  This diagram
represents a $4N-4$-scalar vertex in which $2N-4$ of the scalars on the
external lines are assigned VEV's, and its contribution is proportional
to $\lambda^4$.  At high energies, when supersymmetry is effectively unbroken, this
diagram is cancelled by the contribution from the diagram in
Fig.~\protect\ref{superscotch}.}
\label{butterscotch}
\vskip 0.2 truein
\centerline{
   \epsfxsize 3.0 truein \epsfbox {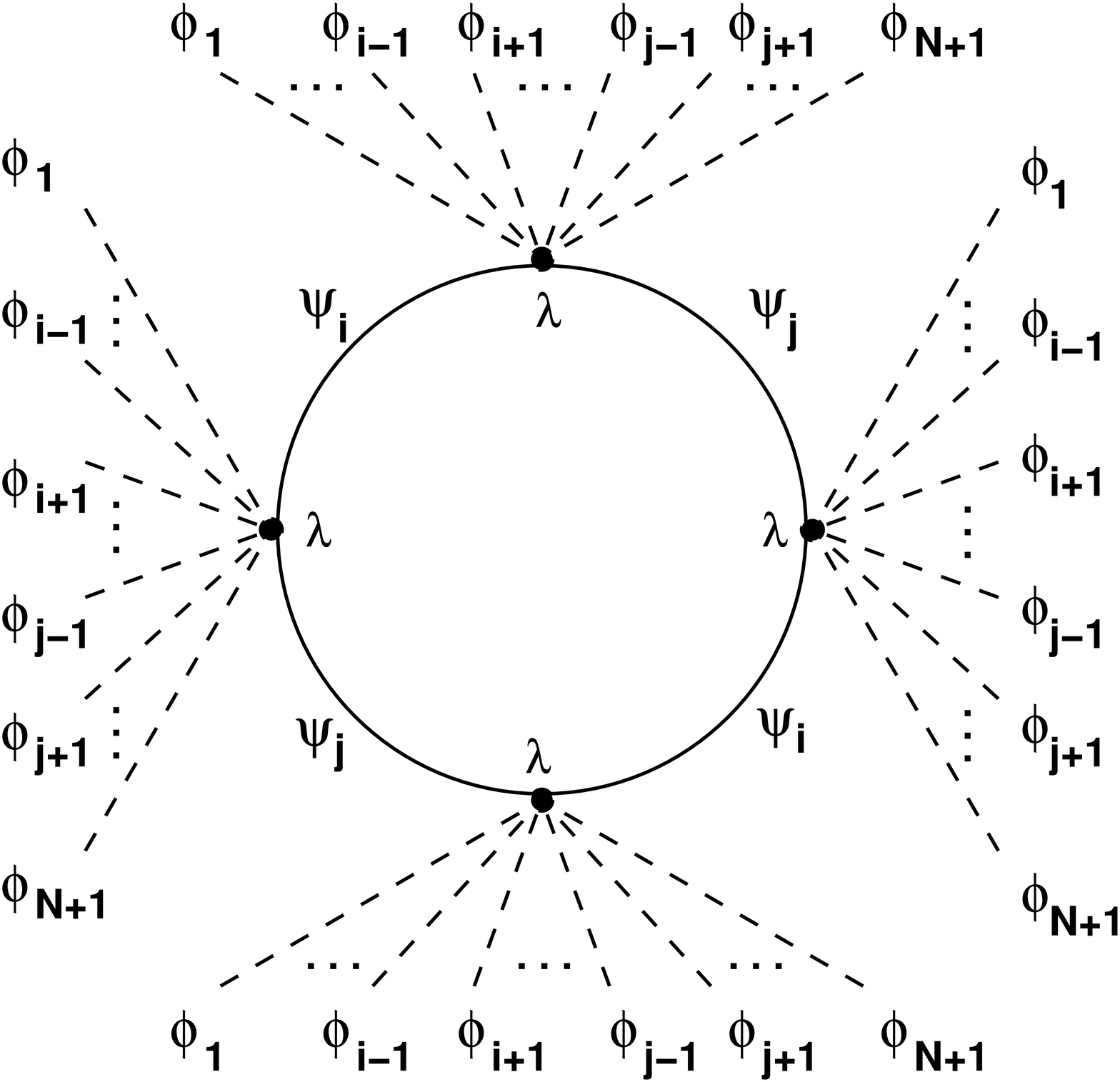} }
\caption{A representative of the class of diagrams which correct the effective
$2N$-scalar vertex at one loop.  The net contribution from this class of
diagrams would exactly cancel the contribution from the class of diagrams
depicted in Fig.~\protect\ref{butterscotch} if supersymmetry were unbroken.}
\label{superscotch}
\end{figure}  
%======================================================================== 

The final category of radiative corrections we must consider are corrections 
to the effective $2N$-scalar couplings, each of which has the tree-level 
coefficient $\lambda^2$.  Thus, these are essentially corrections to the
superpotential coefficient $\lambda$  
which themselves depend on $\lambda$.
The leading contribution arises at one-loop order from 
diagrams of the form shown in Fig.~\ref{butterscotch}, in which $2N-4$ fields 
are replaced by their VEV's, chosen appropriately for a given vacuum.
However, along with these contributions we must also include the contributions from  
diagrams of the form shown in Fig.~\ref{superscotch}, also with VEV's assigned to an appropriate
number of external fields.  Again, these contributions cancel in the limit of unbroken 
supersymmetry, but their contributions can be expected to survive below the 
supersymmetry-breaking scale.

In the $n$-vacuum, the 
one-loop contribution 
to the $(\prod_{i=1}^{N+1}|\phi|^2)/|\phi_{\ell}|^2$ vertex 
from a given diagram of the sort shown in Fig.~\ref{butterscotch} 
is roughly
\begin{equation}
    \frac{\lambda^4}{16\pi^2}\,
    P_{ijq}\, f(m_i^2,m_{j}^2)\,
    \frac{\prod_{k=1}^{N+1}v_k^2}{v_{N+1}^2v_{N-n+1}^2v_q^2}~.
\label{eq:lamcorslam}
\end{equation}
Here $f(m_j^2,m_{j}^2)$ is a function of the 
masses $m_i^2$ and $m_{j}^2$ whose dependence on these masses is essentially 
logarithmic, 
while $q$ denotes the index of the third scalar field $\phi$ whose
VEV is missing from the contribution arising from
the particular diagram in question.
Likewise, $P_{jkq}$ is a combinatorial factor representing the number 
of ways of assigning the appropriate VEV's to the external fields.  Roughly 
speaking, we find that $P_{jkq}\sim 2^{N+1}$ for all $(j,k,q)$.  
Similarly, there are $\sim(N-2)(N-1)$ diagrams which
contribute at one-loop order to any given $2N$-scalar vertex.  

Since the product of VEV's in Eq.~(\ref{eq:lamcorslam}) 
also appears in Eq.~(\ref{truelambdastar}), 
we may rewrite our results directly in tems of 
$\lambda^{\ast}_{N,n}$.  Thus, 
we find that this correction will be small compared to the tree-level 
term in any particular $n$-vacuum 
as long as
\begin{equation}
   \frac{g^2}{16\pi^2}\, 2^{N+1}\, (N-1)(N-2)\, 
         \frac{\lambda^4}{(\lambda^{\ast})^2}\, c_f ~\ll~\lambda^2~.
\end{equation} 
Here $c_f$ is an $\mathcal{O}(1)$ coefficient which embodies the 
effect of including the various
$f(m_i^2,m_{j}^2)$ functions.  

We recall, however, that we must also satisfy Eq.~(\ref{range2}) in order
to guarantee the stability of our entire vacuum tower. 
Thus, combining these two results, we find 
that the conditions under which both perturbativity and vacuum-stability 
constraints are simultaneously satisfied are given by 
\begin{equation}
  1 ~<~ \left(\lambda\over\lambda_N^\ast\right)^2 ~\ll~ 
     \frac{16\pi^2}{2^{N+1} g^2 (N-1)(N-2) c_f}~. 
\label{eq:LambdaRange}   
\end{equation}
Thus, there is no problem satisfying these two inequalities simultaneously
provided
\beq
                   g^2 ~\ll~ {16\pi^2 \over 2^{N+1} (N-1)(N-2) c_f}~.
\label{gconstraint}
\eeq
We observe that this is also consistent with our previous constraint that 
$g\ll 4\pi$. 

We conclude, then, that as long as the gauge coupling $g$ satisfies Eq.~(\ref{gconstraint})
and $\lambda$ lies within the range specified in Eq.~(\ref{eq:LambdaRange}), our
model will remain perturbative for arbitrary $N$ 
without compromising the stability of the vacuum tower.  
As a result, the tree-level results we have presented 
above are robust against quantum corrections.
Of course, we observe that the required values of $g$ tend to be rather small
when $N$ becomes large.  However, $N$ need not necessarily be taken large
for all possible phenomenological applications.  Moreover, situations in which
$N$ is taken to be extremely large tend to be higher-dimensional deconstruction-type 
scenarios in which we would naturally expect our four-dimensional gauge coupling to take
an extreme value.  Indeed, it is not unnatural to expect that the fine-tuning
inherent in whatever drives $N\to\infty$ can also simultaneously drive $g\to 0$.
Unfortunately, the details of such a mechanism 
lie within the full physical framework into which such
a model is ultimately embedded, and thus requires a UV completion before they
can be adequately addressed.  

Our purpose here, however, has been to demonstrate
that there exists a window in which both perturbativity and stability constraints
can be simultaneously satisfied for any value of $N$.
As we see from the above discussion, this is indeed the case.

Given these observations,
it is interesting to investigate the degree 
to which our model can be considered natural. 
For a given choice of model parameters,
and for all $N>2$, this model contains two
dimensionful parameters: the Wilson-line coefficient $\lambda$ and the
Fayet-Iliopoulos term $\xi$.  Clearly, we can associate a mass scale with each of
these parameters, defining $\mu_{\lambda}$ and $\mu_\xi$ such that
$\lambda\equiv \mu_\lambda^{2-N}$ and $\xi\equiv \mu_\xi^2$.  If we assume
that  $\xi$ and $\lambda$ are generated at the same underlying scale $\mu$
by the same physics, then our model can be considered natural from an effective
field theory point of view as long as $\lambda=c_{\lambda}\mu^{2-N}$ and
$\xi=c_{\xi}\mu^2$, where $c_{\lambda}$ and $c_{\xi}$ are both
$\mathcal{O}(1)$ coefficients.
We shall take this to be our definition of naturalness from an effective
field theory point of view~\cite{BiraEFT}.

The question that arises, then, is whether our model meets this criterion.
Recall that in our model, the particular values chosen for $\lambda$ and $\xi$ are constrained by
the vacuum stability and perturbativity requirements embodied in
Eq.~(\ref{eq:LambdaRange}), which in turn depend on the underlying model parameters
primarily through $\lambda_N^\ast$.  When written in terms of the
scales $\mu_\lambda$ and $\mu_\xi$, this quantity (in rescaled
variables) is proportional to
\begin{equation}
  \lambda_{N}^{\ast}~\propto ~\frac{1}{g}\left(\frac{\mu_{\xi}}{\mu_{\lambda}}\right)^{N-2}
    ~=~
   \frac{1}{g}\, c_{\lambda}\, c_{\xi}^{N/2-1}~.
\end{equation}
Note, in particular,
that this expression contains $c_{\xi}$
taken to the $N/2-1$ power.  As a result, the extremely large values of our rescaled
dimensionless $\lambda$ which are required for vacuum stability 
are not in conflict with either perturbativity constraints or naturalness
considerations.  Indeed, all that is required is that $c_\xi$ be slightly larger 
than (but still of order) one.

%======================================================================================

\section{Dynamics on the Vacuum Tower:
Cascades, Collapses, Great Walls, and Forbidden Cities\label{sec:Lifetimes}}

We now turn to the issue of dynamics within the metastable vacuum
tower.  What will be the pattern of tunneling-induced vacuum decays along
the entire length of this tower?

Let us begin by recalling the simpler situation that arises if we have
only two vacua separated by a single saddle-point barrier, with one vacuum 
state having higher vacuum energy than the other.
In such a situation, the state with higher energy
can decay to the state with lower energy 
via instanton transitions,
the rate (per unit volume) for which may be parametrized as~\cite{Instantons}
\begin{equation}
        \frac{\Gamma_{\rm inst}}{{\rm Vol}} ~=~ A\, e^{-B}~.
\label{decayrate}
\end{equation}
We will not be particularly concerned with the form of the coefficient $A$.
Instead, we will focus our attention on the exponent $B\equiv S_E(\phi_+,\phi_-)$,
usually referred to as the {\it bounce action}\/, which represents the Euclidean
action evaluated along the classical path in field space which
connects $\phi_+$, 
the field-space location of the higher-energy vacuum state,
to $\phi_-$, 
the field-space location of the lower-energy vacuum state,
through 
the field-space location of the saddle point between them.
In general, one can evaluate $B$ as in Ref.~\cite{Triangles} by approximating the
potential along the classical path between the two vacua as a triangle.
In this approximation, the bounce action depends
on four parameters: 
$\Delta \phi_{\pm}$, which is the distance in field space between 
the top of the potential
barrier and the higher-energy ($+$) or lower-energy ($-$) vacuum;
and $\Delta V_{\pm}$, which is the potential difference between
the top of the barrier and each respective vacuum state.
Note that in a multi-dimensional field space, 
the use of these results also intrinsically embodies a further
approximation, namely that 
the classical path of least action follows
a trajectory in field space consisting of two straight-line segments 
(one from the higher-energy vacuum to 
the saddle point, and the second from the saddle point to the lower-energy vacuum).
However, this turns out to be a reasonably good approximation.  

Calculating $B$ is then relatively straightforward~\cite{Triangles}.
When
\begin{equation}
  \frac{\Delta\phi_-}{\Delta\phi_+}~\geq~\frac{\sqrt{1+c}+1}{\sqrt{1+c}-1}~
\label{eq:DecayIneq}
\end{equation}
with $c\equiv(\Delta V_-/\Delta V_+)(\Delta\phi_+/\Delta\phi_-)$, the bounce action
is given by~\cite{Triangles}
\begin{equation}
  B ~=~\frac{32\pi^2}{3 g^2}
      \frac{1+c}{(\sqrt{1+c}-1)^4}\left(\frac{\Delta \phi_+^4}{\Delta V_+}\right)~.
\label{forma}
\end{equation}
By contrast, when the inequality in Eq.~(\ref{eq:DecayIneq}) is not satisfied,
the appropriate expression is instead given by~\cite{Triangles}
\begin{eqnarray}
   B ~&=&~\frac{\pi^2}{96 g^2}\left(\frac{\Delta V_+}{\Delta\phi_+}\right)^2 R_T^3\nonumber\\
        && ~~ \times
          \left(-\beta_+^3+3c\beta_+^2\beta_-+3c\beta_-^2\beta_+-c^2\beta_-^3\right)~,
\label{Bform1}
\end{eqnarray}
where $\beta_{\pm}$ and $R_T$ are given by
\begin{equation}
     \beta_{\pm}\equiv \left(\frac{8\Delta \phi_{\pm}^2}{\Delta V_{\pm}}\right)^{1/2}~,~~~~~
       R_T\equiv \half\left(\frac{\beta_+^2+c\beta_-^2}{c\beta_- -\beta_+}\right)~.
\label{Bform2}
\end{equation}
It can be verified that these solutions match smoothly at the point where Eq.~(\ref{eq:DecayIneq})
is saturated.

Note that the factors of $g^2$ which appear in the denominators of Eqs.~(\ref{Bform1}) 
and (\ref{Bform2}) arise from the fact that we are using rescaled energies $\Delta V_\pm$
and field-space distances $\Delta \phi_\pm$ in these expressions, in accordance
with the discussion in Sect.~II.~
In the following, we shall take $g=1$ for simplicity in all numerical evaluations of
the bounce action.

This is the situation that emerges when there are only two vacua to consider.
However, in this paper we face a situation in which we have a whole tower of metastable vacua
with many possible pairwise saddle-point solutions. 
The situation we face is therefore significantly more complicated than
that sketched above.

In order to approach this situation, therefore, 
we begin with some preliminary observations.
First, we observe that 
if we are interested in transitions between an initial vacuum $n_i$
and a final vacuum $n_f$, we need only consider the leading
quantum-mechanical transition amplitude $\langle n_f|n_i\rangle$,
corresponding to the bounce action $B(n_i,n_f)$.
Although
higher-order quantum-mechanical contributions
of the form $\sum_n \langle n_f|n\rangle\langle n|n_i\rangle$
can appear when there are more than two vacua,
such contributions are all exponentially suppressed.
It is therefore sufficient to examine the bounce action $B(n_i,n_f)$ 
itself in order to determine the transition rate between two specified 
vacua $n_i$ and $n_f$.

Second, we observe that in general, a given initial vacuum state $n_i$
can decay into all possible final vacuum states $n_f$, where $n_i < n_f \leq N-1$.
As a quantum-mechanical issue, of course, all of these transitions
take place simultaneously, with rates determined by the corresponding
bounce actions $B(n_i,n_f)$.  However, once again, these transition rates
will typically experience huge, exponential variations as functions of the possible
value $n_f$.
Indeed, this will be the situation for all $0<\chi<1/2$.
As a result, we shall make a ``classical'' approximation in which
each vacuum state $n_i$ is assumed to decay to the unique final vacuum $n_f$
for which $B(n_i,n_f)$ is minimized, with a rate determined by $B(n_i,n_f)$.

Third, in order to evaluate $B(n_i,n_f)$,
we shall need explicit expressions for the energies of the $n_i$- and $n_f$-vacua
as well as the height of the saddle-point barrier which connects
them.  We shall also require the field-space separations between the two
vacua and the saddle point. 
While analytical expressions for these vacuum configurations exist (and
were given in Sect.~II),
we do not have analytical expressions for the saddle-point configurations
except in the ``asymptotic'' $\lambda\to\infty$ limit.
Therefore, although we will {\it not}\/ assume that $\lambda$ is actually infinite in what follows,
we {\it shall}\/ assume that $\lambda$ is sufficiently large that the asymptotic
saddle-point solutions given in Sect.~II may be utilized without significant error.
As we have already seen in Sect.~IID, this assumption is not necessarily in conflict with the
presumed perturbativity of our model;  indeed, the approach to the asymptotic limit
was sketched for the $N=3$ case in Fig.~\ref{fig:Saddle}, whereupon we observe
that the large-$\lambda$ asymptotic behavior emerges even for
relatively small values of $\lambda$.
The assumption of the large-$\lambda$ limit 
will also have the added advantage of removing the free variable
$\lambda$ from the subsequent analysis.

Finally, we shall deem a metastable vacuum to be ``stable'' if its lifetime
exceeds the age of the universe.  
More precisely, we demand a lifetime of such magnitude that
no decay event would be expected within our Hubble volume over a duration equal to 
the known age of the universe $\approx 13.7$~Gyr.  
Using Eq.~(\ref{decayrate}),
we can package this requirement as a constraint on the bounce action:
\beq
    B ~\gsim~ 471 \, +\,  4\, \ln \left(\frac{M_{\mathrm{inst}}}{M_{\mathrm{Planck}}}\right)~,
\label{eq:BUniverse}  
\eeq
where $M_{\mathrm{inst}}\equiv A^{1/4}$.  When this constraint is satisfied, the corresponding
vacuum in question is stable on cosmological time scales;  when it is not, the vacuum 
is assumed to have decayed sometime in the past.  In what follows, we will adopt 
the most conservative assumption
that $M_{\mathrm{inst}}\approx M_{\mathrm{Planck}}$, whereupon the logarithmic contribution
in Eq.~(\ref{eq:BUniverse}) can be ignored.
Thus, $B\approx 471$ shall serve as our critical bounce action for stability.

Given these assumptions, it is then possible to examine the corresponding
decay patterns along our entire metastable tower.
To do this, we need to understand the behavior of the bounce actions $B(n_i,n_f)$
as functions of $n_i$ and $n_f$ as we vary our two remaining variables, $\chi$ and $N$.
As we shall see, there are two principal modes of possible behavior patterns 
(``collapse'' and ``cascade'') which will emerge.

\subsection{Collapse behavior}

\begin{figure}[b!]
\centerline{
   \epsfxsize 3.3 truein \epsfbox {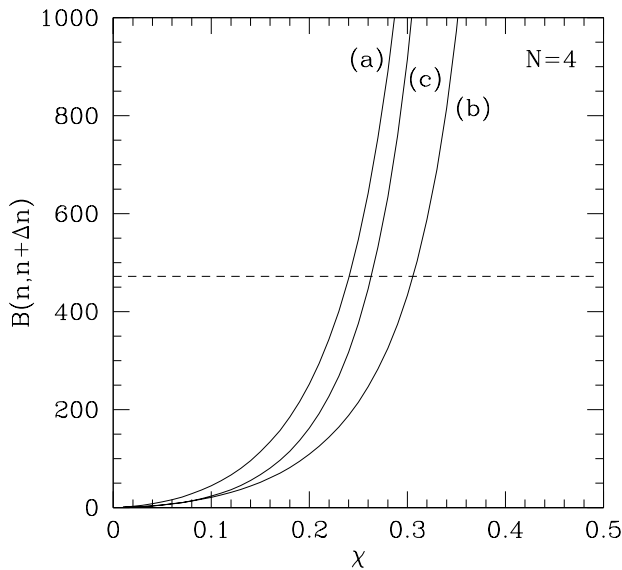}
 }
\caption{ Bounce actions for the $N=4$ model, plotted as functions of $\chi$.
In each case we have plotted $B(n,n+\Delta n)$ 
where (a)  $(n,\Delta n)=(1,1)$,
(b)  $(n,\Delta n)=(1,2)$,
and
(c)  $(n,\Delta n)=(2,1)$.
Also shown (dotted line) is the bounce action corresponding a
decay lifetime approximating the age of the universe. 
In general, all bounce actions increase with increasing $\chi$,
ultimately diverging as $\chi\to 1/2$.
However, for each value of $\chi$, we see that the bounce actions
corresponding to the greatest $\Delta n$ are smaller than 
those corresponding to smaller $\Delta n$.
 } 
\label{fig:LifetimesChi}
\end{figure}

To understand how these two different patterns arise,
let us begin our analysis by revisiting the
simple $N=4$ case discussed in detail in Sect.~IIB.
For $N=4$, the vacuum tower contains
three minima (the $n=1$, $n=2$, and $n=3$ vacua), and hence three 
different decay transitions are possible.
The bounce actions
$B(n_i,n_f)$ 
associated with these three transitions are plotted in Fig.~\ref{fig:LifetimesChi} as functions 
of $\chi$.  

%================== FIGURE ============================================
%======================================================================== 

This figure illustrates several general trends.
First, we observe the
\begin{itemize}
\item {\it General feature \#1:}\/ ~All of our bounce actions vanish as $\chi\to 0$ and
diverge as $\chi\to \infty$.  
\end{itemize}
This feature is easy to understand.  As $\chi\to 0$,
our entire vacuum tower becomes unstable, whereupon all of the possible decays out of any given
vacuum state become essentially instantaneous.  
Likewise, as $\chi\to 1/2$, the 
vacuum energy differences between any two vacua in the tower vanish.  
There is thus no ``driving force'' for decays in the $\chi\to 1/2$ limit, 
whereupon the lifetime of any given
metastable state approaches infinity and the states become truly stable
with respect to instanton-tunnelling transitions.

Second, we observe
from Fig.~\ref{fig:LifetimesChi}
that $B(2,3)<B(1,3)<B(1,2)$  for all $\chi$.
This implies that the $n=1$ vacuum decays preferentially 
 {\it not}\/ to the $n=2$ metastable state immediately below it,
but directly to the $n=3$ ground state. 
Moreover, if both the $n=1$ and $n=2$ vacua were somehow initially occupied
(\eg, in the different regions of the universe),
we find  that the $n=2$ region would decay to the ground state
before the $n=1$ vacuum region does.

\begin{figure}[thb!]
\centerline{
   \epsfxsize 3.0 truein \epsfbox {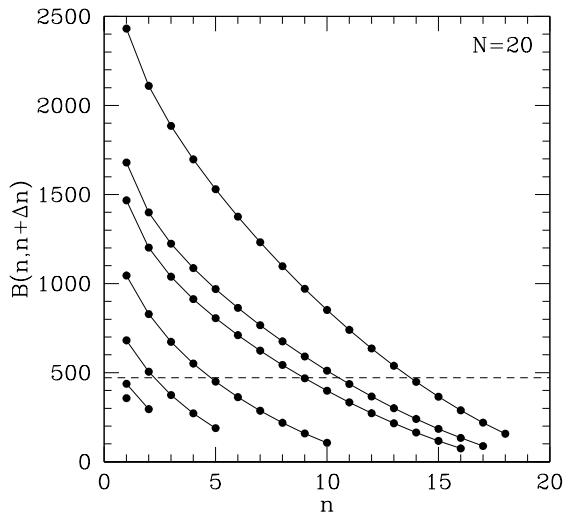}
 }
\caption{ Bounce actions for the $N=20$ theory with $\chi=1/5$.
      For each vacuum $1\leq n \leq 18$, we plot $B(n,n+\Delta n)$ 
      where the different ``curves'' (from top to bottom) correspond 
      to $\Delta n= 1, 2, 3, 9, 14, 17, 18$ respectively.
      Note that each plot is truncated when $n+\Delta n$ would exceed
      $n_{\rm max}=19$.
      We observe that for fixed $n$, the bounce action $B(n,n+\Delta n)$
      generally decreases with increasing $\Delta n$.
      As a result, for fixed $n$, transitions which maximize
      $\Delta n$ are generally favored.  } 
\label{fig:Lifetimes2}
\end{figure}

These observations are examples of two additional general features:
\begin{itemize}
\item {\it General feature~\#2:}\/~
    A given bounce action $B(n,n+\Delta n)$ tends to decrease
    with increasing $n$ if $\Delta n$ is held fixed.
\item {\it (Nearly) general feature~\#3:}\/~
      A given bounce action $B(n,n+\Delta n)$ tends to decrease 
      with increasing $\Delta n$ if $n$ is held fixed.
      (Important exceptions will be discussed below.)
\end{itemize}
Both of these features are illustrated for the $N=20$ model
in Fig.~\ref{fig:Lifetimes2}, where we plot the values of
the bounce actions $B(n,n+\Delta n)$ as functions of $n$
for a variety of fixed $\Delta n$.
               
%================== FIGURE ============================================
%======================================================================== 

These features also have direct physical consequences.
For example, Feature~\#3
implies that each metastable vacuum in our tower tends to decay directly 
to the ground state of the theory rather than to any other
metastable state of lower energy.
We shall refer to this type of behavior as a ``collapse'':
each state, one at a time,  suddenly drops directly to the ground state with $n=N-1$.
As we shall discuss below, Feature~\#3 (and thus the ensuing collapse behavior)
tends to hold for most values of $N$ and $\chi$;
indeed, the only exceptions tend to arise in the 
$\chi\to0$, $N\to\infty$ limit, with both $n, \Delta n \ll N$.
Thus, collapse behavior tends to dominate along the full length of
most metastable vacuum towers, and along the lower portions of 
all towers even when $\chi\to 0$ and $N\to\infty$.

If we imagine situations in which all vacuum states are initially
populated ({\it e.g.}\/, in different regions of the universe),
the specific collapse pattern along the metastable vacuum
tower becomes of particular interest.
To address this issue, we cannot hold $n$ or $\Delta n$ fixed;
we must vary both simultaneously in order to hold $n_f=n+\Delta n=N-1$ fixed.
In other words, we wish to examine
the bounce action $B(n,N-1)$ as a function of $n$.

\begin{figure}[thb!]
\centerline{
   \epsfxsize 3.0 truein \epsfbox {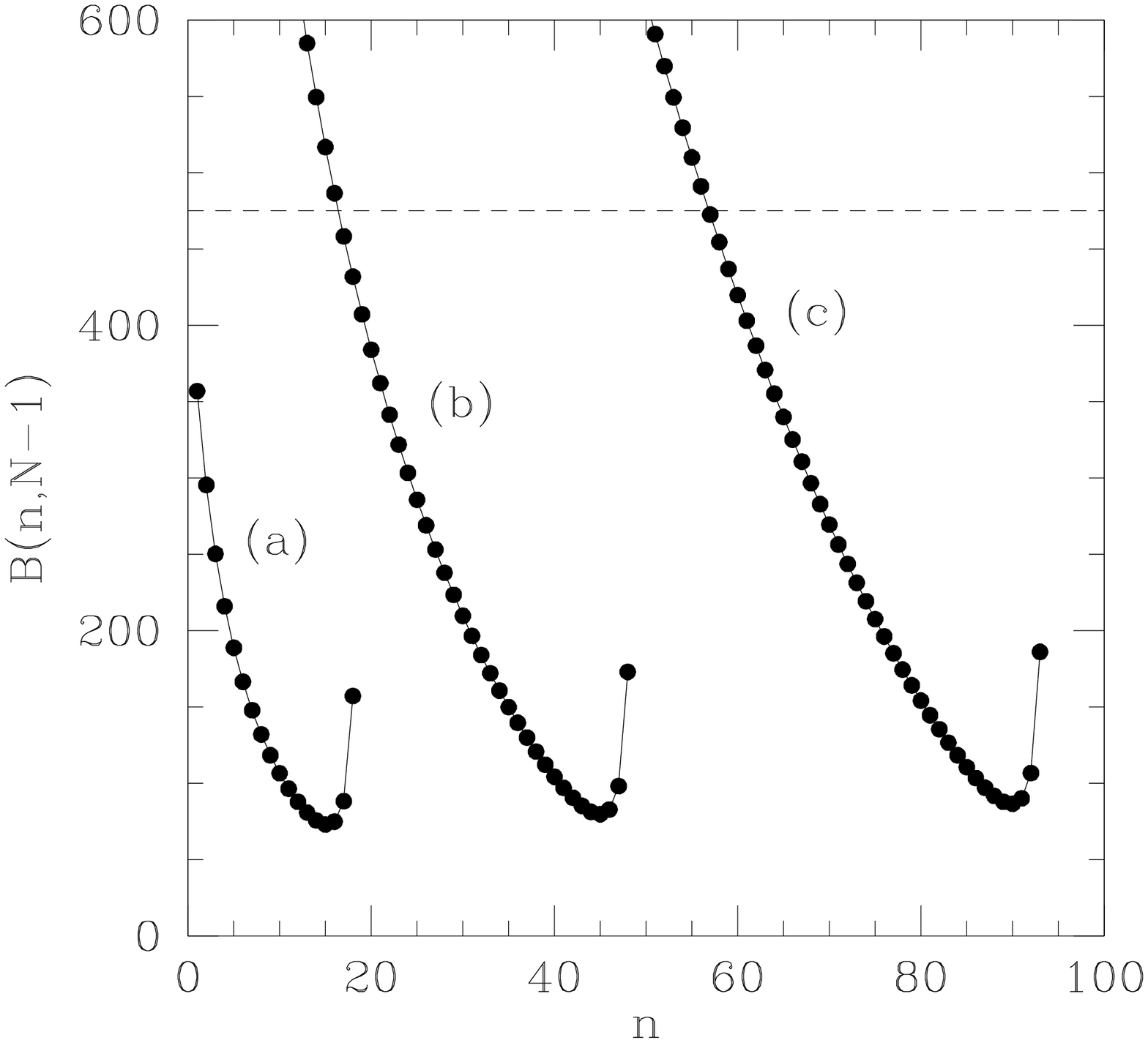}
 }
\vskip -0.38 truein
\caption{ Bounce actions $B(n,N-1)$ with $\chi=1/5$,
    plotted as functions of $n$ for 
    (a)  $N=20$,
    (b)  $N=50$,
    and
    (c)  $N=95$.
    In each case, the vacua which populate the lower
    portions of each vacuum tower
    eventually decay by tunneling directly to 
    the true ground state. 
    However, the first metastable vacuum to 
    decay in this  manner is {\it not}\/ the first excited vacuum;  
    instead, these decays follow a complicated collapse 
    pattern, with different portions of the vacuum tower
    decaying at different times.
    By contrast, for sufficiently large $N$, the vacua populating
    the upper portions of
    our vacuum towers have lifetimes exceeding the age of the
    universe, corresponding to the critical bounce action 
    $B\approx 471$ (dotted line).  
       }
\label{avalanchefig}
\vskip 0.25 truein
\centerline{
   \epsfxsize 3.0 truein \epsfbox {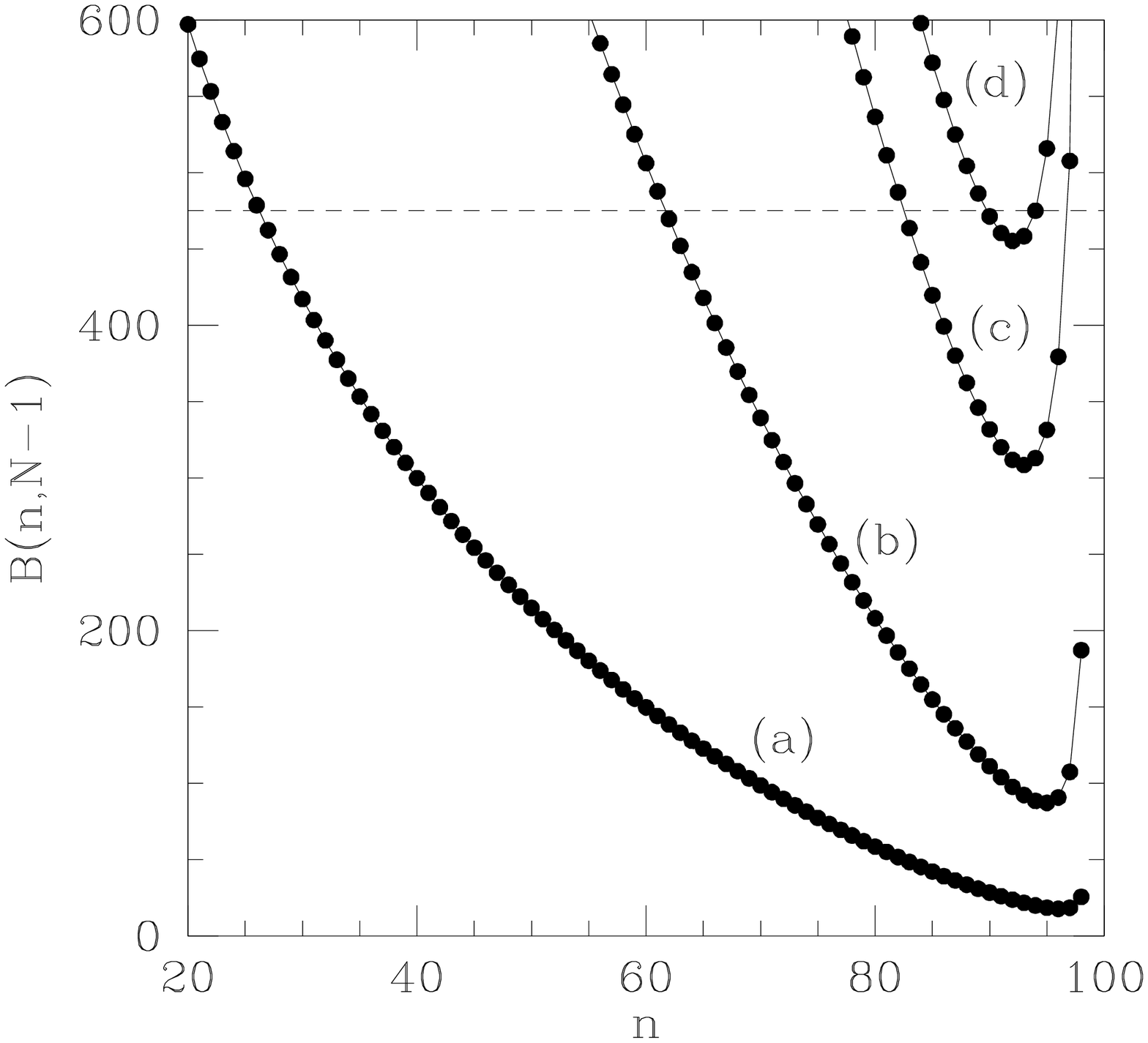}
 }
\vskip -0.38 truein
\caption{ Bounce actions $B(n,N-1)$ for $N=100$,
     plotted as functions of $n$ for
     (a) $\chi=0.1$,
     (b) $\chi=0.2$,
     (c) $\chi=0.3$,
     and 
     (d) $\chi=0.33$.
     The critical bounce action (dotted line) corresponds
     to lifetimes exceeding the age of the universe.
     In general, 
     we see that only a narrow band of metastable vacua
     within the vacuum tower
     will take part in the collapse pattern
     and decay to the ground state;
     by contrast,
     vacua above or below this band
     in the vacuum tower will generally remain stable. 
     As the kinetic mixing parameter $\chi$ increases
     towards its maximum value $1/2$,
     this collapse band grows increasingly narrow
     and ultimately disappears.  }
\label{avalanchechifig}
\vskip -0.25 truein
\end{figure}

This behavior is shown in Fig.~\ref{avalanchefig}
for $N=20$, $50$, and $95$.
In each case, we see that the vacua which populate the lower
portions of each vacuum tower
tend to decay first.
However, we also see that the first metastable vacuum to 
decay is {\it not}\/ the first excited vacuum
with $n=N-2$;  
instead, these decays follow a complicated collapse
pattern, with different portions of the vacuum tower
decaying at different times.
For example, in the case with $N=20$ and $\chi=1/5$ shown in Fig.~\ref{avalanchefig}(a),
we see that the first vacuum to decay into the ground state with $n=19$
is actually the $n=15$ vacuum.
The vacua then decay sequentially, with decreasing values of $n$, 
except that the $n=16$ vacuum decays between the $n=15$ and $n=14$
vacua, the $n=17$ vacuum decays between the $n=12$ and $n=11$
vacua, and the $n=18$ vacuum decays between the $n=6$ and $n=5$ vacua.

%================== FIGURE ============================================
%======================================================================== 

As $N$ increases, this behavior persists.  Ultimately,
however, we find that our collapse pattern develops a new feature:  
an {\it upper critical location}\/ on the tower above which our vacua remain stable
on cosmological time scales.
For example, we see from Fig.~\ref{avalanchefig}(b) that
for $N=50$,
the collapse pattern begins with the $n=45$ vacuum decaying first.  
The avalanche of collapses ultimately spreads up the tower until it reaches
the $n=16$ vacuum, at which point it stops.
Thus, the vacua with $n\leq 16$ remain stable. 

The above results apply for relatively small values of $\chi$.
However, as $\chi$ increases, our wave of collapses
develops a {\it lower limit}\/ as well.
This behavior is shown in Fig.~\ref{avalanchechifig}.
For example, we see from 
Fig.~\ref{avalanchechifig}(c)
that for $N=100$ and $\chi=0.3$,
the avalanche begins with the $n=93$ vacuum
and spreads up the tower to the $n=83$ vacuum
before stopping.  However, it only spreads down
the tower to the $n=96$ vacuum, where it stops
as well.  Indeed, vacua with $n\leq 82$ and
$n=97,98$ are stable.
In general, 
as the kinetic mixing parameter $\chi$ increases
towards its maximum value $1/2$,
this avalanche band grows increasingly narrow
and ultimately disappears. 
This is consistent with Feature~\#1 that
lifetimes along the vacuum tower diverge
as $\chi\to 1/2$. 

We also note from Fig.~\ref{avalanchefig} that
\begin{itemize}
\item    {\it General feature \#4:}\/ ~For fixed $n_i$ and $n_f$, 
    a given bounce action $B(n_i,n_f)$ tends to
     increase with $N$.
\end{itemize}
This is a direct consequence of the fact that 
even when $n_i$ and $n_f$ are held fixed, the
corresponding distance in field space between the $n_i$-
and $n_f$-vacua increases with $N$ because the dimensionality
of the field space itself increases with $N$. 
Since the energies of the $n_i$- and $n_f$-vacua are $N$-independent, 
the lifetime of the $n_i$-vacuum increases.

\subsection{Cascade behavior}

As we have seen, it is Feature~\#3 which is directly
responsible for the ``collapse'' behavior  
in which each metastable vacuum state decays
directly into the true ground state of the theory
rather than into another metastable ground state.
However, as we shall now discuss,
Feature~\#3 is not generally valid, and indeed
the resulting behavior tends to 
change rather dramatically in the $\chi\to 0$, $N\to\infty$ limit. 

To understand the emergence of this qualitatively new behavior,
let us consider the 
$\chi\to 0$, $N\to \infty$ limit analytically.
We then find that the quantities
which determine our bounce actions have the leading behavior 
\beqn
\Delta V_+ &\sim& {1\over 2 n_i^2 (n_f-n_i)} \,\chi^2 + {\cal O}(\chi^3)~\nonumber\\  
\Delta V_- &\sim& {n_f-n_i\over 2 n_i n_f }  + {\cal O}(\chi)~\nonumber\\  
\Delta \phi_+ &\sim& {N\over n_i} \, \chi + {\cal O}(\chi^{3/2})~\nonumber\\ 
\Delta \phi_- &\sim& c_0 + c_1 \chi^{1/2} + {N\over n_f} \, \chi + {\cal O}(\chi^{3/2})~\nonumber\\ 
\label{Taylor}
\eeqn
where we have assumed that $n_i,n_f\ll N$, 
and where $c_0,c_1$ are ${\cal O}(1)$ coefficients which generally decrease with
$n_i$ but increase approximately linearly with $n_f$.
For example, $c_0=(n_f-1)/2$ for $n_i=1$.

The results in Eq.~(\ref{Taylor}) are easy to understand.
As $\chi\to 0$, the saddle point between the $n_i$- and
$n_f$-vacua shifts to join (and thereby destabilize)
the $n_i$-vacuum;  as a result, $\Delta V_+$ and
$\Delta \phi_+$ both vanish in this limit.
However, as expected, 
$\Delta V_-$ and $\Delta \phi_-$ remain non-zero,
even in this limit.
We also observe from the above results that 
$\Delta V_\pm$ are $N$-independent, while
all $N$-dependence lies within $\Delta \phi_\pm$.
This too is expected.

Given these results, it is relatively
straightforward to understand the leading behavior
of $B(n_i,n_f)$ as $\chi\to 0$, $N\to\infty$.
We see from Eq.~(\ref{Taylor}) that
there are two limiting cases to consider, depending on the
value of $N\chi$.
For $N\chi/n_f\ll c_0$,
the leading term $c_0$ in $\Delta \phi_-$
dominates.
In such cases, we find that the
inequality in Eq.~(\ref{eq:DecayIneq})
is always satisfied, whereupon 
the bounce action $B(n_i,n_f)$ takes the 
approximate analytical form
\beq
   B(n_i,n_f)~\approx~ {64 c_0 \pi^2 \over 3 g^2}\, {n_f\over n_i^2 (n_f-n_i)}\, 
                    (N \chi)^3~.
\label{form1}
\eeq
By contrast, for 
$N\chi/n_f\gg c_0$,
it is the $N\chi/n_f$ term in $\Delta\phi_-$ which dominates.
We then find that the 
inequality in Eq.~(\ref{eq:DecayIneq})
is never satisfied,
in which case
the bounce action $B(n_i,n_f)$ takes the 
approximate analytical form
\beq
    B(n_i,n_f)~\approx~ {4\pi^2\over 3g^2}\, 
           \left({3 n_f-n_i\over n_f-n_i}\right)\, 
           \left( {n_i+n_f\over n_i n_f}\right)^3\, (N\chi)^4~.
\label{form2}
\eeq

Thus, for any value of $N \chi$, we see that the choice of
whether Eq.~(\ref{form1}) or Eq.~(\ref{form2}) applies
depends on how $N\chi$ compares with $c_0 n_f$.
However, since $c_0$ itself grows approximately linearly with $n_f$,
we see that the first case will apply for $n_f\gg n_f^\ast$,
where
\beq
         n_f^\ast ~\equiv~ \sqrt{N\chi}~.
\label{nfast}
\eeq
By contrast, the second case will apply for $n_f\ll n_f^\ast$. 
Also note that in either case, the actual value of $B(n_i,n_f)$
depends on the product $N\chi$.  This is not unexpected.
According to Feature~\#1, the bounce
action should vanish as $\chi\to 0$, while it should diverge as $N\to\infty$ according
to Feature~\#4.  Thus, when both limits are taken simultaneously, it is not unreasonable
that the bounce action depend on the product.

%================== FIGURE ============================================
\begin{figure}[thb!]
\centerline{
\epsfxsize 3.0 truein \epsfbox {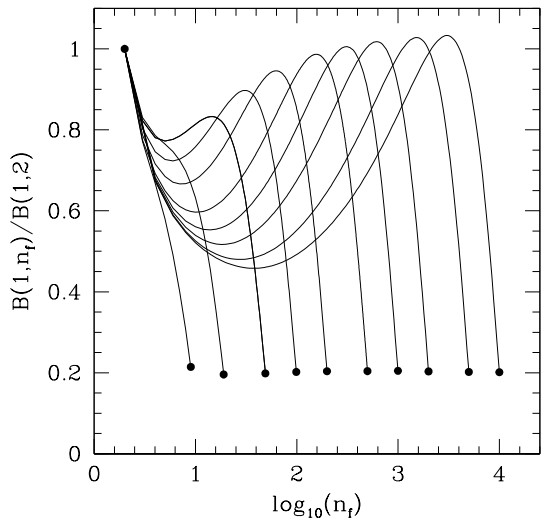}
}
\caption{
   Normalized bounce actions $B(1,n_f)$ with $\chi=1/10$, plotted as functions 
   of $n_f$ in the allowed range $2\leq n_f\leq N-1$.
   The curves whose lower portions progress from left
   to right correspond to ordered increasing values of 
             $N\in \lbrace 1,2,5\rbrace \times 10^{\lbrace 1,2,3\rbrace}$, ending with $N=10^4$. 
   We observe that as $N$ increases, the bounce actions $B(1,n_f)$ develop a
   deepening ``trough'' whose minimum is approximately located at 
   $n_f\approx n_f\ast\equiv \sqrt{N\chi}$. 
   The depth of the trough asymptotes to a finite value as $N\to\infty$
   in such a way that the minimum bounce action along any such curve corresponds
   to $n_f=N-1$.  } 
\label{dippy}
\end{figure}  
%========================================================================

Given these results, we can now consider how 
$B(n_i,n_f)$ varies with $n_f$ for fixed $n_i$ in the 
limit as $\chi\to 0$, $N\to \infty$.
In general, Eq.~(\ref{form1}) is a {\it rising}\/ function of $n_f$,
while Eq.~(\ref{form2}) is a {\it falling}\/ function.
As a result, we expect that as $B(n_i,n_f)$ for fixed $N\chi$ should
generally develop a dip (or ``trough'') centered around
$n_f^\ast$
as $\chi\to 0$, $N\to \infty$.  
This behavior is shown in Fig.~\ref{dippy}, where we plot
the normalized bounce actions $B(1,n_f)/B(1,2)$ 
as functions of $n_f$ in the allowed range $2\leq n_f\leq N-1$.
In this figure, we have taken $\chi=1/10$  and we vary $N$ from 
$N=10$ to $N=10^4$.
As $N$ increases, we see from Fig.~\ref{dippy} that
our bounce-action function indeed develops a deepening 
trough whose minimum is approximately located at
$n_f\ast\equiv \sqrt{N\chi}$.
This, then, is a counter-example to Feature~\#3.

This does not, however, eliminate the resulting collapse behavior. 
As long as the minimum bounce action along any individual curve 
continues to occur at the maximum value $n_f=N-1$, 
the $n=1$ vacuum (and indeed all $n_i$-vacua)
will continue to preferentially decay directly to the $n_f=N-1$ ground state.
At first glance, one might suspect that taking even larger values of $N$
in Fig.~\ref{dippy} would produce an even deeper trough which would eventually
become deeper than the minimum bounce action at $n_f=N-1$.
However, as we see from Fig.~\ref{dippy}, the depth of the trough actually
approaches an asymptote as $N\to\infty$, and this asymptotic depth continues to 
exceed the $n_f= N-1$ bounce action.  
Thus our collapse behavior remains undisturbed.

%================== FIGURE ============================================
\begin{figure}%[thb!]
\centerline{
\epsfxsize 3.0 truein \epsfbox {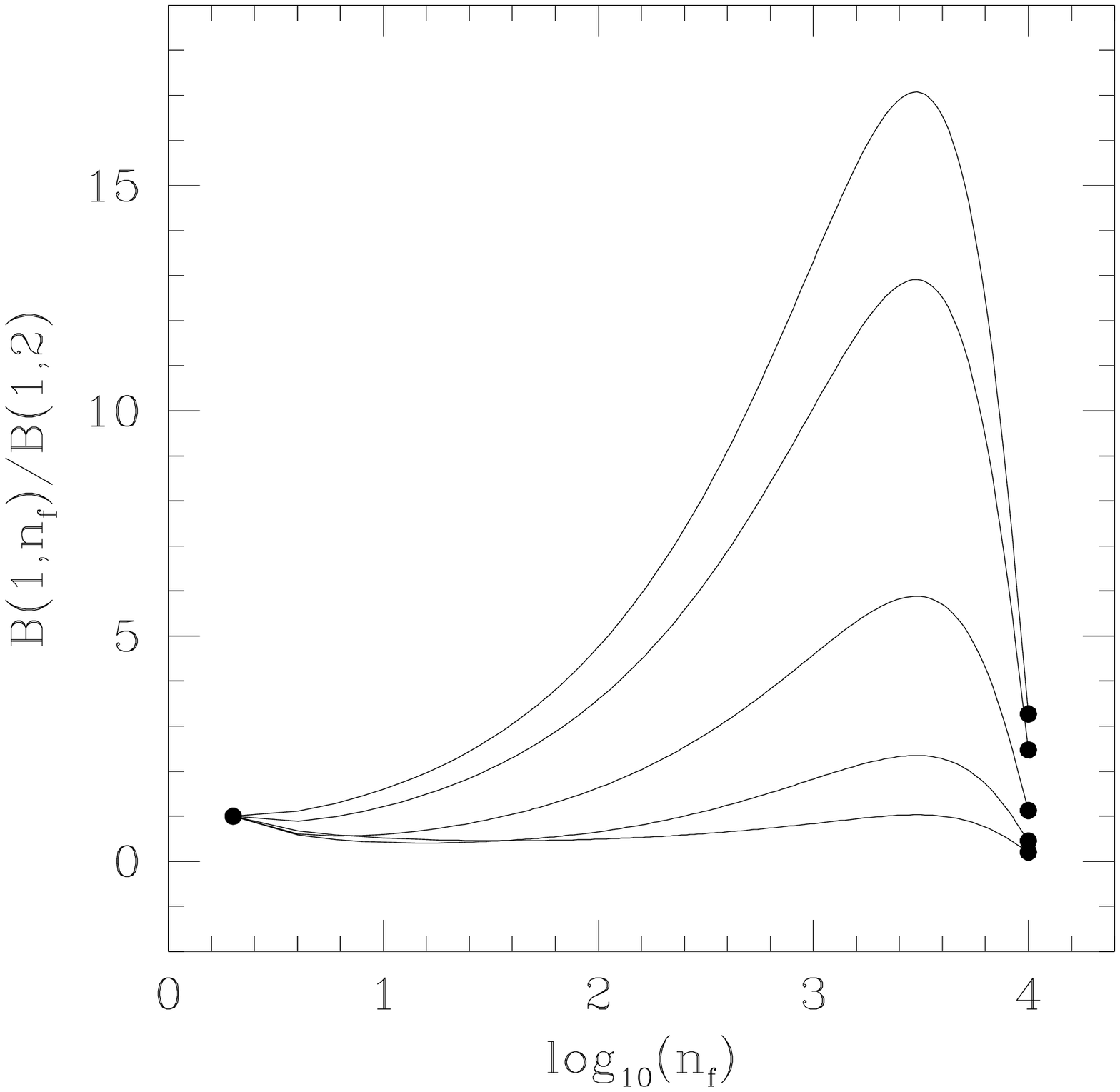} }
\vskip -0.2 truein
\caption{
   The normalized bounce actions $B(1,n_f)$, plotted as functions of
   $n_f$ in the allowed range $2\leq n_f\leq N-1$, with $N=10^4$.
   The different curves progressing from lowest to highest
   correspond to $\chi=10^{-1}, 10^{-2}, 10^{-3}, 10^{-4}$, and $10^{-5}$;
   note that the lowest curve here is the same as the $N=10^4$ curve shown
   in Fig.~\protect\ref{dippy}.
   As $\chi\to 0$, we see that $n_f^\ast\to 0$;  thus the bounce
   actions increasingly tend to {\it grow}\/ as functions of $n_f$,
   at least for sufficiently small $n_f$.
   Also important is the fact that for sufficiently small $\chi$,
   the bounce action at $n_f=N-1$ begins to {\it exceed}\/ that at
   $n_f=2$.  For each such curve, the minimum bounce action therefore
   occurs {\it not}\/ for $n_f=N-1$, but for $n_f=2$.
   This triggers the onset of cascade (rather than collapse) behavior
   for our vacuum towers.} 
\label{dippy2}
\vskip 0.2 truein
\centerline{
\epsfxsize 3.0 truein \epsfbox {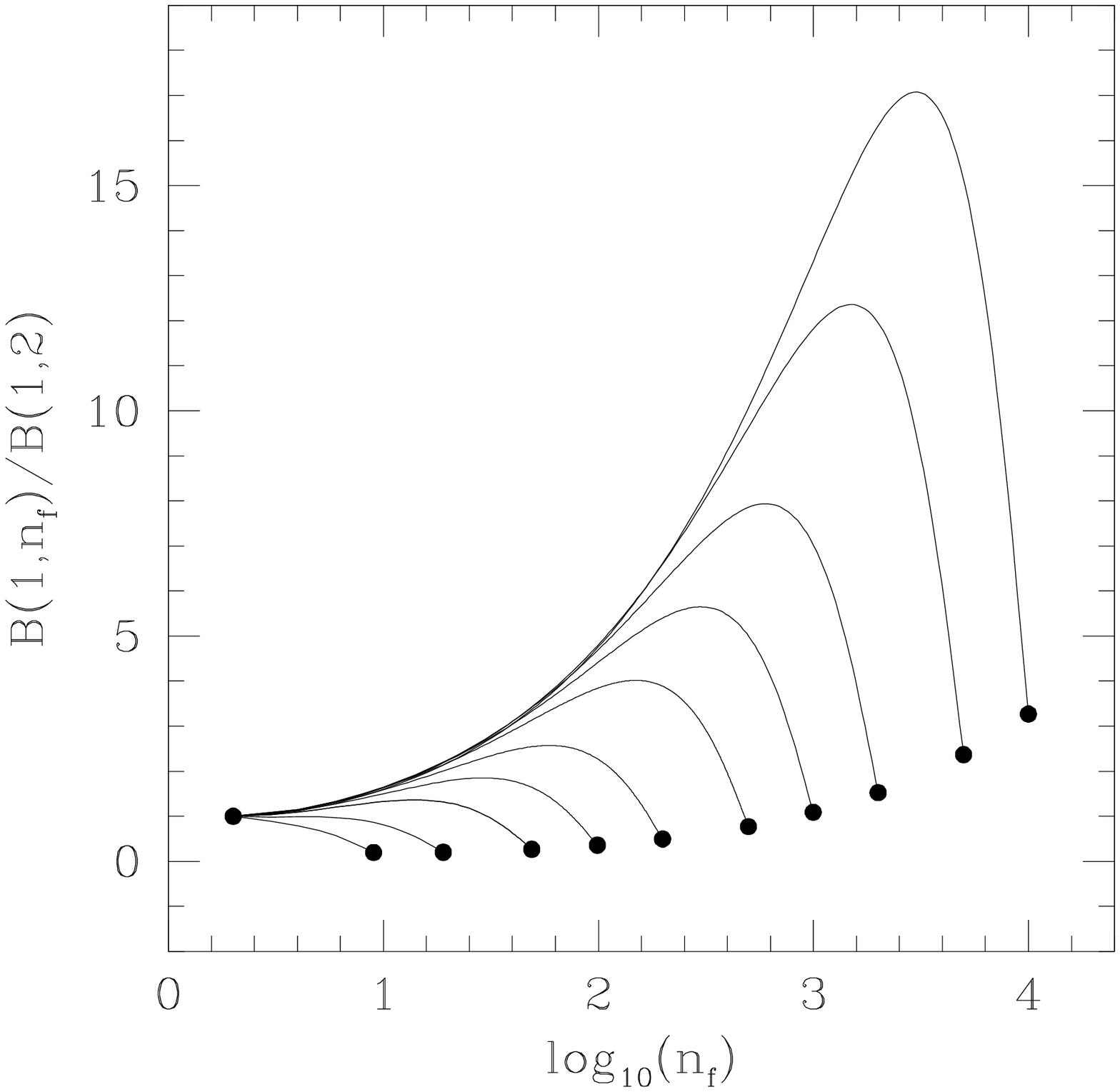} }
\vskip -0.2 truein
\caption{
   Same as Fig.~\protect\ref{dippy}, except plotted for $\chi=10^{-5}$ rather than $\chi=1/10$.
   We observe that as $N$ increases (approaching the true $N\to\infty$ limit), 
   the minimum bounce action begins to occur for relatively small values of $n_f$
   rather than for $n_f=N-1$.  As in Fig.~\protect\ref{dippy2}, this triggers
   the onset of cascade (rather than collapse) behavior
   for our vacuum towers.} 
\label{dippy3}
\end{figure}  
%========================================================================

Of course, the above comments pertain to the situation in Fig.~\ref{dippy} 
for which we took $\chi=1/10$.  
Such a value for $\chi$ --- while suitable for illustrating the emergence of the
trough --- is still not yet small enough to alter the collapse behavior.
The situation changes dramatically, however, if we
take $\chi$ even smaller and enter the true $\chi\to 0$ regime.   
For any fixed $N$, we have seen from the above analysis
that taking the $\chi\to 0$ limit has the effect of sliding 
$n_f^\ast\to 0$.
We therefore expect that as $\chi\to 0$, the initially falling portion of
the $B(n_i,n_f)$ curve should disappear, and $B(n_i,n_f)$ should actually 
begin to {\it rise}\/ as a function of $n_f$ (as long as $n_f\ll N$ and $N$ is held fixed).
This behavior is shown in Fig.~\ref{dippy2},
where we have 
held $N$ fixed and plotted $B(1,n_f)$ as a function of $n_f$ while we
reduce $\chi$ from $10^{-1}$ to $10^{-5}$.
Thus, as we see from Fig.~\ref{dippy2}, reducing $\chi$ has the geometric
effect of ``uplifting'' each of the curves in Fig.~\ref{dippy}. 
Indeed, this uplifting is the generic behavior that occurs 
as we enter the true $\chi\to 0$ limit.

Given these results, we see that the net 
effect of reducing $\chi$ is that the behavior shown in Fig.~\ref{dippy} 
smoothly shifts to become the behavior shown in Fig.~\ref{dippy3}.
Indeed, these two figures plot exactly the same bounce actions $B(1,n_f)$; 
the only difference between them is that the former is calculated with
$\chi=1/10$ while the latter is calculated with $\chi=10^{-5}$.
However, we now see that the effect of reducing $\chi$ has been dramatic:  while 
the smallest bounce action $B(1,n_f)$ along any curve in Fig.~\ref{dippy}
occurs for the {\it maximum}\/ allowed value $n_f=N-2$, 
the smallest bounce actions $B(1,n_f)$ along the curves with large $N$
in Fig.~\ref{dippy3} now occur for the {\it minimum}\/ allowed value $n_f=2$.
Thus, while the top vacuum in our tower prefers to decay directly to
the ground state when $\chi=1/10$, it prefers to decay  
to merely the metastable vacuum immediately below it when $N$ is sufficiently
large and $\chi=10^{-5}$!

Since the top vacuum has only decayed to another metastable state 
in the vacuum tower rather than to the ground state, the cycle can then repeat.
In this case, for example, the nature of the subsequent vacuum decay 
now depends on the behavior of $B(2,n_f)$ as a function of $n_f$ for 
$3\leq n_f\leq N-1$.
However, as long as $n_i\ll N$, the same situation persists:  
the preferred subsequent decay is into the next-lowest vacuum,
and the cycle repeats yet again.
Thus, what emerges is not a 
collapse into the ground state, but rather a sequential {\it cascade}\/ 
from vacuum to lower vacuum.

\begin{figure}[thb!]
\centerline{
\epsfxsize 3.0 truein \epsfbox {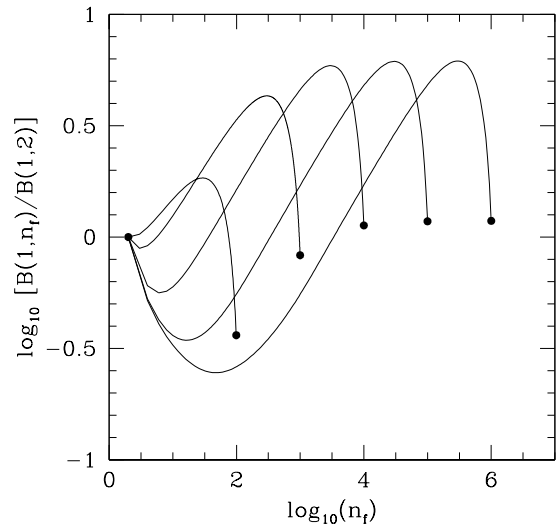}
}
\caption{
   Behavior of $B(1,n_f)$ as a function of
   of $n_f$ in the allowed range $2\leq n_f\leq N-1$, plotted
   for different values of $N\chi$.
   The curves whose end positions progress from left
   to right correspond to ordered increasing values of 
   $N\chi=10^{\lbrace -1,0,1,2,3\rbrace}$, where we have held $\chi=10^{-3}$ fixed.
   As $N\chi$ increases, the bounce actions $B(1,n_f)$ develop a
   deepening ``trough'' whose minimum is approximately located at 
   $n_f\approx n_f^\ast=\sqrt{N\chi}$.  As a result, the vacuum cascade
   step size increases as a function of $N\chi$.}
\label{dippy4}
\end{figure}

It should be stressed that not all vacuum cascades necessarily proceed
through single-vacuum hops.
Indeed, although this was the result emerging for the curves with the largest values of $N$
plotted in Fig.~\ref{dippy3}, 
these curves still all correspond to cases with $N\chi<1$.
However, in the true $\chi\to 0$, $N\to\infty$ limit,
the size of the cascade hops generally depends on the product $N\chi$.
This is illustrated in Fig.~\ref{dippy4}, where we have plotted the values of $B(1,n_f)$
for different values of $N\chi$, all while remaining completely
within the double $\chi\to0$, $N\to\infty$ limit.
As expected from our above analysis, increasing the value of $N\chi$
has the effect of increasing $n_f^\ast$ and thereby shifting the minimum 
of the curve towards larger values of $n_f$.
In such cases, the value of $n_f$ which minimizes $B(n_i,n_f)$ will
occur for $n_f=n_i+\Delta n$ where $\Delta n>1$. 
As a result, for $N\chi\gg {\cal O}(1)$, cascades can often proceed through 
larger, multiple-vacuum hops, and indeed non-trivial
cascade trajectories can easily develop.
An explicit example of such a non-trivial cascade pattern
will be presented below.

%================== FIGURE ============================================
%========================================================================

In general, 
even in the double $\chi\to 0$, $N\to\infty$ limit,
vacuum cascades do not continue all 
the way down the vacuum tower.
Instead, beyond a certain point, 
$n_i$ becomes sufficiently large that
the behavior of $B(n_i,n_f)$ as a function of $n_f$ reverts
back to the collapse pattern, with the minimum bounce action
occurring for the ground state $n_f=N-1$.
Thus, a system which starts at the top of the tower eventually cascades 
down the tower to a critical level
$n^\ast$ at which 
collapse behavior takes over.  The subsequent (and final) decay will then be
directly to the ground state.

The critical value $n^\ast$ at
which cascade behavior becomes collapse behavior
generally increases with $N$.
In fact, we have found numerically that
\beq
    n^\ast ~\approx~  c N
\label{linearrelation}
\eeq
where $c\approx 2.32\times 10^{-3}$. 
Remarkably, this relationship holds with increasing precision as $N$ grows large.
Moreover, this relationship is independent of $\chi$.
Of course, as discussed above,
the mere existence of a cascade region already presupposes
that we are in the $\chi\to 0$ limit.
However, once $\chi$ is sufficiently
small as to produce cascade behavior and establish a non-zero value
of  $n^\ast$, 
reducing $\chi$ still further will not increase $n^\ast$ beyond this
value.
Indeed, only with a simultaneous increase in $N$
can an increase in $n^\ast$ be accomplished.

The result in Eq.~(\ref{linearrelation}) 
implies that for large $N$ and small $\chi$, {\it no more than the 
uppermost $0.23\%$ 
of any given vacuum tower can experience cascade behavior
as opposed to collapse behavior}\/.
This small value presumably reflects the overall numerical coefficients which appear
in the expressions for our bounce actions in Eq.~(\ref{forma});  indeed,
we observe that $c\sim {\cal O}(1/32\pi^2)$ as far as overall scales are concerned. 
Despite its small relative size, however, this cascade region of the vacuum tower
is of tremendous importance for two reasons:
\begin{itemize}
\item   For large $N$, this region gives rise to the vast majority of the vacuum 
        transitions that a given system can undergo --- indeed
        this cascade region of the tower gives rise to literally
        {\it all}\/ possible vacuum transitions for our system, except for the
        final transition to the ground state.
\item   Likewise, for any $N$, these vacuum transitions involve the greatest shifts in
        vacuum energy that the system can experience.   Indeed, as we have
        seen in Sect.~II, the differences in vacuum energy between neighboring
        vacua grow increasingly small as we move down the vacuum tower.
        Thus, for large $N$, essentially all of the original vacuum energy of
        our system is dissipated through vacuum transitions occurring
        in this region of the tower. 
\end{itemize}

Thus, to summarize:
in the double $\chi\to 0$, $N\to \infty$ limit, 
we find that a vacuum cascade can emerge near the top of
our metastable vacuum towers.
In general, the value of the product $N\chi$ governs
the size of the hops taken in each step of this cascade.
Eventually, the cascade proceeds down the tower until a
critical value $n^\ast$ is reached, at which point the cascade
behavior reverts back to collapse behavior.
This value of $n^\ast$ increases with $N$,  but is
essentially independent of $\chi$ so long as we remain in
the $\chi\to 0$ limit which was required to produce
the cascade region in the first place.
Moreover, $n^\ast$ is always
significantly less than $N$.  
As a result, there is always a substantial collapse
region in the lower portion of any vacuum tower.

Finally, of course, we recall Feature~\#2 which asserts
that bounce actions which govern the decays of vacua near the top 
of any vacuum tower generally exceed those near the bottom.   
As a result, it is possible that such bounce actions near the
tops of our metastable vacuum towers will exceed the critical
value $B\approx 471$
required for stability on cosmological time scales.
As a result, it is possible that the uppermost vacua in any vacuum
tower will be essentially stable.

\subsection{An explicit example}

To illustrate all of these decay patterns and features
simultaneously within a single
model, let us consider the specific example 
with $N=5000$ and $\chi=2.8\times 10^{-4}$.
Since this choice has $\chi\ll 1$ and $N\gg 1$,
we see that it is precisely such a choice which
will yield both a cascade region as well as a collapse region.
Moreover, since $N\chi\sim {\cal O}(1)$, we expect that
our vacuum cascade will proceed through relatively small steps. 

%================== FIGURE ============================================
\begin{figure*}%[thb!]
\centerline{ \epsfxsize 3.8 truein \epsfbox{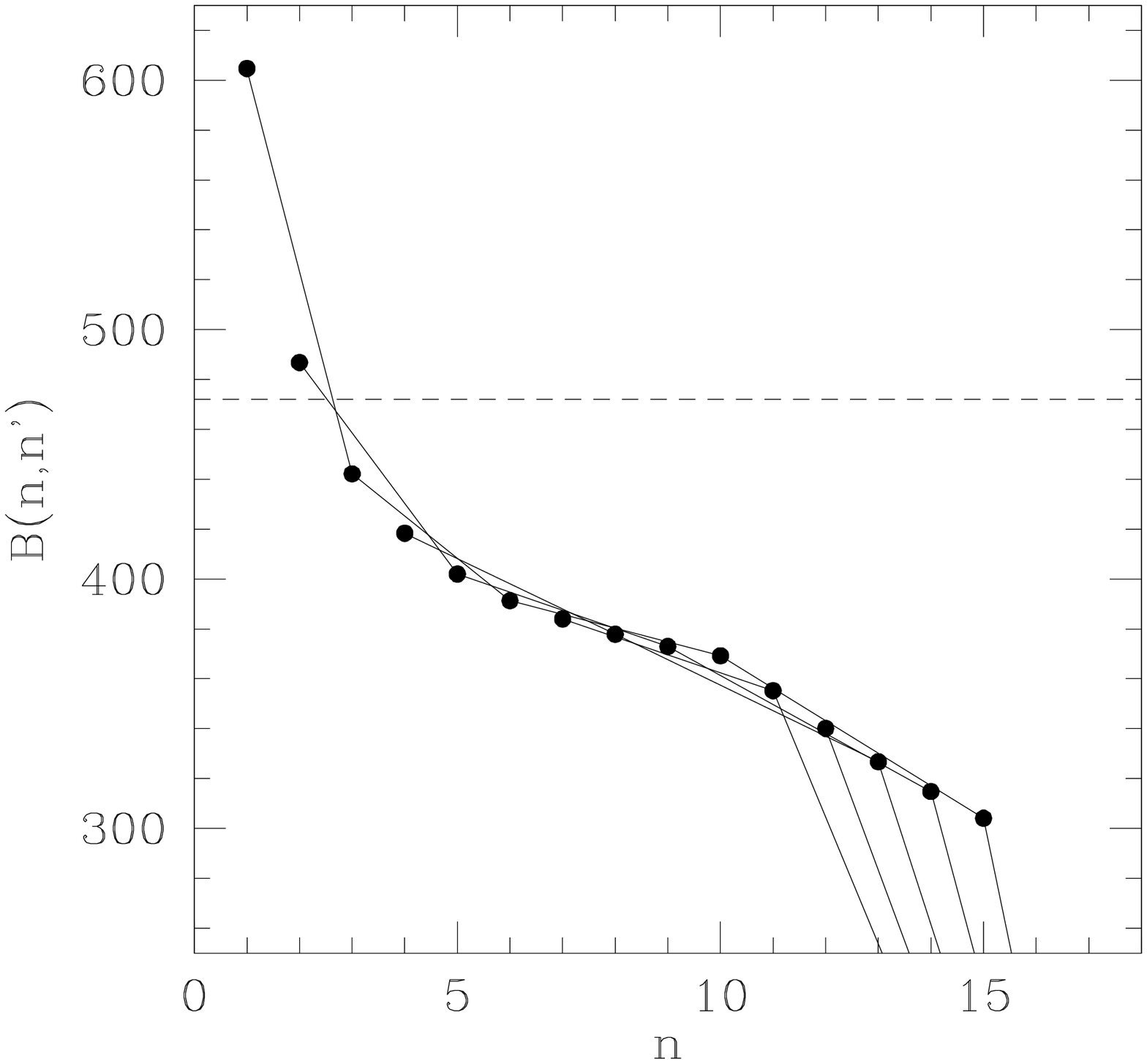} }
\vskip -0.3 truein
\caption{
    Lifetimes along the different cascade trajectories,
    calculated for $N=5000$ and $\chi=2.8\times 10^{-4}$. 
    For each vacuum~$n$,  
    we plot $B(n,n')$ for that value of $n'$ ($n<n'\leq N-1$)  
    which minimizes $B(n,n')$ and which thereby indicates the
    next vacuum along the corresponding cascade trajectory.
    Points are connected according to their sequential trajectories.
    We also show the Great Wall (dotted line) signifying
    the critical bounce action 
    corresponding to the age of the universe.
    We see from this plot that in this case
    the top two vacua are stable, while the $n=3$ through $n=10$ vacua cascade
    down along four distinct trajectories with decreasing lifetimes
    until they pass the $n=11$ vacuum.
    At this stage, each trajectory decays directly to the 
    ground state of our vacuum tower.
    By contrast, vacua beyond the $n=15$ vacuum
    populate a ``Forbidden City'':  it is impossible to tunnel into these
    vacua from any other locations along the vacuum tower, 
    and the universe can inhabit such a vacuum (for a brief time only)
    if and only if it is initially born there.} 
\label{fig:CascadeB}
\vskip 0.5 truein
\centerline{ \epsfxsize 7.0 truein \epsfbox{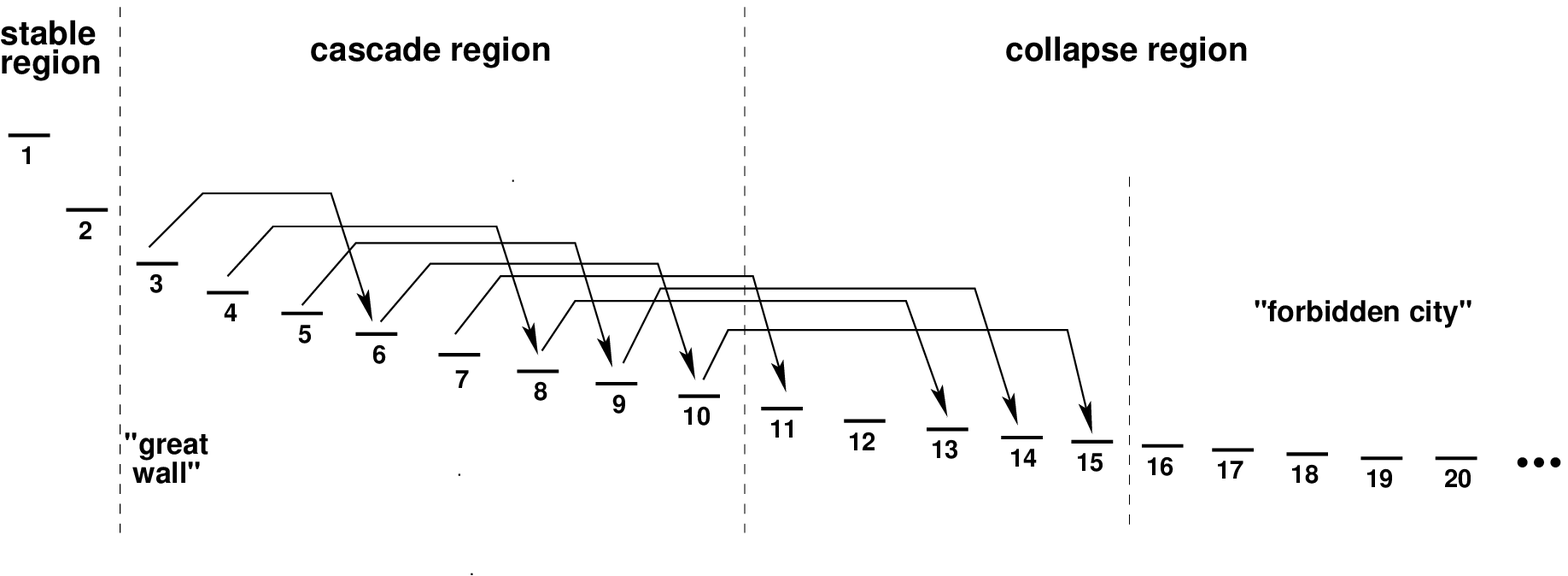} }
\caption{
    A schematic of the vacuum
    structure of the model shown in Fig.~\ref{fig:CascadeB}.  Vacua in the stable region
    to the left of the Great Wall have lifetimes exceeding the age of the universe, while vacua 
    in the cascade region
    decay to other (lower) metastable vacua in the vacuum tower.  By contrast, vacua in the 
    collapse region decay directly to the ground state
    of the vacuum tower.  Finally, vacua which populate
    the ``Forbidden City'' cannot be reached from outside the Forbidden City:  such vacua
    can be populated only as an initial condition at the birth of the vacuum configuration.} 
\label{fig:Cascade}
\end{figure*}  
%========================================================================

A plot detailing the vacuum dynamics for this choice of parameters 
is shown in Fig.~\ref{fig:CascadeB}.  
For each $n$, we have plotted $B(n,n')$ for that value of $n'$ ($n<n'\leq N-1$)  
which minimizes $B(n,n')$ and which thereby indicates the
next vacuum along the corresponding cascade trajectory.
For example, we see from this figure that the $n=3$ vacuum decays into the $n=6$ 
vacuum, which in turn
decays (even more rapidly) into the $n=10$ vacuum;
this in turn decays (even more rapidly) into the $n=15$ vacuum, 
which in turn decays directly into the ground state.
There are, of course, limits to this cascade region, both at the top
and at the bottom. 
For example, the bounce actions for the top two vacua 
exceed our critical bounce action $B\approx 471$;  these vacua,
if initially populated, are consequently deemed stable on cosmological time scales.
Likewise, at the $n=11$ vacuum and beyond, we enter the collapse region
in which all subsequent decays automatically proceed directly to the ground state.  

Nevertheless, within the cascade region between the $n=3$ and $n=10$ vacua,
we see that this model contains four independent potential cascade trajectories,
each of which unfolds with increasing speed (\ie, decreasing lifetimes):
\beqn
   && \bullet ~~3 \to 6 \to 10 \to 15 \to {\rm GS} \nonumber\\
   && \bullet ~~4 \to 8 \to 13 \to {\rm GS}\nonumber\\
   && \bullet ~~5 \to 9 \to 14 \to {\rm GS}\nonumber\\
   && \bullet ~~7 \to 11 \to {\rm GS}
\eeqn
where `GS' signifies the ground state.
It is therefore only an initial condition that determines which trajectory a
given system ultimately follows.

Given these results, we can separate the vacua in our vacuum tower 
into three distinct regions on the basis of their decay phenomenologies.
This is illustrated schematically in Fig.~\ref{fig:Cascade}.   
At the top of the tower is a stable region within which all vacua have lifetimes
longer than the current age of the universe.  
This stable
region is then separated from the remaining unstable regions 
by a ``Great Wall'' which 
denotes the border between the stable ``civilized'' world 
and the remaining ``barbarian'' regions
which are afflicted with the omnipresent threat of sudden and spontaneous vacuum decay.
Moving past the Great Wall, the second region 
is a cascade region 
in which there develops a complex pattern of decays of metastable vacua 
into other metastable vacua. 
Finally, moving further down the tower, the third region is a collapse region
consisting of vacua which decay directly to the ground state of the theory.  

Note that this last region may, in turn, be subdivided into two distinct 
(but overlapping) subregions.  The first consists
of vacua (such as the $n=\{11,13,14,15\}$ vacua in the current example)
which can be reached at the end of one or more 
decay chains beginning in the cascade region.  By contrast,
the second (which here technically includes the $n=3,4,5,7,12$ vacua as well as all $n>15$ vacua) 
consists of vacua which can never be reached through any instanton-tunnelling 
decay chain.  These vacua, which may only be
populated by some initial condition, collectively form a ``Forbidden City'' 
into which one cannot enter from the outside.  Indeed, the universe
can inhabit such a Forbidden City only if it was initially ``born'' there.

Thus, we see that constructions of this sort possess not only a great number of 
metastable vacua in addition to their ground states, but also the 
potential for a highly nontrivial set of vacuum dynamics involving complicated
vacuum cascade/collapse behavior.  
In this vein, it is worth emphasizing that the example illustrated 
in Figs.~\ref{fig:CascadeB} and~\ref{fig:Cascade} represents only one of 
many possible vacuum-cascade scenarios that can be realized in
scenarios of this sort.  Furthermore, if we were to relax some of the
simplifying assumptions inherent in the model outlined in Section~\ref{sec:framework} --- 
for example, our assumption that all nearest-neighbor
kinetic-mixing parameters in Eq.~(\ref{eq:XabArbitraryMix}) 
are equal, with all others vanishing ---
an even wider range of possible cascade scenarios would result.    
A similar possibility exists if $\lambda$ is not taken in the asymptotic
region:  decreasing $\lambda$ tends to decrease the lifetimes of our
metastable vacua, and thereby enables more rapid vacuum transitions to occur.

%==============================================================================================

\section{Spectrum of the Model\label{sec:Spectrum}}

We now turn to a discussion of the spectra of physical particles that
arise in the different vacuum states of our model.  
Rather than provide a detailed phenomenological analysis of these particles,
our main interest is in understanding how their mass spectra evolve as
functions of the vacuum index $n$.  This will enable us to trace the 
changes in these particle spectra as our system evolves down the
vacuum tower towards the ground state.

\subsection{Scalar spectrum}

We begin with the scalar sector, which
comprises $2N+2$ degrees
of freedom: the real and imaginary components of the $\phi_i$.  
In any given vacuum,
the squared masses of these states are the eigenvalues  
of the mass matrix defined in Eq.~(\ref{matrix}), with all fields
replaced by their expectation values appropriate for that vacuum.  
However, we immediately observe from Eq.~(\ref{vacsolns})
that $N-1$ of the  $\phi_i$ receive nonzero VEV's in any 
vacuum state.  As a result,
$N-1$ global $U(1)$ symmetries (corresponding to global  
phase rotations of these fields) are spontaneously broken, and
we expect the  spectrum to contain $N-1$ Nambu-Goldstone bosons.
Thus $N-1$ of the eigenvalues of Eq.~(\ref{vacsolns}) will vanish,
and the squared masses for the
remaining $N+3$ scalar degrees of freedom will be positive (by definition,
since each vacuum is presumed stable). 
Moreover, in any $n$-vacuum, four of these remaining $N+3$ degrees of freedom
are the real and imaginary components of the two
complex scalars $\phi_{N-n+1}$ and $\phi_{N+1}$ --- the fields whose VEV's
vanish in that vacuum --- with masses proportional to $\lambda^2\xi^{N-1}$.  
The other $N-1$ degrees of freedom are real scalar fields 
(the real components of the $\phi_i$ for which $v_i^2\neq 0$),
and will have squared masses proportional to $\xi$.

\subsection{Gauge-boson spectrum}

As we have noted above,
$N-1$ of the $N$ different $U(1)$ gauge-group factors in our model 
are spontaneously broken in each vacuum state 
by the VEV's of the $\phi_i$ fields.  
As a result, the $N-1$ Nambu-Goldstone scalar modes
discussed above are ``eaten'' by the gauge bosons associated with these   
broken $U(1)$'s, resulting in a gauge-boson mass matrix of the form
\begin{equation}
    (M_{\rm gauge}^2)_{ab} ~=~ 
       g^2\, \sum_i^{N+1}\, \hat{Q}_{ai}\hat{Q}_{bi}\, v_i^2~.
\end{equation}
Here we have explicitly written the hats on the charge matrices, as in Sect.~II, to    
indicate that they are defined in the basis in which gauge-kinetic terms  
take their canonical forms.  Of course, this matrix (and indeed the physical mass    
spectrum dictated by its eigenvalues) is different for each vacuum in the
tower.  However, for all $n$, it has precisely one zero eigenvalue which
corresponds to the linear combination of gauge bosons associated with the
single remaining unbroken $U(1)$ gauge group.  
This linear combination, which we will call $U(1)'$ with gauge field $B_{\mu}$, 
turns out to be a uniform admixture of the gauge fields
$A^{\mu}_{N-n+1},\ldots,A^{\mu}_{N}$ in the {\it unhatted}\/ basis:
\beq
     B^{\mu}~=~\frac{1}{\sqrt{n}}\, \sum_{a=N-n+1}^N \, A_{a}^{\mu}~.
\label{Bmu}
\eeq
Note that this massless gauge boson 
couples only to the complex scalars $\phi_{N-n+1}$ and   $\phi_{N+1}$.  

The other $N-1$ linear combinations of gauge fields are nothing
but massive $Z'$ vector bosons with masses proportional to $\xi$ (and independent of  
$\lambda$).  Note that the mass of each such gauge boson is identical to that of one
of the $N-1$ real, massive scalars mentioned above.  This occurs because
all non-vanishing quartic couplings among the various $\phi_i$ are given
(in the unhatted basis) by $g^2$.
This results in the massive gauge bosons having the same masses
as the massive scalars.  Indeed, such a situation arises as a result
of the structure of the $D$-term potential in any supersymmetric
theory in which there is no $F$-term contribution to the quartic
couplings of the scalars.

\subsection{Fermion spectrum}

We now turn to the fermions in our model.
These consist of the superpartners $\psi_i$ of the complex scalars $\phi_i$
$(i=1,...,N+1)$,
as well as the gauginos $\lambda_{a}$ associated with the $U(1)_a$ gauge    
groups $(a=1,...,N)$.  As a result, the fermion sector of the 
theory contains a total of $2N+1$ 
Weyl spinor degrees of freedom.  

Since supersymmetry is broken in each vacuum in the
tower, one of these spinors must play the role of the Goldstino.  Moreover,
since supersymmetry is broken solely by $D$-terms in each vacuum, we can
identify this particle with the gaugino superpartner 
of $B^{\mu}$ in Eq.~(\ref{Bmu}).  
In supergravity scenarios, this particle is ``eaten'' by the gravitino
according to the super-Higgs mechanism.  

All other fermions
are physical and acquire Dirac masses.  In the $n$ vacuum, we find that
$\psi_{N-n+1}$ and $\psi_{N+1}$
(the superpartners of $\phi_{N-n+1}$ and $\phi_{N+1}$) acquire masses
proportional to $\lambda\xi^{(N-1)/2}$, stemming from superpotential
contributions.  The rest of the fermions acquire masses through terms of the form
$\sqrt{2}v_iQ_{ai}\psi_i\lambda_a + {\rm h.c.}$ that arise from 
the supersymmetrization of the gauge interactions, and the 
mass eigenstates of this latter group are often highly mixed.

\subsection{Evolution of spectra under vacuum transitions}

Given these results, we now seek to understand
how these different mass spectra 
evolve as our system undergoes vacuum transitions
from one vacuum to the next.  In order to do this, it will be convenient to
separate the physical, massive particles in the theory into two distinct  
classes based on whether they do or do not couple to the 
massless gauge field $B^\mu$ in Eq.~(\ref{Bmu}).

As we have already seen, the particles which couple to $B^\mu$
are the component fields associated with the
two chiral superfields
$\Psi_{N-n+1}$ and $\Psi_{N+1}$.  These particles have masses proportional
to $\lambda \xi^{1-N/2}$ and mix only with each other.
Because these states couple to our only massless gauge boson $B^\mu$ in the theory,
we shall refer to such states as ``couplers''.

By contrast, the remaining states in the theory couple only to the
extra, massive $Z'$ gauge bosons which are orthogonal to $B^\mu$.
As such, they do not couple to the massless gauge field $B^\mu$ itself,
and we shall refer to these states as ``non-couplers''.
These states are massive and highly mixed, with
masses and mixings
which are rather complicated functions of $\chi$.
Broadly speaking, however, these masses fill out a
closely-spaced ``band'' which approaches a continuum as
$N\rightarrow\infty$.  
Note that the masses of the non-couplers are $\lambda$-independent.

%================== FIGURE ============================================
\begin{figure}[thb!]
\centerline{
   \epsfxsize 3.0 truein \epsfbox {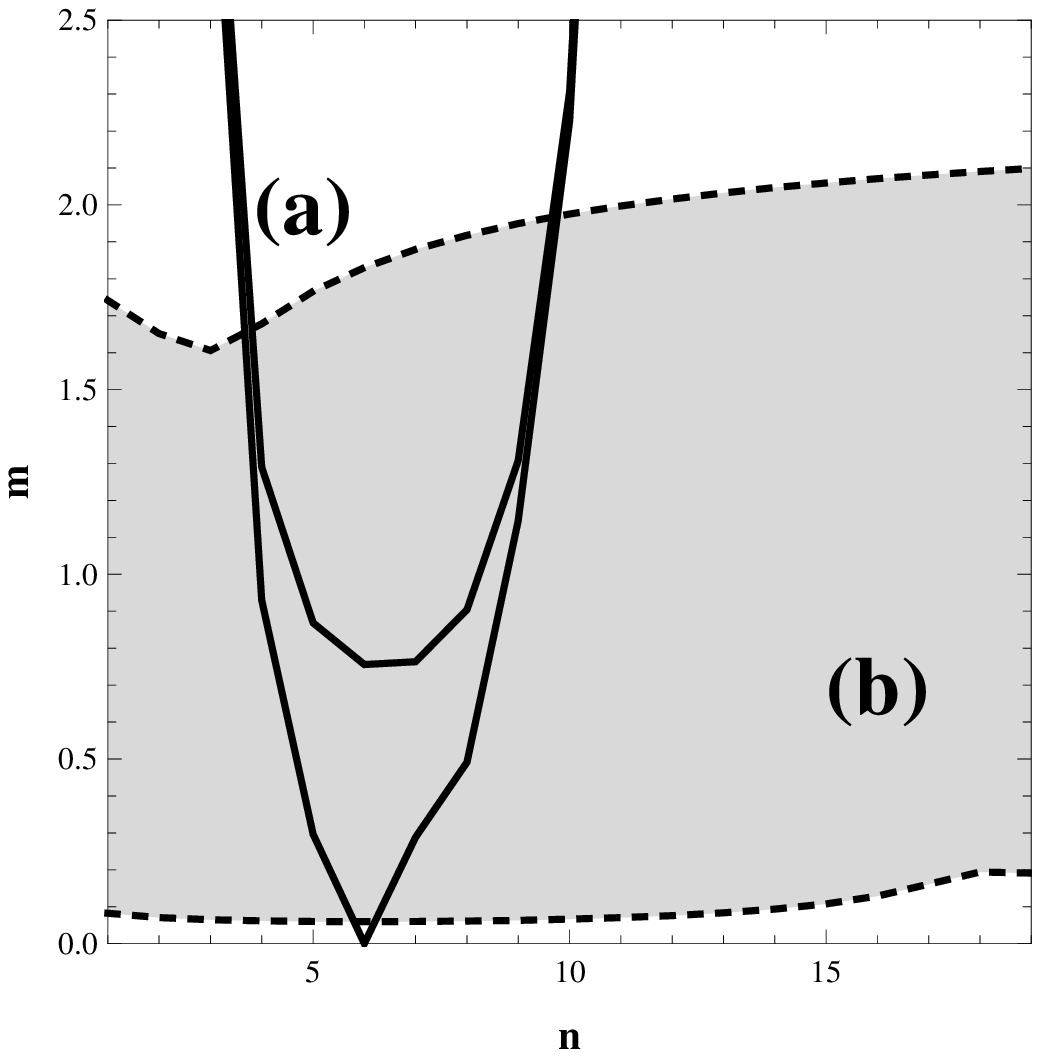} 
 }
\vskip -0.2truein
\caption{
   The spectrum of massive scalars, plotted as a function of the
   vacuum index $n$ for $N=20$, $\chi=1/4$, 
   and $\lambda=\lambda_{20}^{\ast}\approx 1.05\times 10^8$. 
   (a) The solid lines indicate the masses of the two complex
   scalar ``couplers'', while (b) the dotted lines demarcate 
   the edges of the shaded band within which the masses 
   of the remaining scalars lie.  Note that the lighter coupler
   actually becomes massless for the $n=6$ vacuum;  it is in
   this manner that the $n=6$ vacuum begins to destabilize
   for $\lambda=\lambda_{20}^\ast$, in accordance with the
   results shown in Fig.~\protect\ref{criticalns} for $\chi=1/4$.
   Note that the masses plotted are dimensionless rescaled
   masses, as discussed in Sect.~II.
} 
\label{ScalSpec}
\vskip 0.3 truein
\centerline{
   \epsfxsize 3.0 truein \epsfbox {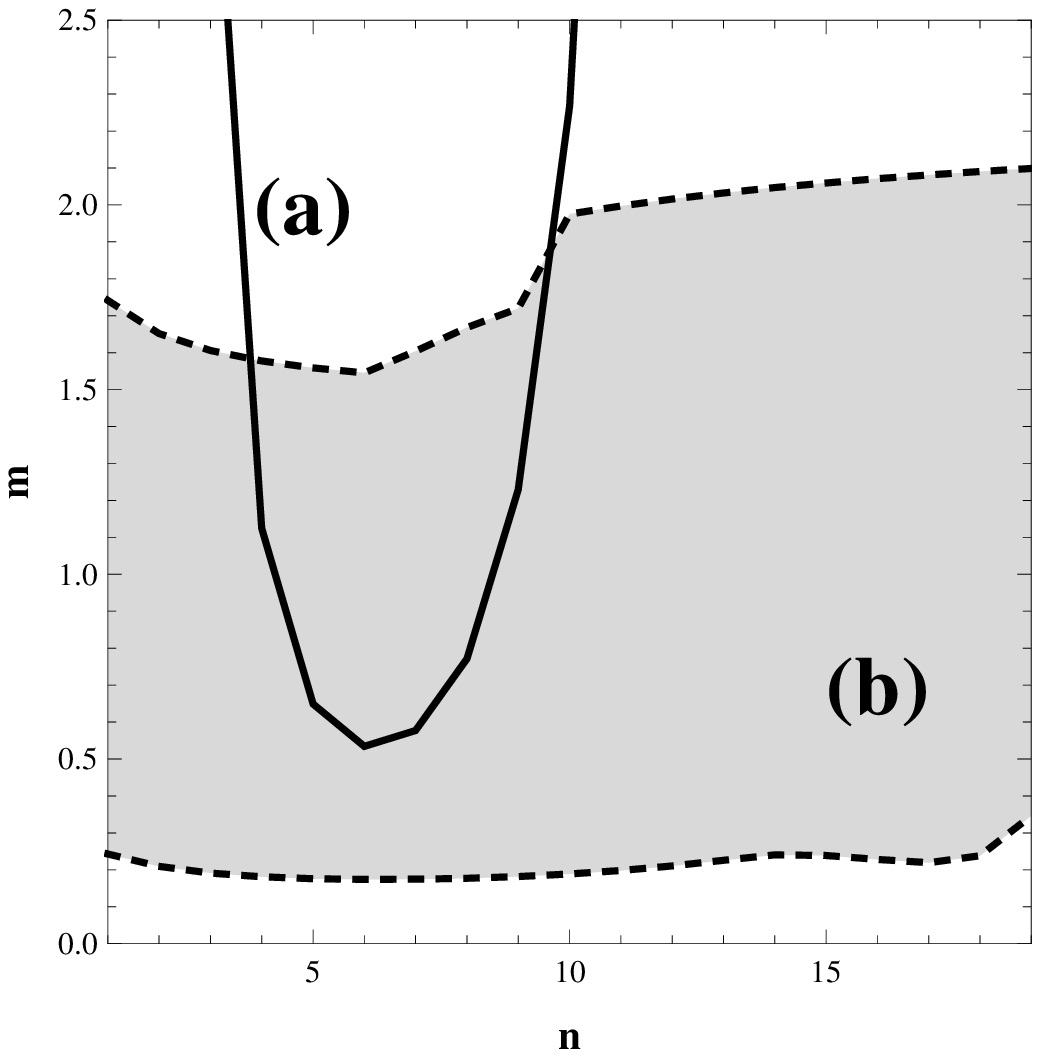} 
 }
\vskip -0.2truein
\caption{ The spectrum of massive fermions,
  plotted as a function of the vacuum index $n$
   for $N=20$, $\chi=1/4$, and $\lambda=\lambda_{20}^{\ast}\approx 1.05\times 10^8$.
    (a)  The solid line indicates the single
    fermionic ``coupler'', while (b) the dotted lines demarcate the 
    edges of the the shaded band within which the masses of the remaining
    fermions lie. 
   As in Fig.~\ref{ScalSpec}, the masses plotted are dimensionless rescaled
   masses discussed in Sect.~II.  } 
\label{FermSpec}
\end{figure}  
%======================================================================== 

%================== FIGURE ============================================
\begin{figure}[thb!]
\centerline{
   \epsfxsize 3.0 truein \epsfbox {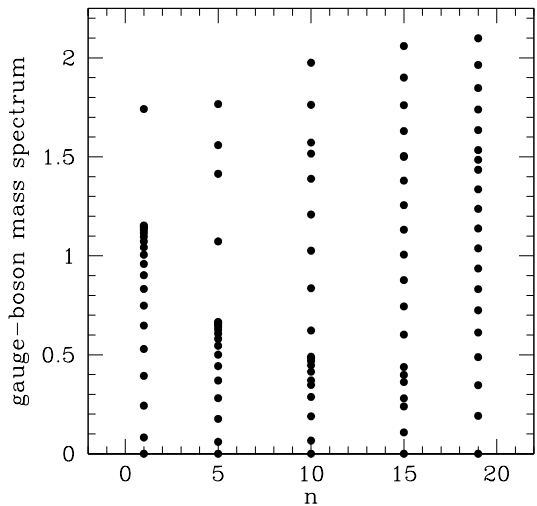} 
 }
\caption{ Gauge-boson mass spectra, plotted for the $n=\{1,5,10,15,19\}$ vacua.
    For this plot, we have taken $N=20$ and $\chi=1/4$, and the (rescaled) masses 
   are displayed on the vertical axis.  Note that when $n$ is small, many of
   the masses tend to cluster
   around a particular value.  As $n$ increases, however, this value drops;
   fewer of the masses are clustered around this value, and the masses
   of the remaining gauge bosons become more widely spaced.  
   By contrast, the mass of the lightest massive gauge boson increases with
   increasing $n$.  Thus, as our system tumbles down the vacuum tower,
   the lightest non-zero gauge boson tends to become increasingly heavy.}
\label{spectrum2}
\end{figure}  
%======================================================================== 

To help illustrate these features, let us consider the spectra of
scalars and fermions
that arise for the parameter assignments $N=20$, $\chi=1/4$, 
and $\lambda=\lambda_{20}^{\ast}\approx 1.05\times 10^8$.
These scalar and fermionic spectra are shown in Figs.~\ref{ScalSpec} 
and \ref{FermSpec}, respectively.
In each plot, the masses of the ``coupler''
fields are indicated by the solid curves.  By contrast, the minimum and maximum masses 
of the remaining fields are indicated by dashed lines, and the spectral 
``band'' of masses which they demarcate has been shaded.  

These figures highlight several significant features of the particle spectra.  
First, it is evident from these figures that the masses of the couplers
are highly dependent on the choice of vacuum.  {\it In particular, 
it is the lightest coupler whose mass vanishes and then 
becomes tachyonic when a given vacuum is destabilized.}\/  
In this particular example, $\lambda$ has 
been set to the critical value $\lambda_{20}^{\ast}$.
Since $\chi=1/4$, we see from Fig.~\ref{criticalns}
that it is the $n=6$ vacuum which is destabilized at this value of $\lambda$,
and hence it is in the $n=6$ vacuum that the lightest coupler becomes massless
for this choice of parameters.
By contrast, the other vacua near the top and bottom of the vacuum tower
are more comfortably stable for this value of $\lambda$.
As a result, the masses of the coupler fields become quite large
for these vacua.  

The situation is quite different for the non-coupler fields. 
For these fields,
we see from Figs.~\ref{ScalSpec} and \ref{FermSpec} that
the boundaries of the non-coupler spectral band 
remain roughly constant as one transitions from vacuum to vacuum.  
However, it will be noted that both the width of the band and the mass 
of the lightest particle in it increase slightly when $n$ is near $N-1$.  
It is also apparent from these diagrams that the lightest massive particles 
in any particular vacuum are generally the lightest non-coupler scalar 
and the lightest massive gauge field, both with precisely the same mass.    

Figs.~\ref{ScalSpec} and \ref{FermSpec} are calculated for $\lambda$
sitting precisely at the critical value $\lambda_{20}^\ast$.
However, it is easy to see what happens as we increase $\lambda$:
the solid curves corresponding to the coupler masses rescale with $\lambda$,
while the shaded bands corresponding to the non-coupler masses
remain invariant.

It is important to note that although the maximum and minimum of the
non-coupler ``bands'' in  Figs.~\ref{ScalSpec} and \ref{FermSpec}
remain roughly constant as a function of vacuum index $n$,
there nevertheless exists a rich $n$-dependence for the masses of the
individual states within the band.
This is shown in Fig.~\ref{spectrum2} for the gauge-boson spectrum.
   When $n$ is small, many of
   the gauge-boson masses tend to cluster
   around a particular value.  As $n$ increases, however, 
   the masses of the remaining gauge bosons becomes more widely spaced.
   By contrast, the mass of the lightest massive gauge boson increases with
   increasing $n$.  Thus, as our system tumbles down the vacuum tower,
   the lightest non-zero gauge boson tends to become increasingly heavy.

Given these results, several intriguing phenomenological possibilities emerge.
As we have seen, for any vacuum in our metastable vacuum tower,
there is only one physical, massless field:  this is the gauge boson in
Eq.~(\ref{Bmu}), associated with the single remaining unbroken $U(1)'$ gauge group.  
If this $U(1)'$ gauge boson resides in a hidden sector --- 
or if it is identified with the photon --- there should be 
no difficulties in making this model compatible with present experimental observations.   
In particular, if we imagine that all of our broken $U(1)$'s are
hidden or broken at mass scales which exceed accessible energies,
then only the coupler fields will be readily observable.
Since the masses of these fields scale linearly with $\lambda$,
we can easily adjust them to be unobservable as well.
Thus, the presence of such large numbers of fields need not 
necessarily pose phenomenological difficulties.

On the other hand, it remains true that
the spectrum contains a large number of massive gauge 
bosons, scalars, and Dirac fermions whose properties depend on the 
particular vacuum state in question.  If such states are observable,
such a spectrum could lead to 
a rich and interesting phenomenology, with a variety of potential 
implications for both collider physics and cosmology.

%======================================================================================================
\section{Degenerate Vacua and Bloch Waves\label{sec:BlochWave}}

Thus far, we have analyzed our model in the range 
$0< \chi< 1/2$, and investigated
the vacuum towers and tunnelling dynamics which result.
Indeed, each successive vacuum in the tower has a lower energy than
the previous one, and consequently there exists a net direction
for dynamical flow.

All of this changes, however, if we consider the endpoint where $\chi=1/2$.
Note from Fig.~\ref{fig:ChiMax}
that for any $N$, the point at which $\chi$ is strictly equal to $1/2$
is still within our set of allowed values of $\chi$. 
In this case, we find that $R_n=2$ for all $n$,
and the expressions for the vacuum energies $V_n$ and saddle-point energies $V_{n,n'}$
become independent of their vacuum indices $n,n'$.
As a result, the
vacuum structure
in the asymptotic-$\lambda$ regime discussed in Sect.~II 
consists of $N-1$ {\it degenerate}\/ 
vacua with energies $\langle V_n \rangle= 1/2$ for all $n$.
These in turn are 
 separated from each other by a 
set of equivalent, saddle-point potential barriers of 
uniform height $\langle V_{n,n'} \rangle= 2/3$ for all $n,n'$.
Furthermore, when $\chi=1/2$, we find that the field-space
configurations for our vacua and saddle points in Eqs.~(\ref{vacsolns}) and (\ref{saddlesolns})
respectively become essentially identical,
differing from one another only insofar as a different  VEV is set to zero in each case.
This implies that the field-space distance between any pair of vacua is equal to 
that between any other such pair.

\begin{figure}[thb!]
\centerline{
   \epsfxsize 3.0 truein \epsfbox {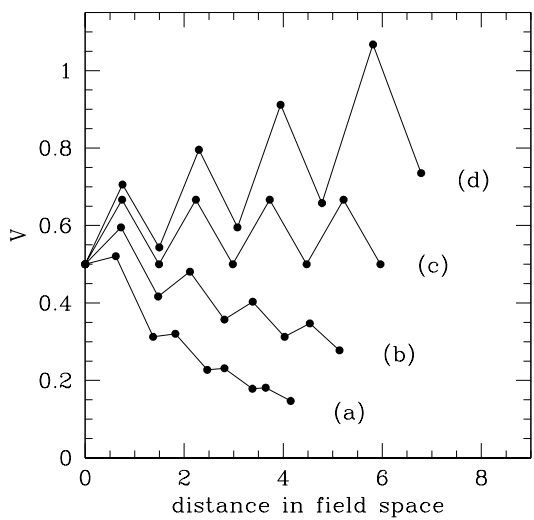}
 }
\caption{
Vacuum structure in the $N=6$ model, 
plotted for (a) $\chi=0.2$, (b) $\chi=0.4$, 
(c) $\chi=0.5$, and (d) $\chi=0.54$.
As $\chi$ increases from zero, we see that
the ``slope'' of our vacuum ``staircase''
decreases, ultimately becoming completely flat 
at $\chi=0.5$.  For $\chi>0.5$, the vacuum staircase
inverts, with the lightest states now becoming the
heaviest states.
 } 
\label{fig:StairsChi}
\end{figure}

This behavior is shown in Fig.~\ref{fig:StairsChi}, where we plot
the vacua and saddle points of our vacuum towers as a function of $\chi$.
For $\chi<1/2$, we see that our vacuum tower ``staircase'' has a non-zero
slope.  However, as $\chi\to 1/2$, this slope becomes less and less until
our vacuum tower becomes degenerate,
with all vacua and saddle points having equal energies and distances in 
field space. 
This situation is shown in Fig.~\ref{fig:StairsChi}(c).

%================== FIGURE ============================================
%======================================================================== 

Taken together, these results imply that  
that the transition rates between all vacua
in the ``tower'' are identical for $\chi=1/2$.
As a result, all vacuum transitions in this setup will occur with identical rates,
and it is no longer appropriate to employ the classical approximations discussed
at the beginning of Sect.~III in which we assume that only one decay channel
dominates the tunneling dynamics.  Instead, our system must be treated quantum-mechanically.

The result, however, is clear:  the true ground state of such a theory is no
longer any of the individual $n$-vacua by itself.  Instead, what results is an
infinite set of {\it Bloch waves}\/ across the entire set of degenerate vacua.
Moreover, the vacuum energies associated with such Bloch waves fill out a continuous
band.  As a result, the vacuum energy of the true ground state of the theory
will be smaller than the vacuum energy of any individual vacuum.     

The phenomenological implications of such a Bloch-wave vacuum structure
will be discussed in more detail in Ref.~\cite{toappear}.
Nevertheless, it is interesting to speculate that the vacuum energy of this
true ground state might actually vanish.
If this is possible, we would have a situation in which our Bloch-wave
vacuum structure actually {\it restores}\/ supersymmetry. 
Another possibility is that the vacuum energy of this true ground
state only {\it approaches}\/ zero as $N\to\infty$.  In this case,
one could potentially obtain a ground state with a very small cosmological
constant, in a manner reminiscent of the proposal in Ref.~\cite{Gordy}.
These and other issues will be explored more fully in Ref.~\cite{toappear}.

Needless to say, the existence of such Bloch-wave ground states is yet another
example of a non-trivial vacuum structure associated with the moose.
In some sense, the ``translational''  symmetry 
that shifts us from one degenerate vacuum to the next is nothing but a reflection
of the underlying translational symmetry of the original moose.
We caution, however, that it is not merely {\it any}\/ moose which gives rise
to this sort of band structure;  it is only one with a precise value for
the kinetic mixing between nearest-neighbor $U(1)$ gauge factors, as
well as non-trivial Fayet-Iliopoulos terms.  
Together, it is these ingredients which conspire to produce a vacuum 
structure consisting of an infinite number of degenerate vacua, 
each of which is stable (lacking either tachyonic or flat directions),
with a non-zero transition probability between them.
However, once these features are achieved,
the ensuing Bloch-wave vacuum structure is inevitable.

%================================================================================================

\section{Discussion\label{sec:Conclusion}}

In this paper, we have presented explicit examples of two new highly
non-trivial vacuum structures that can arise in supersymmetric
field theories.  In particular, we have shown that when kinetic mixing   
among its $U(1)$ gauge groups is permitted, an $N$-site Abelian  
moose construction gives rise to a tower of $N-2$ metastable vacua,
each of which involves nonzero VEVs for all but two of the ``link'' fields 
on the moose, in   addition to a stable ground state. 
As $N$ is increased, the   
energies of the existing vacua remain unchanged, while new vacua 
appear with smaller and smaller energies, the lowest of which
asymptotically approaches zero as $N\rightarrow\infty$.  We investigated
the dynamics associated with transitions between these vacua, and found that
different regions of the vacuum tower could manifest very different instanton-induced
vacuum-decay patterns.  These include ``collapse'' regions, in which all vacua decay   
directly to the ground state in a particular order;  ``cascade'' regions, in
which a potentially complicated decay chain from metastable vacuum to
metastable vacuum develops;  stable regions in which the vacuum states have
lifetimes exceeding cosmological time scales;  and regions into which instanton-induced
tunnelling cannot occur.    We also explored the particle   
spectra associated with these vacua, and showed that in each vacuum state there exists
only one physical, massless mode --- the gauge boson associated with the
lone unbroken $U(1)$ in that vacuum --- along with a tower of massive   
gauge bosons, scalars, and Dirac fermions.  The massless state is coupled
to the rest of the states in the tower only by highly-suppressed
interactions involving extremely heavy particles, and thus can be viewed
as part of a decoupled hidden sector.

Needless to say, there are many additional avenues through
which additional properties of our model might be explored.
For example, we have already seen in Sect.~V that the $\chi\to 1/2$
limit of our general framework produces a Bloch-wave structure
for the true ground state of the theory.  This is another unique
vacuum structure which has not been explored in the literature
thus far, but which clearly emerges in models of this type.
Models exhibiting Bloch-wave ground states can be expected
to have phenomenologies which differ markedly from those based
on single vacuum states, much as the strong CP problem of QCD  
is a feature uniquely associated with the QCD $\theta$-vacuum.
A detailed examination of the implications of such a vacuum
structure is forthcoming~\cite{toappear}.

Another salient issue worthy of investigation
concerns the {\it thermal}\/ properties of such an infinite tower
of metastable vacua.
In this connection, there are two issues which are of paramount
importance:  that of initially populating the landscape, and that
governing its decay patterns.  

The former issue is
critical for determining whether the
universe tends to start out in the cascade region near the
top of the tower, in the collapse region near the middle of the tower, or at the bottom,
at or near the ground state.  Broadly speaking, vacua which contain more
light degrees of freedom tend to be preferred in a thermal context.
Occasionally, this means that the universe is far more likely to end up in
a metastable state than in the ground state of a given theory.  In the scenarios
discussed in Ref.~\cite{ISS},
for example, it is has been shown~\cite{ThermalISS} that thermal effects
prefer the metastable vacuum.  By contrast, in our model, the number of light
degrees of freedom in any given vacuum along the tower is essentially
vacuum-independent.
Thus it is not {\it a priori}\/ obvious into which
vacuum state the universe would prefer to settle, or whether a (potentially large)
number of states would emerge as equally likely candidates.  
This would be an interesting area for future research.

The latter issue is also extremely
important, for in this paper we have limited our attention
to vacuum decays which proceed through instanton-induced tunneling.
While this is indeed the whole story at zero temperature, finite-temperature
scenarios contain additional {\it sphaleron}\/-like processes through 
which such vacua might also decay.  This has the potential to
significantly modify the vacuum dynamics we discussed in Sect.~III.
For example, while the Forbidden City cannot be entered
through instanton-tunnelling transitions, it can nevertheless be
entered through thermal excitations.  

A final question worth exploring concerns the theoretical interpretation of
our model.  As we have seen in Sect.~II, our model has as its core an $N$-site
moose theory of a sort which is familiar to deconstructionists~\cite{Decon}. 
This suggests that our model should have a natural five-dimensional interpretation
in the $N\to\infty$ limit.
However, our moose theory is complicated by the fact that 
we must introduce non-zero kinetic mixings between nearest-neighbor $U(1)$'s
in order to achieve vacuum stability.  One wonders, then, about the extent
to which this modifies our previous five-dimensional interpretation.
One possible clue comes from the fact that we can ``rotate'' our $U(1)$ basis
in such a way as to eliminate the kinetic mixing entirely;  this occurs at the
cost of introducing three or more non-zero $U(1)$ charge assignments 
for each of our chiral superfields as well as the introduction of non-zero ``bulk''
Fayet-Iliopoulos terms which are located off the moose endpoints.
In general, it is known~\cite{WarpedDecon} that introducing non-uniform gauge charges
along the moose is one way of realizing warped or other non-trivial geometries. 
It therefore remains to be seen which, if any, warped five-dimensional geometry 
might effectively emerge from our model
in the $N\to\infty$ limit.
This could be important for understanding the set of possible
UV completions of our model.
Indeed, some of these completions might also potentially include gravitational effects,
as have been considered in other metastable models~\cite{DDG,Lalak},

Clearly, there are also a number of possible applications for a vacuum 
structure of this sort.  

Perhaps the application which most immediately springs to mind concerns
a potential solution to the cosmological-constant problem.
Over the past decade, several scenarios have been proposed 
in which a small cosmological constant emerges as a consequence of a large 
number of vacua~\cite{BoussoPolchinski,banks,Gordy,tye}.  Scenarios of this sort 
tend to posit the existence of a ``landscape'' of vacua with certain gross 
properties, including a vacuum state whose energy is nearly vanishing.   
One then imagines that the universe either dynamically tumbles down
to this special vacuum state, or is somehow born there. 

However, to the best of our knowledge, no explicit model with a vacuum 
structure exhibiting such properties has ever been constructed.  
Moreover, most of the existing scenarios have been proposed in the context of 
the string-theory landscape, where the different vacua correspond not to 
different minima of the same theory, but to separate theories characterized 
by distinct parameter assignments, gauge structures, particle contents, {\it etc}\/.  
In such a context, it is not clear that the instanton methods of 
Ref.~\cite{BrownTeitelboim} apply.
In fact, it is not at all obvious how (or even if) 
transitions between models can occur in such a framework.   As a result, it is not clear
how the landscape of vacua can be populated even qualitatively, much less quantitatively. 

By contrast, the towers of metastable vacua in our model 
exist at a single point in parameter space --- that is, for a 
single choice of $\chi$, $\xi$, $M$, and $\lambda$ --- and correspond 
to different vacuum states {\it within the same theoretical model}\/.  
The method by which transitions 
occur between one vacuum state of the theory and another is therefore 
well understood, both at zero temperature and at finite temperature.  
As a result, reliable statements can be made concerning both the initial, statistical 
population mechanism for these vacua as well as the dynamics associated with 
transitions between them.  

Regardless of the particular scenario envisioned, it is important to note
that a solution to the cosmological-constant problem requires not only 
a small energy for the vacuum we inhabit, but
also that the additional particles present in the model
be thus far experimentally observable.
Thus, either the masses
of all additional particles appearing in that particular vacuum must be  
heavy enough to have thus far avoided detection, or else these additional
fields must decouple from those of the Standard Model and form a hidden 
sector.  Whether the former criterion is met in any given vacuum depends 
sensitively on the values of our underlying model parameters.

It is also important to note that we can achieve a true ground state with
nearly vanishing vacuum energy only by fine-tuning the parameter $N$ 
in our model.  Specifically, if we wish our ground state to have  a
very small vacuum energy, we will require a large number of sites on the moose.
This is not surprising, since we cannot expect to have a fine-tuned
ground state without introducing our fine-tuning in some other fashion.
Indeed, while most previous scenarios merely posit a large number of vacua
as an initial condition for obtaining a small cosmological constant,
we are explicitly realizing this large number of vacua as the result of
a different, equally large number:  the number of $U(1)$ gauge factors
in an underlying moose. 
However, we stress that this is a relatively small price to pay,
since we are obtaining a calculable tower of metastable vacua
in the process, each of which is free of both tachyonic and 
flat directions, along with a true, stable ground state with the
small vacuum energy we desire.

There are other potential applications of our model as well.
For example, one of the major thrusts in recent string-theoretic research
has been a statistical study~\cite{douglas} of the string landscape~\cite{Susskind}.
Through such statistical studies, one hopes to uncover hidden
correlations which may be taken as predictions from (or evidence of)
an underlying string structure at high energy scales. 
However, one important issue that needs to be addressed in this
context concerns the proper definition of a {\it measure}\/ 
across the landscape:  in what manner are the different string 
theories to be weighted relative to each other?
Clearly, the most na\"\i ve approach is to count each string
model equally, interpreting each as contributing a single vacuum
state to the landscape as a whole.  However, it is entirely natural
that moose theories of the sort we have been examining in this paper
can appear as the actual low-energy (deconstructed) limits of 
flux compactifications~\cite{DDG}, and as we have seen, such 
theories give rise to infinite towers of metastable vacua.
Indeed, for certain choices of the underlying parameters in 
these models, literally all of these vacua may be rendered
stable on cosmological time scales.  The question then arises
as to whether such theories should be weighted according to
the infinitely large number of vacua which they contribute
to the landscape as a whole.  In fact, following this line
further, it becomes crucial to determine whether the landscape
measure should be defined in terms of different {\it theories}\/, or in
terms of the different {\it vacua}\/ they contain.
Indeed, if the true underlying landscape measure is based 
on vacua, then a theory with infinite towers of vacua is likely
to dominate any statistical study of the string landscape.
As such, the phenomenological properties of these sorts of models 
will dominate the properties of the landscape as a whole.

Another potential application of the model described here is as a
possible hidden sector in a fully developed
model of supersymmetry breaking. 
The advantage of this would be that a
hierarchy between the Planck scale and the supersymmetry-breaking scale
could arise dynamically, as a consequence of vacuum tumbling dynamics along the
tower.  In order for this to be viable, however, a number of
phenomenological issues would need to be addressed.  For example, while
supersymmetry is indeed broken in each vacuum state in the tower, it is
broken only by background values for the $D$-terms associated with the
various $U(1)$ gauge groups in the model.  Consequently, R-symmetry is
left unbroken in each vacuum.  If the dominant source of
supersymmetry-breaking is to come from a sector of this sort, then the
introduction of additional matter will be required in order to obtain
realistic masses for the gaugino superpartners of the Standard-Model gauge
fields.  Likewise, it is possible that an $F$-term component to supersymmetry
breaking could be engineered via a a modification of this scenario to
include additional, vector-like matter in a manner similar to that
discussed in Ref.~\cite{Nest}.

Needless to say, these are only some of the many applications such
a vacuum structure might have.
There exist, however, numerous additional possibilities.
For example, because our construction relies directly on the presence of extra
$U(1)$ gauge symmetries, the low-energy limit of our setup could
have significant implications for $Z'$ phenomenology.
Indeed, if the coupling between our setup and the Standard Model 
is properly engineered,
the collider signatures of such a scenario could be rather
dramatic.  Note that preliminary analyses of the mass spectra associated
with each vacuum in the tower can be found in Sect.~IV.

Likewise, this scenario could also have a number of astrophysical
and cosmological implications.
The vacuum cascade we have discussed in Sect.~III
involves numerous first-order phase transitions,
and is therefore likely to generate
topological defects.
In particular,
given that our model contains numerous Abelian gauge factors,
there is a specific likelihood of generating
a network of cosmic strings~\cite{cosmicstrings}.
Moreover, different regions of the universe could potentially
exist in different vacuum states along the tower, giving rise
to domain walls.  It therefore becomes an important phenomenological
question as to how the constraints associated with such domain
walls might be alleviated.

In summary, then, it is clear that a number of potential 
extensions and 
applications exist for the new vacuum structures presented in this work.  
%  This is not surprising, since (as we indicated in the Introduction)
%  the structure of the vacuum of any model plays a critical role in determining
%  the resulting phenomenology.
Indeed, as the poet William Carlos Williams might well have written,
 {\tt
 \begin{verse}

 So much depends\\
 upon \\
 \bigskip
 the vacuum\\
 structure \\
 \bigskip
 glazed with\\
 excitations\\
 \bigskip
 forming white\\
 chickens.~\cite{Wheelbarrow}\\

\end{verse}
  }

%============================================================================= 
\section*{Acknowledgments}
This work was supported in part
by the Department of Energy under Grant~DE-FG02-04ER-41298.
We are happy to thank E.~Dudas,
T.~Gherghetta, G.~Shiu, and U.~van~Kolck 
for discussions.
Moreover, for Great Walls and Forbidden Cities, 
there is nothing better than the real thing.  
We are therefore also delighted to thank the new Kavli Institute
for Theoretical Physics (KITPC) in Beijing, China, for gracious 
hospitality, an excellent research environment, and 
abundant inspiration of Olympic proportions during the completion of this work.  
Long may she prosper!
%============================================================================= 

\end{document}